\RequirePackage{fix-cm}
\documentclass[smallextended,natbib]{svjour3}       % onecolumn (second format)
        \smartqed  % flush right qed marks, e.g. at end of proof

        \usepackage{scalerel}
        \usepackage{tikz}
        \usetikzlibrary{svg.path}

        \definecolor{orcidlogocol}{HTML}{A6CE39}
        \tikzset{
          orcidlogo/.pic={
            \fill[orcidlogocol] svg{M256,128c0,70.7-57.3,128-128,128C57.3,256,0,198.7,0,128C0,57.3,57.3,0,128,0C198.7,0,256,57.3,256,128z};
            \fill[white] svg{M86.3,186.2H70.9V79.1h15.4v48.4V186.2z}
                         svg{M108.9,79.1h41.6c39.6,0,57,28.3,57,53.6c0,27.5-21.5,53.6-56.8,53.6h-41.8V79.1z M124.3,172.4h24.5c34.9,0,42.9-26.5,42.9-39.7c0-21.5-13.7-39.7-43.7-39.7h-23.7V172.4z}
                         svg{M88.7,56.8c0,5.5-4.5,10.1-10.1,10.1c-5.6,0-10.1-4.6-10.1-10.1c0-5.6,4.5-10.1,10.1-10.1C84.2,46.7,88.7,51.3,88.7,56.8z};
          }
        }

        \newcommand\orcidicon[1]{\href{https://orcid.org/#1}{\mbox{\scalerel*{
        \begin{tikzpicture}[yscale=-1,transform shape]
        \pic{orcidlogo};
        \end{tikzpicture}
        }{|}}}}

        \usepackage{hyperref}
        \usepackage{array, pbox}
        \usepackage{mathtools}
        \usepackage{caption}
        \usepackage{subcaption}
        \usepackage{graphicx}

        \usepackage{CJKutf8}

        \usepackage[page, title]{appendix}

        \usepackage{siunitx}

        \usepackage{ntheorem}
        \theoremseparator{:}
        \newtheorem{rsq}{Research Question}

        \makeatletter
        \newcounter{subrsq} 
        \let\savedc@rsq\c@rsq
        \newenvironment{subrsq}
         {%
          \setcounter{subrsq}{0}%
          \stepcounter{rsq}%
          \edef\saved@rsq{\thersq}% Save the current value of rsq
          \let\c@rsq\c@subrsq     % Now rsq is subrsq
          \renewcommand{\thersq}{\saved@rsq\alph{rsq}}%
         }
         {}
        \newcommand{\normrsq}{%
          \let\c@rsq\savedc@rsq % revert to the old one
          \renewcommand\thersq{\arabic{rsq}}%
        } 
        \makeatother

        \usepackage{marvosym}

        \newcommand*{\affaddr}[1]{#1} % No op here. Customize it for different styles.
        \newcommand*{\affmark}[1][*]{\textsuperscript{#1}}

        \usepackage{multirow}
        \usepackage[normalem]{ulem}
        \useunder{\uline}{\ul}{}

        \usepackage{lscape}

        \journalname{Information Technology \& Tourism}

\begin{document}

\title{Differences in Chinese and Western tourists faced with Japanese hospitality: A natural language processing approach}

% Fold 1
  % Fold 2
    % Fold 3
      % Fold 4
        \titlerunning{Differences in Chinese and Western tourists faced with Japanese hospitality: ...}

        \author{Elisa Claire Alem\'an Carre\'on \protect\affmark[1] \orcidicon{0000-0002-6437-0866} \and
                Hugo Alberto Mendoza Espa\~na                                   \protect\affmark[1] \and
                Hirofumi Nonaka                                                 \protect\affmark[1] \and
                Toru Hiraoka                                                    \protect\affmark[2]
        }

        \authorrunning{E. Alemán Carreón et al.}

        \institute{\Letter \hspace*{0.16em} Elisa Claire Alem\'an Carre\'on  \at
                    \hspace*{1em} \email{elisa.claire.aleman.carreon@gmail.com} \\
                    \hspace*{1em} ORCID: 0000-0002-6437-0866
                \and
                    \hspace*{1em} Hugo Alberto Mendoza Espa\~na \at
                    \hspace*{1em} \email{mendoza.espana@gmail.com}
                \and
                    \hspace*{1em} Hirofumi Nonaka \at
                    \hspace*{1em} \email{nonaka@kjs.nagaokaut.ac.jp}
                \and
                    \hspace*{1em} Toru Hiraoka \at
                    \hspace*{1em} \email{hiraoka@sun.ac.jp}
                \and 
                    \affaddr{\affmark[1] \hspace*{0.15em} Nagaoka University of Technology, Nagaoka, Japan}
                \at 
                    \affaddr{\affmark[2] \hspace*{0.15em} University of Nagasaki, Nagasaki, Japan}
        }

        \date{Received: date / Accepted: date}

        \maketitle

\begin{abstract}

  Since culture influences expectations, perceptions, and satisfaction, a cross-culture study is necessary to understand the differences between Japan's biggest tourist populations, Chinese and Western tourists. 
  However, with ever-increasing customer populations, this is hard to accomplish without extensive customer base studies. There is a need for an automated method for identifying these expectations at a large scale. For this, we used a data-driven approach to our analysis.
  Our study analyzed their satisfaction factors comparing soft attributes, such as service, with hard attributes, such as location and facilities, and studied different price ranges.
  We collected hotel reviews and extracted keywords to classify the sentiment of sentences with an SVC. We then used dependency parsing and part-of-speech tagging to extract nouns tied to positive adjectives.
  We found that Chinese tourists consider room quality more than hospitality, whereas Westerners are delighted more by staff behavior. Furthermore, the lack of a Chinese-friendly environment for Chinese customers and cigarette smell for Western ones can be disappointing factors of their stay. 
  As one of the first studies in the tourism field to use the high-standard Japanese hospitality environment for this analysis, our cross-cultural study contributes to both the theoretical understanding of satisfaction and suggests practical applications and strategies for hotel managers.


\end{abstract}

\section{Introduction}\label{intro}

  Inbound international tourism has been increasingly affecting Japanese economy \cite[][]{jones2009}. A year-on-year growth rate of 19.3\% was observed in 2017, with \num[group-separator={,}]{28691073} inbound tourists \cite[][]{jnto2003-2019}. 

  Japan’s hospitality has been known historically to be of the highest quality. \textit{Omotenashi}, which describes the spirit of Japanese hospitality, with roots in Japanese history and tea ceremony, is celebrated worldwide \cite[][]{al2015characteristics}. Consequently, it would stand to reason that tourists visiting Japan would have this hospitality as their first and foremost satisfaction factor. However, it is known that customers from different countries and cultures have different expectations \cite[][]{engel1990}. Thus, it could be theorized that their satisfaction factors should be different. 

  The Japanese tourist market is gradually becoming diverse because of multicultural tourist populations. This diversity means that the expectations when staying at a hotel will be varied. Cultural backgrounds have a decisive role in aspects of satisfaction and in the perceptions of quality \cite[][]{mattila1999role,winsted1997service}, or behavioral intentions \cite[][]{liu2001relationships}, such as the difference in Westerners and Asians in their willingness to pay more \cite[][]{LEVY2010319}. A difference in cultural background can also heavily influence customers' expectations, as well as their perceptions of quality, and the difference between these two is what expresses itself as satisfaction. This difference in expectations and perceptions of quality can be smaller or larger depending on the culture in reaction to the same service. 

  For a growing industry with increasing cultural diversity, it is essential to identify the cross-culture expectations of customers in order to provide the appropriate services, cater to these expectations to ensure and increase customer satisfaction, maintain a good reputation, and generate positive word-of-mouth. 

  In 2017, Chinese tourists accounted for 25.63\% of the tourist population. On the other hand, Western countries accounted for 11.4\% of the total, and 7.23\% were countries where English is the official or the de facto national language \cite[][]{jnto2003-2019}. The effect of Chinese tourists on international economies is increasing, along with the number of studies on this phenomenon, \cite[][]{sun2017}. Despite this, many tourist-behavior analyses have been performed only involving Western subjects. Yet, it is known that Western and Asian customers are heavily differentiated\cite[][]{LEVY2010319}. As such, a knowledge gap existed until recent decades. Considering the numbers of inbound tourists in Japan and our team's language capabilities, our study focuses on Western and Chinese tourists. 

  In studies involving Asian populations in the analysis, Chinese-tourist behaviors have been evaluated most commonly \cite[e.g.][]{liu2019, chang2010, dongyang2015}. The few studies reporting comparisons between Asian and Western tourists’ behaviors \cite[e.g.][]{choi2000} are typically survey- or interview-based, using small samples. These studies, although valid, can have limitations, namely, the scale and sampling. In the past, survey-based studies have provided a theoretical background for a few specific tourist populations of a single culture or traveling with a single purpose. These studies' limited scope often leads to difficulties in observing cultural and language differences in a single study. This creates a need for large-scale cross-cultural studies for the increasing Asian and Western tourist populations. It could be said that Westerners account for a smaller portion of the tourist population compared to Asians. However, according to \cite{choi2000}, Westerners are known as ``long-haul'' customers, spending more than 45\% of their budget on hotels. In comparison, their Asian counterparts only spend 25\% of their budget on hotels. Therefore, it is essential to study Asian and Western tourist populations, their differences, and the contrast with the existing literature results.

  However, with ever-increasing customer populations, this is hard to accomplish without extensive studies of the customer base. There is a need for an automated method for identifying these expectations at a large scale. Our study intends to answer the need for such a methodology utilizing machine learning and natural language processing of large amounts of data. For this, we used a data-driven approach to our analysis, taking advantage of hotel review data. With this methodology, we explore the expectations and needs for the two most differing cultures currently interacting with the hospitality industry in Japan.

  Owing to the advent of Web 2.0 and customer review websites, researchers realized the benefits of online reviews for research, sales  \cite[][]{ye2009, basuroy2003}, customer consideration \cite[][]{vermeulen2009} and perception of services and products \cite[][]{browning2013}, among other effects of online interactions between customers \cite[e.g.][]{xiang2010, ren2019}. Consequently, information collected online is being used in tourism research for data mining analysis, such as opinion mining \cite[e.g.][]{hu2017436}, predicting hotel demand from online traffic \cite[][]{yang2014}, recommender systems \cite[e.g.][]{loh2003}, and more. Data mining and machine learning technologies can increase the number of manageable samples in a study from hundreds to hundreds of thousands. These technologies can not only help confirm existing theories but also lead to finding new patterns and to knowledge discovery \cite[][]{fayyad1996data}. 

  In this study, we evaluate the satisfaction factors of two essential tourist populations that are culturally different from Japan: Chinese and Western tourists. We take advantage of the wide availability of online reviews of Japanese hotels by both Mainland Chinese tourists posting on \textit{Ctrip} and Western, English-speaking tourists posting on \textit{TripAdvisor}. Based on these data, we can confirm existing theories regarding the differences in tourists’ behavior and discover factors that could have been overlooked in the past. We use machine learning to automatically classify sentences in the online reviews as positive or negative opinions on the hotel. We then perform a statistical extraction of the topics that most concern the customers of each population.

\section{Research objective}\label{research_objective}

  With the knowledge that cultural background influences expectations in customers, which is the basis for satisfaction, it becomes important to know the difference in factors influencing satisfaction and dissatisfaction between the most differing and numerous tourist populations in a given area. 

  This study aims to determine the difference in factors influencing satisfaction and dissatisfaction between Chinese and English-speaking tourists in the context of high-grade hospitality of Japanese hotels across several price ranges. We use machine learning to classify the sentiment in texts and natural language processing to study commonly used word pairings. More importantly, we also intend to measure how hard and soft attributes influence customer groups' satisfaction and dissatisfaction. We define hard attributes as attributes relating to physical and environmental aspects, such as the hotel's facilities, location, infrastructure, and surrounding real estate. In contrast, soft attributes are the hotel's non-physical attributes related to services, staff, or management.

\section{Theoretical background and hypothesis development}\label{theory_hypothesis}

  \subsection{Cultural influence in expectation and satisfaction}\label{theory_expectations}

    Customer satisfaction in tourism has been analyzed since decades past, \cite{hunt1975} having defined customer satisfaction as the realization or overcoming of expectations towards the service. \cite{oliver1981} defined it as an emotional response to the provided services in retail and other contexts, and \cite{oh1996} reviewed the psychological processes of customer satisfaction for the hospitality industry. It is generally agreed upon that satisfaction and dissatisfaction stem from the individual expectations of the customer. As such, \cite{engel1990} states that each customer's background, therefore, influences satisfaction and dissatisfaction. It can also be said that satisfaction stems from the perceptions of quality in comparison to these expectations. 

    These differences in customers' backgrounds can be summed up in cultural differences as well. In the past, satisfaction and perceived service quality have been found to be influenced by cultural differences \cite[e.g.][]{mattila1999role, winsted1997service}. Service quality perceptions have been studied via measurements such as SERVQUAL \cite[e.g.][]{armstrong1997importance}. 

    Previous studies on the dimensions of culture that influence differences in expectations have been performed in the past as well \cite[e.g.][]{MATTILA201910, LEVY2010319, donthu1998cultural}, such as comparing individualism vs. collectivism, high context vs. low context, uncertainty avoidance, among other factors. While culture as a concept is difficult to quantify, some researchers have tried to use these and more dimensions to measure cultural differences, such as the six dimensions described by \cite{hofstede1984culture}, or the nine dimensions of the GLOBE model \cite[][]{house1999cultural}.

    These cultural dimensions are more differentiated in Western and Asian cultures \cite[][]{LEVY2010319}. Our study being located in Japan, it stands to reason that the differences in expectations between Western tourists and Asian tourists should be understood in order to provide a good service. However, even though geographically close, Japanese and Chinese cultures are both very different when it comes to customer service. This is why our study focuses on the difference between Chinese and Western customers in Japan. The contrasting cultural backgrounds between Chinese and Western customers will lead to varying expectations of the hotel services, the experiences they want to have while staying at a hotel, and the level of comfort that they will have. In turn, these different expectations will determine the distinct factors of satisfaction and dissatisfaction for each kind of customer and the order in which they prioritize them. 

    Because of their different origins, expectations, and cultures, it stands to reason Chinese and Western tourists could have completely different factors to one another. Therefore, it could be that some factors do not appear in the other reviews at all. For example, between different cultures, it can be that a single word can express some concept that would take more words in the other language. Therefore, we must measure their differences or similarities at their common ground as well.

    \subsection{Customer satisfaction and dissatisfaction towards individual factors during hotel stay}\label{theory_satisfaction}

    We reviewed the importance of expectations in the development of satisfaction and dissatisfaction and the influence that cultural backgrounds have in shaping these expectations. This is true for overall satisfaction for the service as a whole, as well as individual elements that contribute to satisfaction.

    In this study, we study not overall customer satisfaction but the satisfaction and dissatisfaction that stem from individual-specific expectations, be they conscious or unconscious. For example, if a customer has a conscious expectation of a comfortable bed and a wide shower, and it is realized during their visit, they will be satisfied with this matter. However, suppose that same customer with a conscious expectation of a comfortable bed experienced loud noises at night. In that case, they can be dissatisfied with a different aspect, regardless of the satisfaction towards the bed. Then, the same customer might have packed their toiletries, thinking that the amenities might not include those. They can then be pleasantly surprised with good quality amenities and toiletries, satisfying an unconscious expectation. This definition of satisfaction does not allow us to examine overall customer satisfaction. However, it will allow us to examine the factors that a hotel can revise individually and how a population perceives them as a whole. In our study, we consider the definitions in \cite{hunt1975} that satisfaction is a realization of an expectation, and we posit that customers can have different expectations towards different service aspects. Therefore, in our study, we define satisfaction as the emotional response to the realization or overcoming of conscious or unconscious expectations towards an individual aspect or factor of a service. On the other hand, dissatisfaction is the emotional response to the lack of a realization or under-performance of these conscious or unconscious expectations towards specific service aspects.

    Studies on customer satisfaction \cite[e.g.][]{truong2009, romao2014, wu2009} commonly use the Likert scale \cite[][]{likert1932technique} (e.g. 1 to 5 scale from strongly dissatisfied to strongly satisfied) to perform statistical analysis of which factors relate most to satisfaction on the same dimension as dissatisfaction \cite[e.g.][]{chan201518, choi2000}. The Likert scale's use leads to correlation analyses where one factor can lead to satisfaction, implying that the lack of it can lead to dissatisfaction. However, a binary distinction (satisfied or dissatisfied) could allow us to analyze the factors that correlate to satisfaction and explore factors that are solely linked to dissatisfaction. There are fewer examples of this approach, but studies have done this in the past \cite[e.g.][]{zhou2014}. This method can indeed decrease the extent to which we can analyze degrees of satisfaction or dissatisfaction. However, it has the benefit that it can be applied to a large sample of text data via automatic sentiment detection techniques using artificial intelligence. 

  \subsection{Japanese hospitality and service: \textit{Omotenashi}}\label{theory_omotenashi}

    The spirit of Japanese hospitality, or \textit{Omotenashi}, has roots in the country’s history, and to this day, it is regarded as the highest standard \cite[][]{ikeda2013omotenashi, al2015characteristics}. There is a famous phrase in customer service in Japan: \textit{okyaku-sama wa kami-sama desu}, meaning ``The customer is god.'' Some scholars say that \textit{omotenashi} originated from the old Japanese art of the tea ceremony in the 16th century, while others found that it originates in the form of formal banquets in the 7th-century \cite[][]{aishima2015origin}. The practice of high standards in hospitality has survived throughout the years. Presently, it permeates all business practices in Japan, from the cheapest convenience stores to the most expensive ones. Manners, service, and respect towards the customer are taught to workers in their training. High standards are always followed to not fall behind in the competition. In Japanese businesses, including hotels, staff members are trained to speak in \textit{sonkeigo}, or ``respectful language,'' one of the most formal of the Japanese formality syntaxes. They are also trained to bow differently depending on the situation, where a light bow could be used to say ``Please, allow me to guide you.'' Deep bows are used to apologize for any inconvenience the customer could have faced, followed by a very respectful apology. Although the word \textit{omotenashi} can be translated directly as ``hospitality,'' it includes both the concepts of hospitality and service \cite[][]{Kuboyama2020}. This hospitality culture permeates every type of business with customer interaction in Japan. A simple convenience shop could express all of these hospitality and service standards, which are not exclusive to hotels.  

    It stands to reason that this cultural aspect of hospitality would positively influence customer satisfaction. However, in many cases, other factors such as proximity to a convenience store, transport availability, or room quality might be more critical to a customer.  In this study, we cannot directly determine whether a hotel is practicing the cultural standards of \textit{omotenashi}. Instead, we consider it as a cultural factor that influences all businesses in Japan. We then observe the customers' evaluations regarding service and hospitality factors and compare them to other places and business practices in the world. In summary, we consider the influence of the cultural aspect of \textit{omotenashi} while analyzing the evaluations on service and hospitality factors that are universal to all hotels in any country.

    Therefore, we pose the following research question:

    \begin{subrsq}
    \begin{rsq}
    \label{rsq:hospitality}
    To what degree are Chinese and Western tourists satisfied with Japanese hospitality factors such as staff behavior or service?
    \end{rsq}

    However, Japanese hospitality is based on Japanese culture. Different cultures interacting with it could provide a different evaluation of it. Some might be impressed by it, whereas some might consider other factors more important to their stay in a hotel. This point leads us to a derivative of the aforementioned research question:

    \begin{rsq}
    \label{rsq:hospitality_both}
    Do Western and Chinese tourists have a different evaluation of Japanese hospitality factors such as staff behavior or service?
    \end{rsq}
    \end{subrsq}

  \subsection{Customer expectations beyond service and hospitality}\label{theory_soft_hard}

    Staff behavior, hospitality and service, and therefore \textit{Omotenashi}, are all soft attributes of a hotel. That is, they are non-physical attributes of the hotel, and as such, they are practical to change through changes in management. While it is important to know this, it is not known if the cultural differences between Chinese and Western tourists also influence other expectations and satisfaction factors, such as the hard factors of a hotel.

    Hard factors are attributes uncontrollable by the hotel staff, which can play a part in the customers' choice behavior and satisfaction. Examples of these factors include the hotel's surroundings, location, language immersion of the country as a whole, or touristic destinations, and the hotel's integration with tours available nearby, among other factors. 

    Besides the facilities, many other aspects of the experience, expectation, and perception of the stay in a hotel can contribute to the overall satisfaction, as well as individual satisfactions and dissatisfactions. However, previous research focuses more on these soft attributes, with little focus on hard attributes, if only focusing on facilities \cite[e.g.][]{shanka2004, choi2001}. Because of this gap in knowledge, we decided to analyze the differences in cultures regarding both soft and hard attributes of a hotel.

    This leads to two of our research questions:

    \begin{subrsq}
    \begin{rsq}
    \label{rsq:hard_soft}
    To what degree do satisfaction and dissatisfaction stem from hard and soft attributes of the hotel?
    \end{rsq}

    \begin{rsq}
    \label{rsq:hard_soft_diff}
    How differently do Chinese and Western customers perceive hard and soft attributes of the hotel?
    \end{rsq}
    \end{subrsq}

    The resulting proportions of hard attributes to soft attributes for each population could measure how much the improvement of management in the hotel can increase future satisfaction in customers. 

  \subsection{Chinese and Western tourist behavior}\label{theory_zh_en}

    % Asians vs. Western behavior
    In the past, social science and tourism studies focused extensively on Western tourist behavior in other countries. Recently, however, with the rise of Chinese outbound tourism, both academic researchers and businesses have decided to study Chinese tourist behavior, with rapid growth in studies following the year 2007 \cite[][]{sun2017}. However, studies focusing on only the behavior of this subset of tourists are the majority. To this day, studies and analyses specifically comparing Asian and Western tourists are scarce, and even fewer are the number of studies explicitly comparing Chinese and Western tourists. One example is a study by \cite{choi2000}, which found that Western tourists visiting Hong Kong are satisfied more with room quality, while Asians are satisfied with the value for money. Another study by \cite{bauer1993changing} found that Westerners prefer hotel health facilities, while Asian tourists were more inclined to enjoy the Karaoke facilities of hotels. Both groups tend to have high expectations for the overall facilities. Another study done by \cite{kim2000} found American tourists to be individualistic and motivated by novelty, while Japanese tourists were collectivist and motivated by increasing knowledge and escaping routine.

    One thing to note with the above Asian vs. Western analyses is that they were performed before 2000 and not Chinese-specific. Meanwhile, the current Chinese economic boom is increasing the influx of tourists of this nation. The resulting increase in marketing and the creation of guided tours for Chinese tourists could have created a difference in tourists' perceptions and expectations. In turn, if we follow the definition of satisfaction in \cite{hunt1975}, the change in expectations could have influenced their satisfaction factors when traveling. Another note is that these studies were performed with questionnaires in places where it would be easy to locate tourists, i.e., airports. However, our study of online reviews takes the data that the hotel customers uploaded themselves. This data makes the analysis unique in exploring their behavior compared with Western tourists via factors that are not considered in most other studies. Furthermore, our study is unique in observing the customers in the specific environment of high-level hospitality in Japan.

    More recent studies have surfaced as well. A cross-country study \cite[][]{FRANCESCO201924} using posts from U.S.A. citizens, Italians, and Chinese tourists, determined using a text link analysis that customers from different countries indeed have a different perception and emphasis of a few predefined hotel attributes. According to their results, U.S.A. customers perceive cleanliness and quietness most positively. In contrast, Chinese customers perceive budget and restaurant above other attributes. Another couple of studies \cite[][]{JIA2020104071, HUANG2017117} analyze differences between Chinese and U.S. tourists using text mining techniques and more massive datasets, although in a restaurant context. 

    These last three studies focus on the U.S.A. culture, whereas our study focuses on the Western culture. Another difference with our study is that of the context of the study. The first study \cite[][]{FRANCESCO201924} was done within the context of tourists from three countries staying in hotels across the world. The second study chose restaurant reviews from the U.S.A. and Chinese tourists eating in three countries in Europe. The third study analyzed restaurants in Beijing.

    On the other hand, our study focuses on Western culture, instead of a single Western country, and Chinese culture clashing with the hospitality environment in Japan, specifically. Japan's importance in this analysis comes from the unique environment of high-grade hospitality that the country presents. In this environment, customers could either hold their satisfaction to this hospitality regardless of their culture or value other factors more depending on their cultural differences. Our study measures this at a large scale across different hotels in Japan.

    % Universal behavior
    Other studies have gone further and studied people from many countries in their samples and performed a more universal and holistic (not cross-culture) analysis. \cite{choi2001} analyzed hotel guest satisfaction determinants in Hong Kong with surveys in English, Chinese and Japanese translations, with people from many countries in their sample. \cite{choi2001} found that staff service quality, room quality, and value for money were the top satisfaction determinants. As another example, \cite{Uzama2012} produced a typology for foreigners coming to Japan for tourism, without making distinctions for their culture, but their motivation in traveling in Japan. In another study, \cite{zhou2014} analyzed hotel satisfaction using English and Mandarin online reviews from guests staying in Hangzhou, China coming from many countries. The general satisfaction score was noticed to be different among those countries. However, a more in-depth cross-cultural analysis of the satisfaction factors was not performed. As a result of their research, \cite{zhou2014} thus found that customers are universally satisfied by welcome extras, dining environments, and special food services. 

    % Western behavior 
    Regarding Western tourist behavior, a few examples can tell us what to expect when analyzing our data. \cite{kozak2002} found that British and German tourists' satisfaction determinants while visiting Spain and Turkey were hygiene and cleanliness, hospitality, the availability of facilities and activities, and accommodation services. \cite{shanka2004} found that English-speaking tourists in Perth, Australia were most satisfied with staff friendliness, the efficiency of check-in and check-out, restaurant and bar facilities, and lobby ambiance. 

    % Chinese behavior
    Regarding outbound Chinese tourists, academic studies about Chinese tourists have increased \cite[][]{sun2017}. Different researchers have found that Chinese tourist populations have several specific attributes. According to \cite{ryan2001} and their study of Chinese tourists in New Zealand, Chinese tourists prefer nature, cleanliness, and scenery in contrast to experiences and activities. \cite{dongyang2015} studied Chinese tourists in the Kansai region of Japan and found that Chinese tourists are satisfied mostly with exploring the food culture of their destination, cleanliness, and staff. Studying Chinese tourists in Vietnam, \cite{truong2009} found that Chinese tourists are highly concerned with value for money. According to \cite{liu2019}, Chinese tourists tend to have harsher criticism compared with other international tourists. Moreover, as stated by \cite{gao2017chinese}, who analyzed different generations of Chinese tourists and their connection to nature while traveling, Chinese tourists prefer nature overall. However, the younger generations seem to do so less than their older counterparts. 

    % Studies are not universal.
    Although the studies focusing only on Chinese or Western tourists have a narrow view, their theoretical contributions are valuable. We can see that depending on the study and the design of questionnaires and the destinations; the results can vary greatly. Not only that, but while there seems to be some overlap in most studies, some factors are completely ignored in one study but not in the other. Since our study uses data mining, each factor's definition is left for hotel customers to decide en masse via their reviews. This means that the factors will be selected through statistical methods alone instead of being defined by the questionnaire. Our method allows us to find factors that we would not have contemplated. It also avoids enforcing a factor on the mind of study subjects by presenting them with a question that they did not think of by themselves. This large variety of opinions in a well-sized sample, added to the automatic findings of statistical text analysis methods, gives our study an advantage compared to others with smaller samples. This study analyzes the satisfaction and dissatisfaction factors cross-culturally and compares them with the existing literature.

    % Reviewer comparison table
    Undoubtedly previous literature has examples of other cross-culture studies of tourist behavior and may further highlight our study and its merits. A contrast is shown in Table \ref{tab:lit-rev}. This table shows that older studies were conducted with surveys and had a different study topic. These are changes in demand \cite[][]{bauer1993changing}, tourist motivation \cite[][]{kim2000}, and closer to our study, satisfaction levels \cite[][]{choi2000}. However, our study topic is not the levels of satisfaction but the factors that drive it and dissatisfaction, which is overlooked in most studies. Newer studies with larger samples and similar methodologies have emerged, although two of these study restaurants instead of hotels \cite[][]{JIA2020104071, HUANG2017117}. One important difference is the geographical focus of their studies. While \cite{FRANCESCO201924} , \cite{JIA2020104071} and \cite{HUANG2017117} have a multi-national focus, we instead focus on Japan. The focus on Japan is important because of its top rank in hospitality across all types of businesses. Our study brings light to the changes, or lack thereof, in different touristic environments where an attribute can be considered excellent. The number of samples in other text-mining studies is also smaller than ours in comparison. Apart from that, every study has a different text mining method.

    % \usepackage{graphicx}
    % \usepackage{lscape}
    \begin{landscape}
    \begin{table}[p]
      \centering
      \caption{Comparison between cross-culture or cross-country previous studies and our study.}
      \label{tab:lit-rev}
      \resizebox{\linewidth}{!}{%
      \begin{tabular}{l|l|l|l|l|l|l|l|}
        \cline{2-8}
        \textbf{} &
          \textbf{Bauer et.al (1993)} &
          \textbf{Choi and Chu (2000)} &
          \textbf{Kim and Lee (2000)} &
          \textbf{Huang (2017)} &
          \textbf{Francesco and Roberta (2019)} &
          \textbf{Jia (2020)} &
          \textbf{Our study} \\ \hline
        \multicolumn{1}{|l|}{\textbf{Comparison objects}} &
          \begin{tabular}[c]{@{}l@{}}Asians\\ vs\\ Westerns\end{tabular} &
          \begin{tabular}[c]{@{}l@{}}Asians\\ vs\\ Westerners\end{tabular} &
          \begin{tabular}[c]{@{}l@{}}Anglo-Americans \\ vs \\ Japanese\end{tabular} &
          \begin{tabular}[c]{@{}l@{}}Chinese \\ vs \\ English-speakers\end{tabular} &
          \begin{tabular}[c]{@{}l@{}}USA \\ vs \\ China \\ vs \\ Italy\end{tabular} &
          \begin{tabular}[c]{@{}l@{}}Chinese \\ vs \\ US tourists\end{tabular} &
          \begin{tabular}[c]{@{}l@{}}Chinese \\ vs \\ Westerners\end{tabular} \\ \hline
        \multicolumn{1}{|l|}{\textbf{Study topic}} &
          Changes in demand &
          Satisfaction Levels &
          Tourist Motivation &
          \begin{tabular}[c]{@{}l@{}}Dining experience \\ of Roast Duck\end{tabular} &
          \begin{tabular}[c]{@{}l@{}}Perception and \\ Emphasis\end{tabular} &
          \begin{tabular}[c]{@{}l@{}}Motivation and \\ Satisfaction\end{tabular} &
          \textbf{\begin{tabular}[c]{@{}l@{}}Satisfaction and\\ Dissatisfaction\end{tabular}} \\ \hline
        \multicolumn{1}{|l|}{\textbf{Geographical focus}} &
          Asia Pacific region &
          Hong Kong &
          Global &
          Beijing &
          Multi-national &
          Multi-national &
          \textbf{Japan} \\ \hline
        \multicolumn{1}{|l|}{\textbf{Industry}} &
          Hotels &
          Hotels &
          Tourism &
          Restaurant (Beijing Roast Duck) &
          Hotels &
          Restaurants &
          Hotels \\ \hline
        \multicolumn{1}{|l|}{\textbf{Study subjects}} &
          Hotel managers &
          Hotel customers &
          \begin{tabular}[c]{@{}l@{}}Tourists arriving \\ in airport\end{tabular} &
          \begin{tabular}[c]{@{}l@{}}Diners \\ online reviews\end{tabular} &
          \begin{tabular}[c]{@{}l@{}}Hotel customers \\ online reviews\end{tabular} &
          \begin{tabular}[c]{@{}l@{}}Diners \\ online reviews\end{tabular} &
          \begin{tabular}[c]{@{}l@{}}Hotel customers \\ online reviews\end{tabular} \\ \hline
        \multicolumn{1}{|l|}{\textbf{Sample method}} &
          surveys &
          surveys &
          survey &
          text mining &
          text mining &
          text mining &
          text mining \\ \hline
        \multicolumn{1}{|l|}{\textbf{Number of samples}} &
          185 surveys &
          540 surveys &
          \begin{tabular}[c]{@{}l@{}}165 Anglo-American\\ 209 Japanese\end{tabular} &
          \begin{tabular}[c]{@{}l@{}}990 Chinese reviews\\ 398 English reviews\end{tabular} &
          9000 reviews (3000 per country) &
          \begin{tabular}[c]{@{}l@{}}2448 reviews\\ (1360 Chinese)\\ (1088 English)\end{tabular} &
          \textbf{\begin{tabular}[c]{@{}l@{}}89,207 reviews\\ (48,070 Chinese)\\ (41,137 English)\end{tabular}} \\ \hline
        \multicolumn{1}{|l|}{\textbf{Study method}} &
          statistics &
          VARIMAX &
          MANOVA &
          \begin{tabular}[c]{@{}l@{}}Semantic \\ Network \\ Analysis\end{tabular} &
          Text Link Analysis &
          \begin{tabular}[c]{@{}l@{}}Topic modeling \\ (LDA)\end{tabular} &
          \textbf{\begin{tabular}[c]{@{}l@{}}SVM, \\ Dependency Parsing\\ and POS tagging\end{tabular}} \\ \hline
        \multicolumn{1}{|l|}{\textbf{Subject nationality}} &
          \begin{tabular}[c]{@{}l@{}}Asians: \\ China,\\ Fiji,\\ Hong Kong,\\ Indonesia,\\ Malaysia,\\ Singapore,\\ Taiwan,\\ Guam,\\ Tahiti,\\ Thailand \\ \\ Westerners: Australia,\\ New Zealand\end{tabular} &
          \begin{tabular}[c]{@{}l@{}}Asians:\\ China,\\ Taiwan,\\ Japan,\\ South Korea,\\ South-East Asia\\ \\ Westerners:\\ North America,\\ Europe,\\ Australia,\\ New Zealand\end{tabular} &
          USA, Japan &
          \begin{tabular}[c]{@{}l@{}}English-speakers: \\ U.K., U.S., Australia,\\ New Zealand, Canada,\\ Ireland\\ \\ Chinese-speakers: China\end{tabular} &
          USA, China, Italy &
          USA, China &
          \begin{tabular}[c]{@{}l@{}}Chinese-speakers:\\ China\\ \\ English-speakers:\\ (U.K., U.S.,\\ Australia,\\ New Zealand,\\ Canada, Ireland)\end{tabular} \\ \hline
        \end{tabular}%
        }
    \end{table}
    \end{landscape}

  \subsection{Data mining, machine learning, knowledge discovery and sentiment analysis}\label{theory_data}

    % Explain the discovery of theory inside the data
    In the current world, data is presented to us in larger and larger quantities. Today's data sizes were commonly only seen in very specialized large laboratories with supercomputers a couple of decades ago. However, they are now standard for market and managerial studies, independent university students, and any scientist connecting to the Internet. Such quantities of data are available to study now more than ever. Nevertheless, it would be impossible for researchers to parse all of this data by themselves. As \cite{fayyad1996data} summarizes, data by itself is unusable until it goes through a process of selection, preprocessing, transformation, mining, and evaluation. Only then can it be established as knowledge. With the tools available to us in the era of information science, algorithms can be used to detect patterns that would take researchers too long to recognize. These patterns can, later on, be evaluated to generate knowledge. This process is called Knowledge Discovery in Databases. 

    % Text mining 
    Now, there are, of course, many sources of numerical data to be explored.  However, perhaps what is most available and interesting to managerial purposes is the resource of customers' opinions in text form. Since the introduction of Web 2.0, an unprecedented quantity of valuable information is posted to the Internet at a staggering speed. Text mining has then been proposed more than a decade ago to utilize this data \cite[e.g.][]{rajman1998text,nahm2002text}. Using Natural Language Processing, one can parse language in a way that translates to numbers so that a computer can analyze it. Since then, text mining techniques have improved over the years. This has been used in the field of hospitality as well for many purposes, including satisfaction analysis from reviews \cite[e.g][]{berezina2016, xu2016, xiang2015, hargreaves2015, balbi2018}, social media's influence on travelers \cite[e.g.][]{xiang2010}, review summarization \cite[e.g.][]{hu2017436}, perceived value of reviews \cite[e.g][]{FANG2016498}, and even predicting hotel demand using web traffic data \cite[e.g][]{yang2014}.

    % Sentiment Analysis
    More than only analyzing patterns within the text, researchers have found how to determine the sentiment behind a statement based on speech patterns, statistical patterns, and other methodologies. This method is called sentiment analysis or opinion mining. A precursor of this method was attempted decades ago \cite[][]{stone1966general}. With sentiment analysis, one could use patterns in the text to determine whether a sentence was being said with a positive opinion, or a critical one. This methodology could even determine other ranges of emotions, depending on the thoroughness of the algorithm. Examples of sentiment analysis include ranking products through online reviews \cite[e.g][]{liu2017149, zhang2011}, predicting political poll results through opinions in Twitter \cite[][]{oconnor2010}, and so on. In the hospitality field, it has been used to classify reviewers' opinions of hotels in online reviews \cite[e.g.][]{kim2017362, alsmadi2018}. 

    % Machine Learning
    Our study used an algorithm for sentiment analysis called a Support Vector Machine (SVM), a supervised machine learning used for binary classification. Machine learning is a general term used for algorithms that, when given data, will automatically use that data to "learn" from its patterns and apply them for improving upon a task. Learning machines can be supervised, as in our study, where the algorithm has manually labeled training data to detect patterns in it and use them to establish a method for classifying other unlabeled data automatically. Machine learning can also be unsupervised, where there is no pre-labeled data. In this latter case, the machine will analyze the structure and patterns of the data and perform a task based on its conclusions. Our study calls for a supervised machine since text analysis can be intricate. Many patterns might occur, but we are only interested in satisfaction and dissatisfaction labels. Consequently, we teach the machine through previously labeled text samples. 

    Machine learning and data mining are two fields with a significant overlap since they can use each other's methods to achieve the task at hand. Machine learning methods focus on predicting new data based on known properties and patterns of the given data. Data mining, on the other hand, is discovering new information and new properties of the data. Our machine learning approach will learn the sentiment patterns of our sample texts showing satisfaction and dissatisfaction and using these to label the rest of the data. We are not exploring new patterns in the sentiment data. However, we are using sentiment predictions for knowledge discovery in our database. Thus, our study is a data mining experiment based on machine learning.

    % Data exploration
    Because the methodology for finding patterns in the data is automatic and statistical, it is both reliable and unpredictable. Reliable in that the algorithm will find a pattern by its nature. Unpredictable in that since it has no intervention from the researchers in making questionnaires, it can result in anything that the researchers could not expect. These qualities determine why, similar to actual mining; data mining is mostly exploratory. One can never be sure that one will find a specific something. However, we can make predictions and estimates about finding knowledge and what kind of knowledge we can uncover. The exploration of large opinion datasets with these methods is essential. The reason is that we can discover knowledge that could otherwise be missed by observing a localized sample rather than taking a holistic view of every user's opinion. In other words, a machine algorithm can find the needles in a haystack that we did not know were there by examining small bundles of hay at a time.

\section{Methodology}\label{method}

\begin{figure}[bp]
  \centering
  \includegraphics[width=\textwidth]{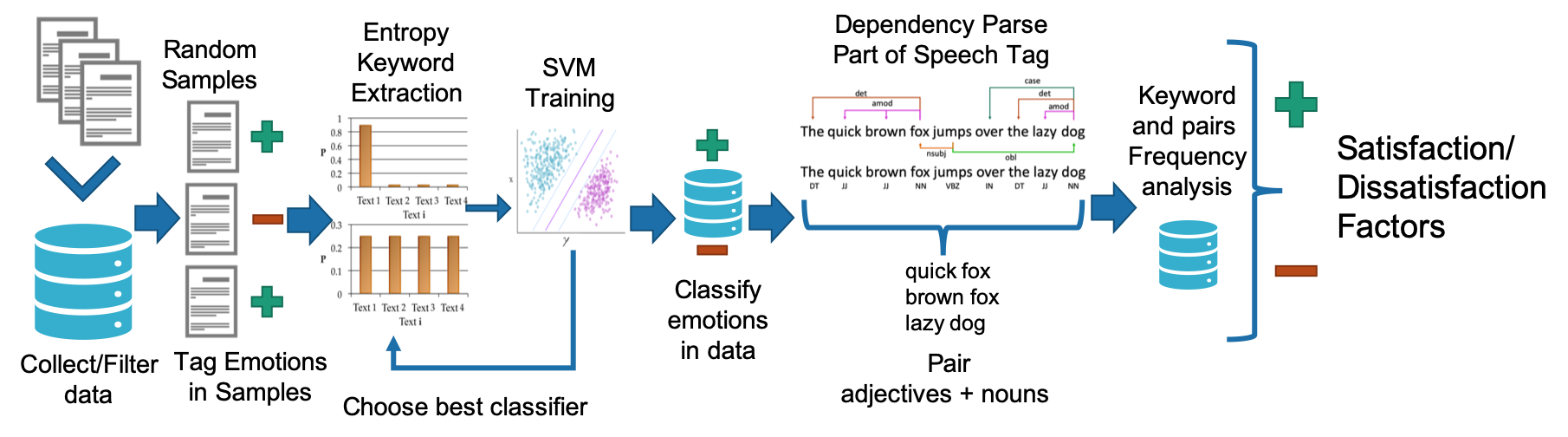}
  \caption{Overview of the methodology to quantitatively rank satisfaction factors.}
  \label{fig:method-overview}
  \end{figure}

  We extracted a large number of text reviews from the site \textit{Ctrip}, with mostly mainland Chinese users, and the travel site \textit{TripAdvisor}. We then determined the most commonly used words that relate to positive and negative opinions in a review. We did this using Shannon's entropy to extract keywords from their vocabulary. These positive and negative keywords allow us to train an optimized Support Vector Classifier (SVC) to perform a binary emotional classification of the reviews in large quantities, saving time and resources for the researchers. We then applied a dependency parsing to the reviews and a Part of Speech tagging (POS tagging) to observe the relationship between adjective keywords and the nouns they refer to. We split the dataset into price ranges to observe the differences in keyword usage between lower-class and higher-class hotels. We observed the frequency of the terms in the dataset to extract the most utilized words in either review. We show an overview of this methodology in Figure \ref{fig:method-overview}, which is an updated version of the methodology used by \cite{Aleman2018ICAROB}. Finally, we also observed if the satisfaction factors were soft or hard attributes of the hotel.

  \subsection{Data collection}\label{datacollection}

    In the \textit{Ctrip} data collection, reviews from a total of \num[group-separator={,}]{5774} hotels in Japan were collected. From these pages, we extracted a total of \num[group-separator={,}]{245919} reviews, from which \num[group-separator={,}]{211932} were detected to be standard Mandarin Chinese. Since a single review can have sentences with different sentiments, we separated sentences using punctuation marks. The Chinese reviews were comprised of \num[group-separator={,}]{187348} separate sentences. 

    In the \textit{TripAdvisor} data collection, we collected data from \num[group-separator={,}]{21380} different hotels. In total, we collected \num[group-separator={,}]{295931} reviews, from which \num[group-separator={,}]{295503} were detected to be in English. Similarly to the Chinese data, we then separated these English reviews into \num[group-separator={,}]{2694261} sentences using the \textit{gensim} python library. For the language detection in both cases we used the \textit{langdetect} python library.

    However, to make the data comparisons fair, we filtered both databases only to contain reviews from hotels in both datasets, using their English names to do a search match. We also filtered them to be in the same date range. In addition, we selected only the hotels that had pricing information available. We extracted the lowest and highest price possible for one night as well. The difference in pricing can be from better room settings, such as double or twin rooms or suites, depending on the hotel. Regardless of the reason, we chose the highest-priced room since it can be an indirect indicator of the hotel's class. After filtering, the datasets contained \num[group-separator={,}]{557} hotels in common. The overlapping date range for reviews was from July 2014 to July 2017. Within these hotels, from \textit{Ctrip} there was \num[group-separator={,}]{48070} reviews comprised of \num[group-separator={,}]{101963} sentences, and from \textit{TripAdvisor} there was \num[group-separator={,}]{41137} reviews comprised of \num[group-separator={,}]{348039} sentences. 

\begin{figure}[ht]
        \centering
        \begin{subfigure}[b]{0.45\textwidth}
            \includegraphics[width=\textwidth]{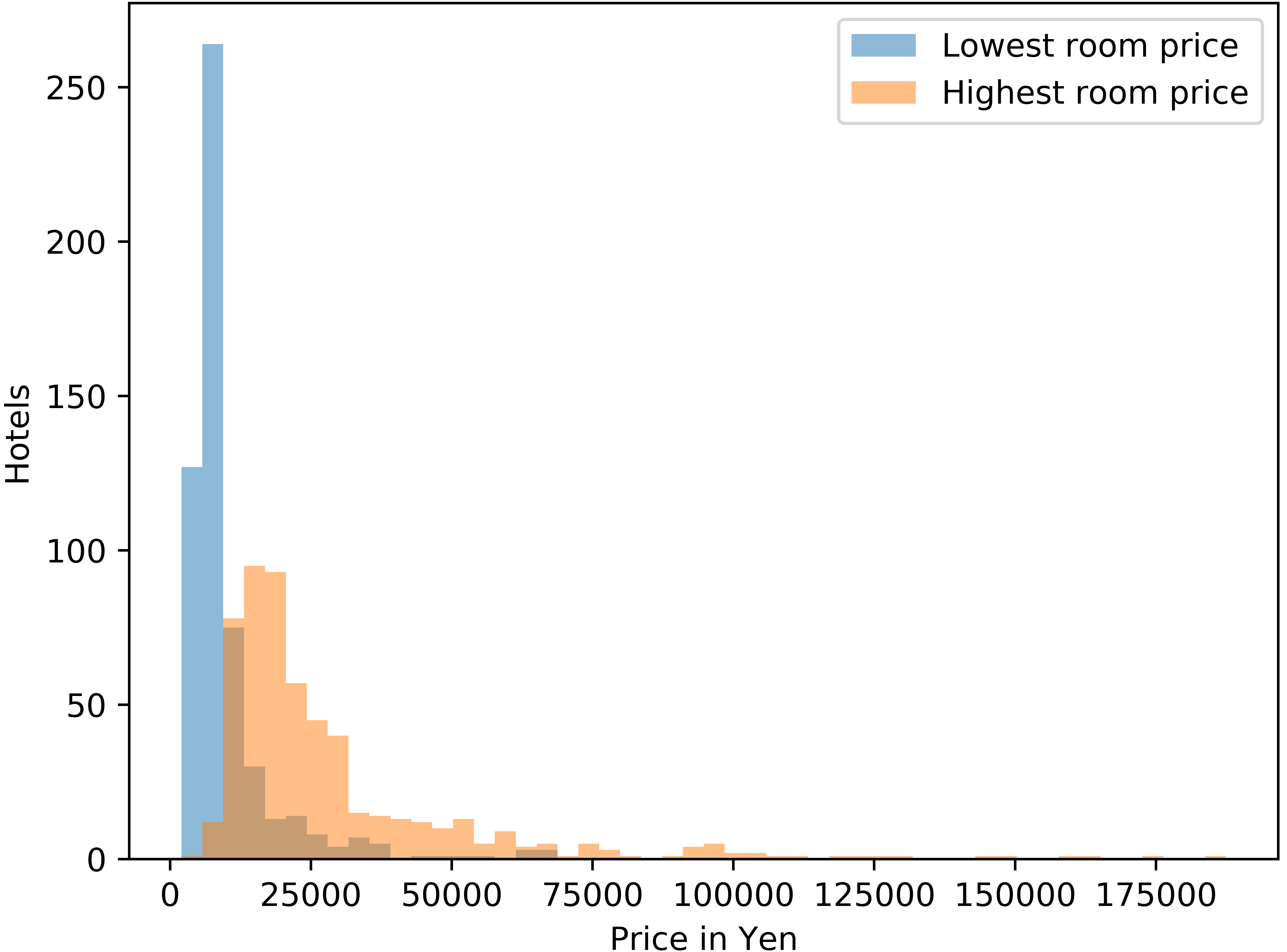}
            \caption{50 equal lenght bins}
        \end{subfigure}
        \begin{subfigure}[b]{0.45\textwidth}
            \includegraphics[width=\textwidth]{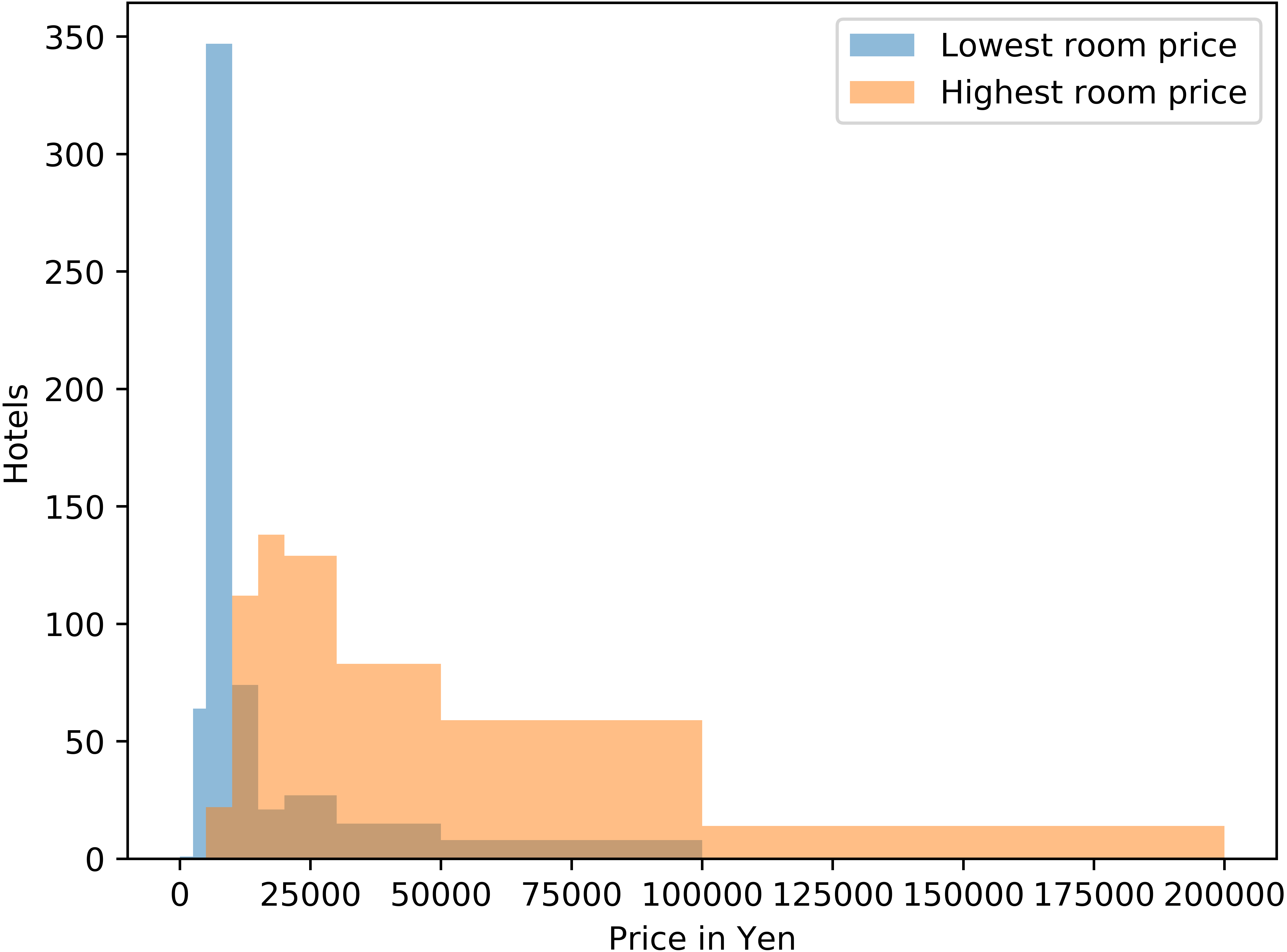}
            \caption{manually set 9 price ranges}
        \end{subfigure}
    \caption{Price for one night distribution, blue: lowest price, orange: highest price.}
    \label{fig:price_dist}
    \end{figure}

    The price for a night in these hotels ranges from cheap capsule hotels at 2000 yen per night to high-end hotels 188,000 yen a night at the far ends of the bell curve. Customers' expectations can vary greatly depending on the pricing of the hotel room they stay at. Therefore, we made observations on the distribution of pricing in our database's hotels and binned the data by price ranges, decided by consideration of the objective of stay. We show these distributions in Figure \ref{fig:price_dist}. The structure of the data after division by price is shown in Table \ref{tab:exp_notes}. This table also includes the results of emotional classification after applying our SVC, as explained in \ref{sentimentanalysis}. The first three price ranges (0 to 2500 yen, 2500 to 5000 yen, 5000 to 10,000 yen) would correspond to low-class hotels or even hostels on the lower end and cheap business hotels on the higher end. Further on, there are business hotels in the next range (10,000 to 15,000 yen). After that, the stays could be at Japanese style \textit{ryokan} when traveling in groups, high-class business hotels, luxury love hotels, or higher class hotels (15,000 to 20,000 yen, 20,000 to 30,000 yen). Further than that is more likely to be \textit{ryokan} or high class resorts or five-star hotels (30,000 to 50,000 yen, 50,000 to 100,000 yen, 100,000 to 200,000 yen). Note that because of choosing the highest price per one night in each hotel, the cheapest two price ranges (0 to 2500 yen, 2500 to 5000 yen) are empty, despite some rooms being priced at 2000 yen per night. Because of this, other tables will omit these two price ranges.

    % \usepackage{multirow}
    % \usepackage{graphicx}
    \begin{table}[ht]
      \centering
      \caption{Collected data and structure after price range categorizing.}
      \label{tab:exp_notes}
      \resizebox{\textwidth}{!}{%
      \begin{tabular}{|l|l|r|r|}
        \hline
        \multicolumn{1}{|c|}{\textbf{Price range}} & \multicolumn{1}{c|}{\textbf{Data collected}} & \textbf{Ctrip database} & \textbf{Tripadvisor database} \\ \hline
        \multirow{5}{*}{0: All Prices}               & Hotels             & 557     & 557     \\
                                                   & Reviews            & 48,070  & 41,137  \\
                                                   & Sentences          & 101,963 & 348,039 \\
                                                   & Positive sentences & 88,543  & 165,308 \\
                                                   & Negative sentences & 13,420  & 182,731 \\ \hline
        \multirow{2}{*}{1: 0 to 2500 yen}            & Hotels             & 0       & 0       \\
                                                   & Reviews            & 0       & 0       \\ \hline
        \multirow{2}{*}{2: 2500 to 5000 yen}         & Hotels             & 0       & 0       \\
                                                   & Reviews            & 0       & 0       \\ \hline
        \multirow{5}{*}{3: 5000 to 10,000 yen}       & Hotels             & 22      & 22      \\
                                                   & Reviews            & 452     & 459     \\
                                                   & Sentences          & 1,108   & 3,988   \\
                                                   & Positive sentences & 924     & 1,875   \\
                                                   & Negative sentences & 184     & 2,113   \\ \hline
        \multirow{5}{*}{4: 10,000 to 15,000 yen}     & Hotels             & 112     & 112     \\
                                                   & Reviews            & 2,176   & 2,865   \\
                                                   & Sentences          & 4,240   & 24,107  \\
                                                   & Positive sentences & 3,566   & 11,619  \\
                                                   & Negative sentences & 674     & 12,488  \\ \hline
        \multirow{5}{*}{5: 15,000 to 20,000 yen}     & Hotels             & 138     & 138     \\
                                                   & Reviews            & 7,043   & 4,384   \\
                                                   & Sentences          & 14,726  & 37,342  \\
                                                   & Positive sentences & 12,775  & 17,449  \\
                                                   & Negative sentences & 1,951   & 19,893  \\ \hline
        \multirow{5}{*}{6: 20,000 to 30,000 yen}     & Hotels             & 129     & 129     \\
                                                   & Reviews            & 11,845  & 13,772  \\
                                                   & Sentences          & 24,413  & 115,830 \\
                                                   & Positive sentences & 21,068  & 55,381  \\
                                                   & Negative sentences & 3,345   & 60,449  \\ \hline
        \multirow{5}{*}{7: 30,000 to 50,000 yen}     & Hotels             & 83      & 83      \\
                                                   & Reviews            & 8,283   & 7,001   \\
                                                   & Sentences          & 17,939  & 58,409  \\
                                                   & Positive sentences & 15,642  & 28,493  \\
                                                   & Negative sentences & 2,297   & 29,916  \\ \hline
        \multirow{5}{*}{8: 50,000 to 100,000 yen}    & Hotels             & 59      & 59      \\
                                                   & Reviews            & 16,670  & 9,646   \\
                                                   & Sentences          & 36,255  & 81,940  \\
                                                   & Positive sentences & 31,638  & 38,217  \\
                                                   & Negative sentences & 4,617   & 43,723  \\ \hline
        \multirow{5}{*}{9: 100,000 to 200,000 yen}   & Hotels             & 14      & 14      \\
                                                   & Reviews            & 1,601   & 3,010   \\
                                                   & Sentences          & 3,282   & 26,423  \\
                                                   & Positive sentences & 2,930   & 12,274  \\
                                                   & Negative sentences & 352     & 14,149  \\ \hline
        \end{tabular}%
      }
    \end{table}
  \subsection{Text processing}\label{textprocessing}

    We needed to analyze the grammatical relationship between words, be it English or Chinese, to understand the connections between adjectives and nouns. For all these processes, we used the Stanford CoreNLP pipeline developed by the Natural Language Processing Group at Stanford University \cite[][]{manning-EtAl:2014:P14-5}. In order to separate Chinese words for analysis, we used the Stanford Word Segmenter \cite[][]{chang2008}. In English texts, however, only using spaces is not enough to correctly collect concepts. The English language is full of variations and conjugations of words depending on the context and tense. Thus, a better segmentation is achieved by using lemmatization, which returns each word's dictionary form. For this purpose, we used the \textit{gensim} library for the English texts.

\begin{figure}[ht]
    \centering
    \includegraphics[width=0.6\textwidth]{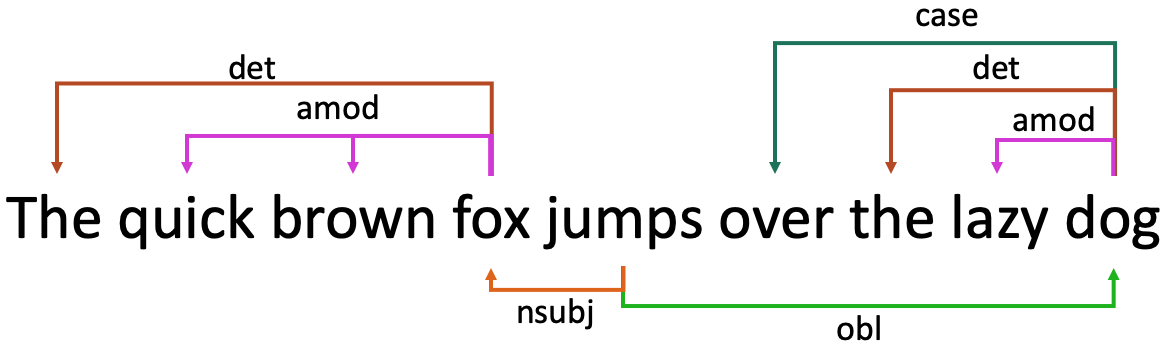}
    \caption{Example of dependency parsing.}
    \label{fig:depparse}
    \end{figure}

    A dependency parser analyzes the grammatical structure, detecting connections between words, and describing the action and direction of those connections. We show an example of these dependencies in Figure \ref{fig:depparse}. This study uses the Stanford NLP Dependency Parser, as described by \cite{chen-EMNLP:2014}. A list of dependencies used by this parser is detailed by \cite{marneffe_manning_2016_depparse_manual}. In more recent versions, they use an updated dependency tag list from Universal Dependencies \cite[][]{zeman2018conll}. In our study, this step was necessary to extract adjective modifiers and their subject. We did that by parsing the database and extracting instances of a few determined dependency codes. One of these dependency codes is ``amod'', which stands for ``adjectival modifier''. This is used when an adjective modifies a noun directly (e.g., A big apple). The other dependency code we used was ``nsubj'', or nominal subject, the class's syntactic subject. We used this one for cases where the adjective is modifying the noun indirectly through other words (e.g., The apple is big). This dependency does not necessarily only include a combination of adjectives and nouns. However, it can also be connected with copular verbs, nouns, or other adjectives. We saw it necessary also to perform a Part of Speech (POS) tagging of these clauses.

\begin{figure}[ht]
    \centering
    \includegraphics[width=0.6\textwidth]{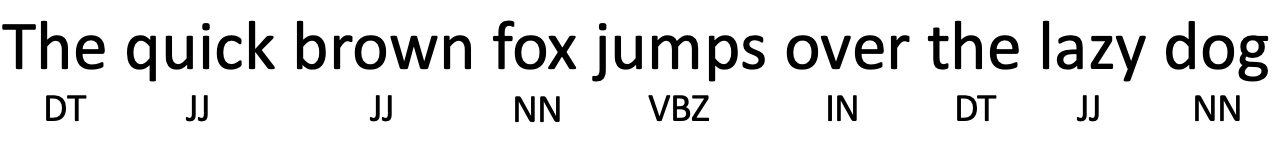}
    \caption{Example of POS tagging with the Penn Treebank tags.}
    \label{fig:postag}
    \end{figure}

    A Part of Speech (POS) tagger is a program that assigns word tokens with tags identifying the part of speech. An example is shown in Figure \ref{fig:postag}. A Part of Speech is a category of lexical items that serve similar grammatical purposes, for example, nouns, adjectives, verbs, or conjunctions. In our study, we used the Stanford NLP POS tagger software, described by \cite{toutanova2000enriching} and \cite{toutanova2003feature}, which uses the Penn Chinese Treebank tags \cite[][]{xia_penntreebank}.

    In this study, we were interested in identifying combinations of adjectives, some verbs, and nouns. We also needed to filter away bad combinations that were brought by the versatility of nominal subject dependencies. For this purpose, we identified the tags for nouns, verbs, and adjectives in Chinese and English, with the English tags being a bit more varied. What would be called adjectives in English corresponds more to stative verbs in Chinese, so we needed to extract those as well. We show a detailed description of the chosen tags in Table \ref{tab:target_postag}. We also show a detailed description of the tags we needed to filter. We selected these tags heuristically by observing commonly found undesired pairs in Table \ref{tab:filter_postag}.

    % \usepackage{multirow}
    \begin{table}[ht]
      \centering
      \caption{Target Parts of Speech for extraction and pairing.}
      \label{tab:target_postag}
      \resizebox{\textwidth}{!}{%
      \begin{tabular}{|c|c|l|l|}
        \hline
        \textbf{Language}                    & \textbf{POS Tag} & \multicolumn{1}{c|}{\textbf{Part of Speech}} & \multicolumn{1}{c|}{\textbf{Examples}}     \\ \hline
        \multirow{4}{*}{Chinese target tags} & NN               & Noun (general)                               & \begin{CJK}{UTF8}{gbsn}酒店\end{CJK} (hotel) \\ \cline{2-4} 
                                             & VA  & Predicative Adjective (verb)       & \begin{CJK}{UTF8}{gbsn}干净 的\end{CJK} (clean)   \\ \cline{2-4} 
                                             & JJ  & Noun modifier (adjectives)         & \begin{CJK}{UTF8}{gbsn}干净\end{CJK} (clean)     \\ \cline{2-4} 
                                             & VV  & Verb (general)                     & \begin{CJK}{UTF8}{gbsn}推荐\end{CJK} (recommend) \\ \hline
        \multirow{9}{*}{English target tags} & NN  & Noun (general)                     & room                                           \\ \cline{2-4} 
                                             & NNS & Noun (plural)                      & beds                                           \\ \cline{2-4} 
                                             & JJ  & Adjective                          & big                                            \\ \cline{2-4} 
                                             & JJS & Adjective (superlative)            & best                                           \\ \cline{2-4} 
                                             & JJR & Adjective (comparative)            & larger                                         \\ \cline{2-4} 
                                             & VB  & Verb (base form)                   & take                                           \\ \cline{2-4} 
                                             & VBP & Verb (single present)              & take                                           \\ \cline{2-4} 
                                             & VBN & Verb (past participle)             & taken                                          \\ \cline{2-4} 
                                             & VBG & Verb (gerund / present participle) & taking                                         \\ \hline
        \end{tabular}%
        }
    \end{table}

    % \usepackage{multirow}
    \begin{table}[ht]
      \centering
      \caption{Filtered out Parts of Speech to aid pairing.}
      \label{tab:filter_postag}
      \resizebox{\textwidth}{!}{%
      \begin{tabular}{|c|l|l|l|}
        \hline
        \textbf{Language} &
          \multicolumn{1}{c|}{\textbf{POS Tag}} &
          \multicolumn{1}{c|}{\textbf{Part of Speech}} &
          \multicolumn{1}{c|}{\textbf{Examples}} \\ \hline
        \multirow{4}{*}{Commonly filtered tags} & DT    & Determiner         & a, an                                      \\ \cline{2-4} 
                                                & PN    & Pronoun            & I, you, they                               \\ \cline{2-4} 
                                                & CD    & Cardinal Number    & 1, 2, 3, 4, 5                              \\ \cline{2-4} 
                                                & PU    & Punctuation        & .!?                                        \\ \hline
        \multirow{5}{*}{Chinese filtered tags} &
          DEV &
          Particle &
          \begin{CJK}{UTF8}{gbsn}地\end{CJK} (Japan) (adverbial particle) \\ \cline{2-4} 
                                                & NR    & Noun (proper noun) & \begin{CJK}{UTF8}{gbsn}日本\end{CJK} (Japan) \\ \cline{2-4} 
         &
          M &
          Measure word &
          \begin{CJK}{UTF8}{gbsn}个\end{CJK} (general classifier), \begin{CJK}{UTF8}{gbsn}公里\end{CJK} (kilometer) \\ \cline{2-4} 
         &
          SP &
          Sentence-final particle &
          \begin{CJK}{UTF8}{gbsn}他\end{CJK} (he), \begin{CJK}{UTF8}{gbsn}好\end{CJK} (good) \\ \cline{2-4} 
                                                & IJ    & Interjection       & \begin{CJK}{UTF8}{gbsn}啊\end{CJK} (ah)     \\ \hline
        \multirow{3}{*}{English target tags}    & NNP   & Noun (proper noun) & Japan                                      \\ \cline{2-4} 
                                                & PRP\$ & Possessive Pronoun & My, your, her, his                         \\ \cline{2-4} 
                                                & WP    & Wh-pronoun         & What, who                                  \\ \hline
        \end{tabular}%
        }
    \end{table}

    Once we had these adjective + noun or verb + noun pairs, we could determine what the customers referred to in their reviews. With what frequency they use those pairings positively or negatively.

  \subsection{Sentiment analysis using a Support Vector Classifier}\label{sentimentanalysis}

    The sentiment analysis was performed using the methodology described by \cite{Aleman2018ICAROB}. Keywords are determined by a comparison of Shannon's entropy \cite[][]{shannon1948} between two classes by a factor of \(\alpha\) for one class and \(\alpha'\) for the other, and then they are used in an SVC \cite[][]{cortes1995}, optimizing keywords to select the best performing classifier using the \(F_1\)-measure \cite[][]{powers2011}. The selected SVC keywords would then clearly represent the user driving factors leading to positive and negative emotions. We also performed experiments to choose the best value of the parameter C used in the SVC. C is a constant that affects the optimization process when minimizing the error of the separating hyperplane. Low values of C give some freedom of error, which minimizes false positives but can also increase false negatives. Inversely, high C values will likely result in minimal false negatives but a possibility of false positives. SVC performance results are displayed in  Tables \ref{tab:svm_f1_zh} and \ref{tab:svm_f1_en}. Examples of tagged sentences are shown in Table \ref{tab:training_examples}. 

    % \usepackage{graphicx}
    \begin{table}[ht]
      \centering
      \caption{Best performing SVC 5-fold cross-validation Chinese text classifiers.}
      \label{tab:svm_f1_zh}
      \resizebox{0.65\textwidth}{!}{%
      \begin{tabular}{|l|l|l|l|l|}
        \hline
        \textbf{Keyword List} &
          \textbf{\begin{tabular}[c]{@{}l@{}}Classifier\\ emotion\end{tabular}} &
          \textbf{C} &
          \begin{tabular}[c]{@{}l@{}}\(F_1\)\\ \(\mu\)\end{tabular} &
          \begin{tabular}[c]{@{}l@{}}\(F_1\)\\ \(\sigma\)\end{tabular} \\ \hline
        \begin{tabular}[c]{@{}l@{}}Satisfaction keywords\\ (\(\alpha = 2.75\))\end{tabular} 
                & Satisfaction              & 2.5             & 0.91              & 0.01          \\ \hline
        \begin{tabular}[c]{@{}l@{}}Negative keywords\\ (\(\alpha' = 3.75\))\end{tabular}    
                & Dissatisfaction           & 0.5             & 0.67              & 0.11          \\ \hline
        \begin{tabular}[c]{@{}l@{}}\textbf{Combined}\\ (\(\alpha=2.75\), \(\alpha'=3.75\))\end{tabular} 
                & \textbf{Satisfaction}     & \textbf{0.5}    & \textbf{0.95}     & \textbf{0.01} \\ \hline
        \end{tabular}
        }
    \end{table}

    % \usepackage{graphicx}
    \begin{table}[ht]
      \centering
      \caption{Best performing SVC 10-fold cross-validation English text classifiers.}
      \label{tab:svm_f1_en}
      \resizebox{0.65\textwidth}{!}{%
      \begin{tabular}{|l|l|l|l|l|}
        \hline
        \textbf{Keyword List} &
          \textbf{\begin{tabular}[c]{@{}l@{}}Classifier\\ emotion\end{tabular}} &
          \textbf{C} &
          \begin{tabular}[c]{@{}l@{}}\(F_1\)\\ \(\mu\)\end{tabular} &
          \begin{tabular}[c]{@{}l@{}}\(F_1\)\\ \(\sigma\)\end{tabular} \\ \hline
        \begin{tabular}[c]{@{}l@{}}Satisfaction keywords \\ (\(\alpha=1.5\))\end{tabular}       
                & Satisfaction              & 1.75            & 0.82              & 0.02          \\ \hline
        \begin{tabular}[c]{@{}l@{}}Dissatisfaction keywords \\ (\(\alpha'=4.25\))\end{tabular}  
                & Dissatisfaction           & 3               & 0.80              & 0.03          \\ \hline
        \begin{tabular}[c]{@{}l@{}}\textbf{Combined} \\ (\(\alpha=1.5\), \(\alpha'=4.25\))\end{tabular} 
                & \textbf{Satisfaction}     & \textbf{2}      & \textbf{0.83}     & \textbf{0.02} \\ \hline
        \end{tabular}
        }
    \end{table}

    % \usepackage{multirow}
    % \usepackage{graphicx}
    % \usepackage[normalem]{ulem}
    \begin{table}[ht]
      \centering
      \caption{Examples of positive and negative sentences used for training SVM.}
      \label{tab:training_examples}
      \resizebox{\textwidth}{!}{%
      \begin{tabular}{|c|c|l|}
        \hline
        \multicolumn{1}{|l|}{\textbf{Language}} &
          \multicolumn{1}{l|}{\textbf{Emotion}} &
          \textbf{Sentences} \\ \hline
        \multirow{4}{*}{Chinese} &
          \multirow{2}{*}{Positive} &
          \begin{tabular}[c]{@{}l@{}}\begin{CJK}{UTF8}{gbsn}酒店 的 服务 很 好 和 我 住 过 的 所有 日本 酒店 一样 各 种 隐形 服务 非常 厉害\end{CJK}\\ (translated as: "The service of the hotel is very good.\\ All the services of the Japanese hotels I have stayed in are extremely good.")\end{tabular} \\ \cline{3-3} 
         &
           &
          \begin{tabular}[c]{@{}l@{}}\begin{CJK}{UTF8}{gbsn}有 一 个 后门 到 地铁站 非常 近 周边 也 算 方便 酒店 服务 和 卫生 都 很 好\end{CJK}\\ (translated as: "There is a back door to the subway station very close to it. \\ The surrounding area is also convenient hotel service and health are very good")\end{tabular} \\ \cline{2-3} 
         &
          \multirow{2}{*}{Negative} &
          \begin{tabular}[c]{@{}l@{}}\begin{CJK}{UTF8}{gbsn}酒店 旁边 很 荒凉 连个 便利 店 都 要 走 很远\end{CJK}\\ (translated as: "The hotel is very bleak, \\ and you have to go very far to go to the nearest convenience store.")\end{tabular} \\ \cline{3-3} 
         &
           &
          \begin{tabular}[c]{@{}l@{}}\begin{CJK}{UTF8}{gbsn}唯一 不 足 是 价格 太高\end{CJK}\\ (translated as: "The only negative is that the price is too high.")\end{tabular} \\ \hline
        \multirow{4}{*}{English} &
          \multirow{2}{*}{Positive} &
          It was extremely clean, peaceful and the hotel Hosts made us feel super welcome \\ \cline{3-3} 
         &
           &
          \begin{tabular}[c]{@{}l@{}}Location is very good, close to a main road with a subway station, a bakery,\\  a 7 eleven and a nice restaurant that is not too expensive but serves good food\end{tabular} \\ \cline{2-3} 
         &
          \multirow{2}{*}{Negative} &
          \begin{tabular}[c]{@{}l@{}}The only downside. Our room was labeled 'non-smoking'\\ but our duvet reeked of smoke.\end{tabular} \\ \cline{3-3} 
         &
           &
          A bit pricey though \\ \hline
        \end{tabular}%
        }
    \end{table}

    Shannon's entropy can be used to observe the probability distribution of each word inside the corpus. A word included in many documents will have a high entropy value for that set of documents. Opposite to this, a word appearing in only one document will have an entropy value of zero. 

    An SVC is trained to classify data based on previously labeled data, generalizing the data's features by defining a separating (p-1)-dimensional hyperplane in p-dimensional space. Each dimension is a feature of the data in this space. The separating hyperplane, along with the support vectors, divides the multi-dimensional space and minimizes classification error. 

    Our study used a linear kernel for the SVC, defined by the formula (\ref{eq:svm1}) below. Each training sentence is a data point, a row in the vector \(x\). Each column represents a feature; in our case, the quantities of each of the keywords in that particular sentence. The labels of previously known classifications (1 for positive, 0 for negative) for each sentence comprise the \(f(x)\) vector. The Weight Vector \(w\) is comprised of the influences each point has had in the training process to define the hyperplane angle. The bias coefficient \(b\) determines its position.

    During the SVC learning algorithm, each data point classified incorrectly alters the weight vector to correctly classify new data. These changes to the weight vector are greater for features close to the separating hyperplane. These features have stronger changes because they needed to be taken into account to classify with a minimal error. Sequentially, the weight vector can be interpreted as a numerical representation of each feature's effect on each class's classification process. Below we show the formula for the weight vector \(w\) (\ref{eq:svm_weight}), where \(x\) is the training data and each vectorized sentence \(x_i\) in the data is labeled \(y_i\). Each cycle of the algorithm alters the value of \(w\) by \(\alpha\) to reduce the number of wrong classifications. This equation shows the last value of \(\alpha\) after the end of the cycle.

    \begin{equation}\label{eq:svm1}
    f(x) = w^\top x + b
    \end{equation}

    \begin{equation}\label{eq:svm_weight}
    w = \sum_{i=1}^N \alpha_i y_i x_i
    \end{equation}

    We tagged 159 Chinese sentences and \num[group-separator={,}]{2357} English sentences as positive or negative for our training data. The entropy comparison factors \(\alpha\) and \(\alpha'\) were tested from 1.25 to 6 in intervals of 0.25. We applied this SVC to classify the rest of our data collection. Subsequently, the positive and negative sentence counts shown in Table \ref{tab:exp_notes} result from applying our SVC for classification.

\section{Data Analysis}\label{dataanalysis}

  \subsection{Frequent keywords in differently priced hotels}\label{svmresults}

    We observed the top 10 satisfaction and dissatisfaction keywords with the highest frequencies of emotionally positive and negative statements to study. The keywords are the quantitative rank of the needs of Chinese and English-speaking customers. We show the top 10 positive keywords for each price range comparing English and Chinese in Table \ref{tab:freq_res_pos}. For the negative keywords, we show the results in Table \ref{tab:freq_res_neg}.

    We can observe that the most used keywords for most price ranges in the same language are similar, with a few changes in priority for the keywords involved. For example, in Chinese, we can see that the customers praise cleanliness first in cheaper hotels, whereas the size of the room or bed is praised more in hotels of higher class. Another example is that in negative English reviews, complaints about price appear only after 10,000 yen hotels. After this, it climbs in importance following the increase in the hotel's price.

    % \usepackage{multirow}
    % \usepackage{graphicx}
    % \usepackage[normalem]{ulem}
    \begin{table}[ht]
      \centering
      \caption{English and Chinese comparison of the top 10 positive keywords.}
      \label{tab:freq_res_pos}
      \resizebox{\textwidth}{!}{%
      \begin{tabular}{|c|lr|lr|}
        \hline
        \textbf{Price range} &
          \multicolumn{1}{c|}{\textbf{Chinese keyword}} &
          \multicolumn{1}{c|}{\textbf{Counts in Ctrip}} &
          \multicolumn{1}{c|}{\textbf{English keyword}} &
          \multicolumn{1}{c|}{\textbf{Counts in Tripadvisor}} \\ \hline
        \multirow{10}{*}{\textbf{0: All Prices}}             & \begin{CJK}{UTF8}{gbsn}不错\end{CJK} (not bad)         & 12892 & good        & 19148 \\  
                                                             & \begin{CJK}{UTF8}{gbsn}大\end{CJK} (big)              & 9844  & staff       & 16289 \\  
                                                             & \begin{CJK}{UTF8}{gbsn}干净\end{CJK} (clean)           & 6665  & great       & 16127 \\  
                                                             & \begin{CJK}{UTF8}{gbsn}交通\end{CJK} (traffic)         & 6560  & location    & 11838 \\  
                                                             & \begin{CJK}{UTF8}{gbsn}早餐\end{CJK} (breakfast)       & 5605  & nice        & 11615 \\  
                                                             & \begin{CJK}{UTF8}{gbsn}近\end{CJK} (near)             & 5181  & clean       & 9064  \\  
                                                             & \begin{CJK}{UTF8}{gbsn}地铁\end{CJK} (subway)          & 4321  & helpful     & 5846  \\  
                                                             & \begin{CJK}{UTF8}{gbsn}购物\end{CJK} (shopping)        & 4101  & excellent   & 5661  \\  
                                                             & \begin{CJK}{UTF8}{gbsn}推荐\end{CJK} (recommend)       & 3281  & comfortable & 5625  \\  
                                                             & \begin{CJK}{UTF8}{gbsn}环境\end{CJK} (environment)    & 3258  & friendly    & 5606  \\ \hline
        \multirow{10}{*}{\textbf{3: 5000 to 10,000 yen}}     & \begin{CJK}{UTF8}{gbsn}不错\end{CJK} (not bad)         & 139   & good        & 206   \\  
                                                             & \begin{CJK}{UTF8}{gbsn}干净\end{CJK} (clean)           & 114   & staff       & 181   \\  
                                                             & \begin{CJK}{UTF8}{gbsn}早餐\end{CJK} (breakfast)       & 112   & clean       & 174   \\  
                                                             & \begin{CJK}{UTF8}{gbsn}大\end{CJK} (big)              & 76    & nice        & 166   \\  
                                                             & \begin{CJK}{UTF8}{gbsn}交通\end{CJK} (traffic)         & 72    & great       & 143   \\  
                                                             & \begin{CJK}{UTF8}{gbsn}地铁\end{CJK} (subway)          & 66    & location    & 91    \\  
                                                             & \begin{CJK}{UTF8}{gbsn}近\end{CJK} (near)             & 55    & comfortable & 79    \\  
                                                             & \begin{CJK}{UTF8}{gbsn}地铁站\end{CJK} (subway station) & 51    & helpful     & 70    \\  
                                                             & \begin{CJK}{UTF8}{gbsn}远\end{CJK} (far)              & 41    & friendly    & 64    \\  
                                                             & \begin{CJK}{UTF8}{gbsn}附近\end{CJK} (nearby)          & 34    & recommend   & 59    \\ \hline
        \multirow{10}{*}{\textbf{4: 10,000 to 15,000 yen}}   & \begin{CJK}{UTF8}{gbsn}不错\end{CJK} (not bad)         & 601   & good        & 1399  \\  
                                                             & \begin{CJK}{UTF8}{gbsn}干净\end{CJK} (clean)           & 455   & staff       & 1165  \\  
                                                             & \begin{CJK}{UTF8}{gbsn}大\end{CJK} (big)              & 348   & great       & 961   \\  
                                                             & \begin{CJK}{UTF8}{gbsn}近\end{CJK} (near)             & 323   & nice        & 808   \\  
                                                             & \begin{CJK}{UTF8}{gbsn}早餐\end{CJK} (breakfast)       & 270   & location    & 800   \\  
                                                             & \begin{CJK}{UTF8}{gbsn}卫生\end{CJK} (health)          & 201   & clean       & 656   \\  
                                                             & \begin{CJK}{UTF8}{gbsn}交通\end{CJK} (traffic)         & 196   & excellent   & 412   \\  
                                                             & \begin{CJK}{UTF8}{gbsn}地铁\end{CJK} (subway)          & 164   & friendly    & 400   \\  
                                                             & \begin{CJK}{UTF8}{gbsn}远\end{CJK} (far)              & 158   & helpful     & 393   \\  
                                                             & \begin{CJK}{UTF8}{gbsn}附近\end{CJK} (nearby)          & 150   & comfortable & 391   \\ \hline
        \multirow{10}{*}{\textbf{5: 15,000 to 20,000 yen}}   & \begin{CJK}{UTF8}{gbsn}不错\end{CJK} (not bad)         & 1925  & good        & 2242  \\  
                                                             & \begin{CJK}{UTF8}{gbsn}干净\end{CJK} (clean)           & 1348  & staff       & 1674  \\  
                                                             & \begin{CJK}{UTF8}{gbsn}大\end{CJK} (big)              & 1277  & great       & 1414  \\  
                                                             & \begin{CJK}{UTF8}{gbsn}交通\end{CJK} (traffic)         & 1058  & clean       & 1204  \\  
                                                             & \begin{CJK}{UTF8}{gbsn}近\end{CJK} (near)             & 1016  & nice        & 1175  \\  
                                                             & \begin{CJK}{UTF8}{gbsn}地铁\end{CJK} (subway)          & 801   & location    & 1109  \\  
                                                             & \begin{CJK}{UTF8}{gbsn}早餐\end{CJK} (breakfast)       & 777   & comfortable & 621   \\  
                                                             & \begin{CJK}{UTF8}{gbsn}地铁站\end{CJK} (subway station) & 639   & friendly    & 615   \\  
                                                             & \begin{CJK}{UTF8}{gbsn}附近\end{CJK} (nearby)          & 572   & free        & 581   \\  
                                                             & \begin{CJK}{UTF8}{gbsn}购物\end{CJK} (shopping)        & 516   & helpful     & 552   \\ \hline
        \multirow{10}{*}{\textbf{6: 20,000 to 30,000 yen}}   & \begin{CJK}{UTF8}{gbsn}不错\end{CJK} (not bad)         & 3110  & good        & 6550  \\  
                                                             & \begin{CJK}{UTF8}{gbsn}大\end{CJK} (big)              & 2245  & staff       & 5348  \\  
                                                             & \begin{CJK}{UTF8}{gbsn}交通\end{CJK} (traffic)         & 1990  & great       & 5074  \\  
                                                             & \begin{CJK}{UTF8}{gbsn}干净\end{CJK} (clean)           & 1940  & location    & 4414  \\  
                                                             & \begin{CJK}{UTF8}{gbsn}近\end{CJK} (near)             & 1433  & nice        & 3451  \\  
                                                             & \begin{CJK}{UTF8}{gbsn}地铁\end{CJK} (subway)          & 1073  & clean       & 3364  \\  
                                                             & \begin{CJK}{UTF8}{gbsn}早餐\end{CJK} (breakfast)       & 1007  & shopping    & 1992  \\  
                                                             & \begin{CJK}{UTF8}{gbsn}购物\end{CJK} (shopping)        & 979   & helpful     & 1970  \\  
                                                             & \begin{CJK}{UTF8}{gbsn}周边\end{CJK} (surroundings)    & 837   & comfortable & 1941  \\  
                                                             & \begin{CJK}{UTF8}{gbsn}附近\end{CJK} (nearby)          & 825   & friendly    & 1915  \\ \hline
        \multirow{10}{*}{\textbf{7: 30,000 to 50,000 yen}}   & \begin{CJK}{UTF8}{gbsn}不错\end{CJK} (not bad)         & 2291  & good        & 3407  \\  
                                                             & \begin{CJK}{UTF8}{gbsn}大\end{CJK} (big)              & 1913  & staff       & 2867  \\  
                                                             & \begin{CJK}{UTF8}{gbsn}干净\end{CJK} (clean)           & 1159  & great       & 2620  \\  
                                                             & \begin{CJK}{UTF8}{gbsn}交通\end{CJK} (traffic)         & 1105  & location    & 2186  \\  
                                                             & \begin{CJK}{UTF8}{gbsn}近\end{CJK} (near)             & 935   & nice        & 2160  \\  
                                                             & \begin{CJK}{UTF8}{gbsn}早餐\end{CJK} (breakfast)       & 846   & clean       & 1750  \\  
                                                             & \begin{CJK}{UTF8}{gbsn}推荐\end{CJK} (recommend)       & 638   & helpful     & 1147  \\  
                                                             & \begin{CJK}{UTF8}{gbsn}购物\end{CJK} (shopping)        & 636   & train       & 1040  \\  
                                                             & \begin{CJK}{UTF8}{gbsn}周边\end{CJK} (surroundings)    & 552   & subway      & 1034  \\  
                                                             & \begin{CJK}{UTF8}{gbsn}环境\end{CJK} (environment)    & 541   & friendly    & 1001  \\ \hline
        \multirow{10}{*}{\textbf{8: 50,000 to 100,000 yen}}  & \begin{CJK}{UTF8}{gbsn}不错\end{CJK} (not bad)         & 4451  & great       & 4425  \\  
                                                             & \begin{CJK}{UTF8}{gbsn}大\end{CJK} (big)              & 3670  & good        & 4350  \\  
                                                             & \begin{CJK}{UTF8}{gbsn}早餐\end{CJK} (breakfast)       & 2422  & staff       & 3777  \\  
                                                             & \begin{CJK}{UTF8}{gbsn}交通\end{CJK} (traffic)         & 2012  & nice        & 2991  \\  
                                                             & \begin{CJK}{UTF8}{gbsn}购物\end{CJK} (shopping)        & 1764  & location    & 2439  \\  
                                                             & \begin{CJK}{UTF8}{gbsn}新\end{CJK} (new)              & 1634  & clean       & 1655  \\  
                                                             & \begin{CJK}{UTF8}{gbsn}棒\end{CJK} (great)            & 1626  & excellent   & 1555  \\  
                                                             & \begin{CJK}{UTF8}{gbsn}地铁\end{CJK} (subway)          & 1604  & helpful     & 1313  \\  
                                                             & \begin{CJK}{UTF8}{gbsn}干净\end{CJK} (clean)           & 1577  & comfortable & 1246  \\  
                                                             & \begin{CJK}{UTF8}{gbsn}近\end{CJK} (near)             & 1354  & friendly    & 1238  \\ \hline
        \multirow{10}{*}{\textbf{9: 100,000 to 200,000 yen}} & \begin{CJK}{UTF8}{gbsn}不错\end{CJK} (not bad)         & 375   & great       & 1488  \\  
                                                             & \begin{CJK}{UTF8}{gbsn}大\end{CJK} (big)              & 315   & staff       & 1277  \\  
                                                             & \begin{CJK}{UTF8}{gbsn}棒\end{CJK} (great)            & 189   & good        & 994   \\  
                                                             & \begin{CJK}{UTF8}{gbsn}早餐\end{CJK} (breakfast)       & 171   & nice        & 864   \\  
                                                             & \begin{CJK}{UTF8}{gbsn}环境\end{CJK} (environment)    & 157   & location    & 799   \\  
                                                             & \begin{CJK}{UTF8}{gbsn}交通\end{CJK} (traffic)         & 127   & excellent   & 631   \\  
                                                             & \begin{CJK}{UTF8}{gbsn}选择\end{CJK} (select)          & 112   & beautiful   & 455   \\  
                                                             & \begin{CJK}{UTF8}{gbsn}推荐\end{CJK} (recommend)       & 109   & large       & 404   \\  
                                                             & \begin{CJK}{UTF8}{gbsn}赞\end{CJK} (awesome)          & 101   & helpful     & 401   \\  
                                                             & \begin{CJK}{UTF8}{gbsn}购物\end{CJK} (shopping)        & 98    & wonderful   & 372   \\ \hline
        \end{tabular}%
        }
    \end{table}

    % \usepackage{multirow}
    % \usepackage{graphicx}
    \begin{table}[ht]
      \centering
      \caption{English and Chinese comparison of the top 10 negative keywords.}
      \label{tab:freq_res_neg}
      \resizebox{\textwidth}{!}{%
      \begin{tabular}{|c|lr|lr|}
        \hline
        \textbf{Price range} &
          \multicolumn{1}{c|}{\textbf{Chinese keyword}} &
          \multicolumn{1}{c|}{\textbf{Counts in Ctrip}} &
          \multicolumn{1}{c|}{\textbf{English keyword}} &
          \multicolumn{1}{c|}{\textbf{Counts in Tripadvisor}} \\ \hline
        \multirow{10}{*}{\textbf{0: All Prices}}             & \begin{CJK}{UTF8}{gbsn}价格\end{CJK} (price)     & 1838 & pricey         & 462 \\  
                                                             & \begin{CJK}{UTF8}{gbsn}一般\end{CJK} (general)   & 1713 & poor           & 460 \\  
                                                             & \begin{CJK}{UTF8}{gbsn}中文\end{CJK} (Chinese)   & 733  & dated          & 431 \\  
                                                             & \begin{CJK}{UTF8}{gbsn}地理\end{CJK} (geography) & 691  & disappointing  & 376 \\  
                                                             & \begin{CJK}{UTF8}{gbsn}距离\end{CJK} (distance)  & 434  & worst          & 327 \\  
                                                             & \begin{CJK}{UTF8}{gbsn}陈旧\end{CJK} (obsolete)  & 319  & minor          & 258 \\  
                                                             & \begin{CJK}{UTF8}{gbsn}老\end{CJK} (old)        & 297  & uncomfortable  & 253 \\  
                                                             & \begin{CJK}{UTF8}{gbsn}华人\end{CJK} (Chinese)   & 15   & carpet         & 240 \\  
                                                             &                                                &      & annoying       & 220 \\  
                                                             &                                                &      & sense          & 220 \\ \hline
        \multirow{10}{*}{\textbf{3: 5000 to 10,000 yen}}     & \begin{CJK}{UTF8}{gbsn}价格\end{CJK} (price)     & 31   & worst          & 6   \\  
                                                             & \begin{CJK}{UTF8}{gbsn}一般\end{CJK} (general)   & 28   & walkway        & 5   \\  
                                                             & \begin{CJK}{UTF8}{gbsn}距离\end{CJK} (distance)  & 11   & unable         & 4   \\  
                                                             & \begin{CJK}{UTF8}{gbsn}地理\end{CJK} (geography) & 10   & worse          & 4   \\  
                                                             & \begin{CJK}{UTF8}{gbsn}中文\end{CJK} (Chinese)   & 9    & annoying       & 3   \\  
                                                             & \begin{CJK}{UTF8}{gbsn}老\end{CJK} (old)        & 2    & dirty          & 3   \\  
                                                             &                                                &      & funny smell    & 3   \\  
                                                             &                                                &      & poor           & 3   \\  
                                                             &                                                &      & renovation     & 3   \\  
                                                             &                                                &      & carpet         & 2   \\ \hline
        \multirow{10}{*}{\textbf{4: 10,000 to 15,000 yen}}   & \begin{CJK}{UTF8}{gbsn}价格\end{CJK} (price)     & 98   & dated          & 40  \\  
                                                             & \begin{CJK}{UTF8}{gbsn}一般\end{CJK} (general)   & 91   & poor           & 29  \\  
                                                             & \begin{CJK}{UTF8}{gbsn}距离\end{CJK} (distance)  & 43   & disappointing  & 26  \\  
                                                             & \begin{CJK}{UTF8}{gbsn}陈旧\end{CJK} (obsolete)  & 34   & worst          & 24  \\  
                                                             & \begin{CJK}{UTF8}{gbsn}地理\end{CJK} (geography) & 31   & uncomfortable  & 23  \\  
                                                             & \begin{CJK}{UTF8}{gbsn}老\end{CJK} (old)        & 30   & cigarette      & 22  \\  
                                                             & \begin{CJK}{UTF8}{gbsn}中文\end{CJK} (Chinese)   & 26   & pricey         & 22  \\  
                                                             &                                                &      & minor          & 21  \\  
                                                             &                                                &      & paper          & 19  \\  
                                                             &                                                &      & unable         & 19  \\ \hline
        \multirow{10}{*}{\textbf{5: 15,000 to 20,000 yen}}   & \begin{CJK}{UTF8}{gbsn}价格\end{CJK} (price)     & 296  & poor           & 57  \\  
                                                             & \begin{CJK}{UTF8}{gbsn}一般\end{CJK} (general)   & 218  & dated          & 41  \\  
                                                             & \begin{CJK}{UTF8}{gbsn}地理\end{CJK} (geography) & 125  & disappointing  & 38  \\  
                                                             & \begin{CJK}{UTF8}{gbsn}中文\end{CJK} (Chinese)   & 93   & annoying       & 36  \\  
                                                             & \begin{CJK}{UTF8}{gbsn}距离\end{CJK} (distance)  & 84   & worst          & 36  \\  
                                                             & \begin{CJK}{UTF8}{gbsn}陈旧\end{CJK} (obsolete)  & 43   & cigarette      & 31  \\  
                                                             & \begin{CJK}{UTF8}{gbsn}老\end{CJK} (old)        & 26   & rude           & 28  \\  
                                                             & \begin{CJK}{UTF8}{gbsn}华人\end{CJK} (Chinese)   & 3    & uncomfortable  & 26  \\  
                                                             &                                                &      & paper          & 25  \\  
                                                             &                                                &      & pricey         & 24  \\ \hline
        \multirow{10}{*}{\textbf{6: 20,000 to 30,000 yen}}   & \begin{CJK}{UTF8}{gbsn}一般\end{CJK} (general)   & 504  & poor           & 136 \\  
                                                             & \begin{CJK}{UTF8}{gbsn}价格\end{CJK} (price)     & 472  & dated          & 131 \\  
                                                             & \begin{CJK}{UTF8}{gbsn}地理\end{CJK} (geography) & 164  & pricey         & 120 \\  
                                                             & \begin{CJK}{UTF8}{gbsn}中文\end{CJK} (Chinese)   & 155  & disappointing  & 112 \\  
                                                             & \begin{CJK}{UTF8}{gbsn}距离\end{CJK} (distance)  & 116  & uncomfortable  & 103 \\  
                                                             & \begin{CJK}{UTF8}{gbsn}陈旧\end{CJK} (obsolete)  & 75   & minor          & 93  \\  
                                                             & \begin{CJK}{UTF8}{gbsn}老\end{CJK} (old)        & 55   & smallest       & 88  \\  
                                                             & \begin{CJK}{UTF8}{gbsn}华人\end{CJK} (Chinese)   & 2    & worst          & 86  \\  
                                                             &                                                &      & cigarette      & 79  \\  
                                                             &                                                &      & annoying       & 70  \\ \hline
        \multirow{10}{*}{\textbf{7: 30,000 to 50,000 yen}}   & \begin{CJK}{UTF8}{gbsn}价格\end{CJK} (price)     & 326  & poor           & 92  \\  
                                                             & \begin{CJK}{UTF8}{gbsn}一般\end{CJK} (general)   & 311  & pricey         & 92  \\  
                                                             & \begin{CJK}{UTF8}{gbsn}地理\end{CJK} (geography) & 110  & dated          & 65  \\  
                                                             & \begin{CJK}{UTF8}{gbsn}中文\end{CJK} (Chinese)   & 94   & worst          & 64  \\  
                                                             & \begin{CJK}{UTF8}{gbsn}陈旧\end{CJK} (obsolete)  & 71   & carpet         & 55  \\  
                                                             & \begin{CJK}{UTF8}{gbsn}距离\end{CJK} (distance)  & 68   & uncomfortable  & 55  \\  
                                                             & \begin{CJK}{UTF8}{gbsn}老\end{CJK} (old)        & 45   & dirty          & 51  \\  
                                                             & \begin{CJK}{UTF8}{gbsn}华人\end{CJK} (Chinese)   & 2    & disappointing  & 50  \\  
                                                             &                                                &      & cigarette      & 46  \\  
                                                             &                                                &      & unable         & 43  \\ \hline
        \multirow{10}{*}{\textbf{8: 50,000 to 100,000 yen}} &
                                                               \begin{CJK}{UTF8}{gbsn}价格\end{CJK} (price)     & 561  & pricey         & 163 \\ 
                                                             & \begin{CJK}{UTF8}{gbsn}一般\end{CJK} (general)   & 510  & dated          & 150 \\  
                                                             & \begin{CJK}{UTF8}{gbsn}中文\end{CJK} (Chinese)   & 337  & disappointing  & 129 \\  
                                                             & \begin{CJK}{UTF8}{gbsn}地理\end{CJK} (geography) & 239  & poor           & 124 \\  
                                                             & \begin{CJK}{UTF8}{gbsn}老\end{CJK} (old)        & 134  & worst          & 98  \\  
                                                             & \begin{CJK}{UTF8}{gbsn}距离\end{CJK} (distance)  & 97   & walkway        & 82  \\  
                                                             & \begin{CJK}{UTF8}{gbsn}陈旧\end{CJK} (obsolete)  & 90   & carpet         & 71  \\  
                                                             & \begin{CJK}{UTF8}{gbsn}华人\end{CJK} (Chinese)   & 8    & minor          & 63  \\  
                                                             &                                                &      & sense          & 63  \\  
                                                             &                                                &      & outdated       & 58  \\ \hline
        \multirow{10}{*}{\textbf{9: 100,000 to 200,000 yen}} & \begin{CJK}{UTF8}{gbsn}价格\end{CJK} (price)     & 54   & pricey         & 40  \\  
                                                             & \begin{CJK}{UTF8}{gbsn}一般\end{CJK} (general)   & 51   & sense          & 34  \\  
                                                             & \begin{CJK}{UTF8}{gbsn}中文\end{CJK} (Chinese)   & 19   & minor          & 33  \\  
                                                             & \begin{CJK}{UTF8}{gbsn}距离\end{CJK} (distance)  & 15   & lighting       & 20  \\  
                                                             & \begin{CJK}{UTF8}{gbsn}地理\end{CJK} (geography) & 12   & disappointing  & 19  \\  
                                                             & \begin{CJK}{UTF8}{gbsn}陈旧\end{CJK} (obsolete)  & 6    & poor           & 19  \\  
                                                             & \begin{CJK}{UTF8}{gbsn}老\end{CJK} (old)        & 5    & annoying       & 16  \\  
                                                             &                                                &      & mixed          & 15  \\  
                                                             &                                                &      & disappointment & 14  \\  
                                                             &                                                &      & paper          & 14  \\ \hline
        \end{tabular}%
        }
    \end{table}

  \subsection{Frequently used adjectives and their pairs}\label{adjresults}

    Some keywords in these lists are adjectives, such as the word ``\begin{CJK}{UTF8}{gbsn}大\end{CJK} (big)'' mentioned before. To understand those, we performed the dependency parsing and part of speech tagging explained in section \ref{textprocessing}. While many of these connections, we only considered the top 4 used keyword connections per adjective per price range. We show the most used Chinese adjectives in positive keywords in Table \ref{tab:adj_zh_pos}, and for negative Chinese adjective keywords in Table \ref{tab:adj_zh_neg}. Similarly, for English adjectives used in positive sentences we show the most common examples in Table \ref{tab:adj_en_pos}, and for adjectives used in negative sentences in Table \ref{tab:adj_en_neg}.

    %%%%%%%%%%%%% POSITIVE CHINESE ADJECTIVE PAIRS
    % \usepackage{multirow}
    % \usepackage{graphicx}
    % \usepackage{lscape}
    \begin{landscape}
    \begin{table}[p]
      \centering
      \caption{Top 4 words related to the mainly used adjectives in positive Chinese texts.}
      \label{tab:adj_zh_pos}
      \resizebox{\linewidth}{!}{%
      \begin{tabular}{|c|l|l|l|l|l|l|}
        \hline
        \textbf{Price range} &
          \multicolumn{1}{c|}{\textbf{\begin{CJK}{UTF8}{gbsn}不错\end{CJK} (not bad)}} &
          \multicolumn{1}{c|}{\textbf{\begin{CJK}{UTF8}{gbsn}大\end{CJK} (big)}} &
          \multicolumn{1}{c|}{\textbf{\begin{CJK}{UTF8}{gbsn}干净\end{CJK} (clean)}} &
          \multicolumn{1}{c|}{\textbf{\begin{CJK}{UTF8}{gbsn}近\end{CJK} (near)}} &
          \multicolumn{1}{c|}{\textbf{\begin{CJK}{UTF8}{gbsn}新\end{CJK} (new)}} &
          \multicolumn{1}{c|}{\textbf{\begin{CJK}{UTF8}{gbsn}棒\end{CJK} (great)}} \\ \hline
        \multirow{5}{*}{\textbf{0: All Prices}} &
          \begin{CJK}{UTF8}{gbsn}不错\end{CJK} (not bad) : 12892 &
          \begin{CJK}{UTF8}{gbsn}大\end{CJK} (big) : 9844 &
          \begin{CJK}{UTF8}{gbsn}干净\end{CJK} (clean) : 6665 &
          \begin{CJK}{UTF8}{gbsn}近\end{CJK} (near) : 5181 &
          \begin{CJK}{UTF8}{gbsn}新\end{CJK} (new) : 2775 &
          \begin{CJK}{UTF8}{gbsn}棒\end{CJK} (great) : 3028 \\
         &
          \begin{CJK}{UTF8}{gbsn}不错 酒店\end{CJK} (nice hotel) : 1462 &
          \begin{CJK}{UTF8}{gbsn}大 房间\end{CJK} (big room) : 3197 &
          \begin{CJK}{UTF8}{gbsn}干净 房间\end{CJK} (clean room) : 1224 &
          \begin{CJK}{UTF8}{gbsn}近 酒店\end{CJK} (near hotel) : 453 &
          \begin{CJK}{UTF8}{gbsn}新 设施\end{CJK} (new facility) : 363 &
          \begin{CJK}{UTF8}{gbsn}棒 酒店\end{CJK} (great hotel) : 463 \\
         &
          \begin{CJK}{UTF8}{gbsn}不错 位置\end{CJK} (nice location) : 1426 &
          \begin{CJK}{UTF8}{gbsn}大 床\end{CJK} (big bed) : 772 &
          \begin{CJK}{UTF8}{gbsn}干净 酒店\end{CJK} (clean hotel) : 737 &
          \begin{CJK}{UTF8}{gbsn}近 桥\end{CJK} (near bridge) : 144 &
          \begin{CJK}{UTF8}{gbsn}新 酒店\end{CJK} (new hotel) : 246 &
          \begin{CJK}{UTF8}{gbsn}棒 位置\end{CJK} (great position) : 218 \\
         &
          \begin{CJK}{UTF8}{gbsn}不错 服务\end{CJK} (nice service) : 869 &
          \begin{CJK}{UTF8}{gbsn}大 酒店\end{CJK} (big hotel) : 379 &
          \begin{CJK}{UTF8}{gbsn}干净 卫生\end{CJK} (clean and hygienic) : 464 &
          \begin{CJK}{UTF8}{gbsn}近 地铁站\end{CJK} (near subway station) : 122 &
          \begin{CJK}{UTF8}{gbsn}新 装修\end{CJK} (new decoration) : 116 &
          \begin{CJK}{UTF8}{gbsn}棒 服务\end{CJK} (great service) : 168 \\
         &
          \begin{CJK}{UTF8}{gbsn}不错 环境\end{CJK} (nice environment) : 714 &
          \begin{CJK}{UTF8}{gbsn}大 超市\end{CJK} (big supermarket) : 232 &
          \begin{CJK}{UTF8}{gbsn}干净 环境\end{CJK} (clean environment) : 61 &
          \begin{CJK}{UTF8}{gbsn}近 站\end{CJK} (near station) : 108 &
          \begin{CJK}{UTF8}{gbsn}新 房间\end{CJK} (new room) : 53 &
          \begin{CJK}{UTF8}{gbsn}棒 早餐\end{CJK} (great breakfast) : 164 \\ \hline
        \multirow{5}{*}{\textbf{\begin{tabular}[c]{@{}c@{}}3: 5000 to\\  10,000 yen\end{tabular}}} &
          \begin{CJK}{UTF8}{gbsn}不错\end{CJK} (not bad) : 139 &
          \begin{CJK}{UTF8}{gbsn}大\end{CJK} (big) : 76 &
          \begin{CJK}{UTF8}{gbsn}干净\end{CJK} (clean) : 114 &
          \begin{CJK}{UTF8}{gbsn}近\end{CJK} (near) : 55 &
           &
          \begin{CJK}{UTF8}{gbsn}棒\end{CJK} (great) : 11 \\
         &
          \begin{CJK}{UTF8}{gbsn}不错 酒店\end{CJK} (nice hotel) : 17 &
          \begin{CJK}{UTF8}{gbsn}大 房间\end{CJK} (big room) : 11 &
          \begin{CJK}{UTF8}{gbsn}干净 房间\end{CJK} (clean room) : 21 &
          \begin{CJK}{UTF8}{gbsn}近 酒店\end{CJK} (near hotel) : 4 &
           &
          \begin{CJK}{UTF8}{gbsn}棒 位置\end{CJK} (great position) : 2 \\
         &
          \begin{CJK}{UTF8}{gbsn}不错 位置\end{CJK} (nice location) : 16 &
          \begin{CJK}{UTF8}{gbsn}大 床\end{CJK} (big bed) : 10 &
          \begin{CJK}{UTF8}{gbsn}干净 酒店\end{CJK} (clean hotel) : 10 &
          \begin{CJK}{UTF8}{gbsn}近 地铁\end{CJK} (near subway) : 2 &
           &
           \\
         &
          \begin{CJK}{UTF8}{gbsn}不错 早餐\end{CJK} (nice breakfast) : 12 &
          \begin{CJK}{UTF8}{gbsn}大 超市\end{CJK} (big supermarket) : 5 &
          \begin{CJK}{UTF8}{gbsn}干净 卫生\end{CJK} (clean and hygienic) : 6 &
           &
           &
           \\
         &
          \begin{CJK}{UTF8}{gbsn}不错 服务\end{CJK} (nice service) : 8 &
          \begin{CJK}{UTF8}{gbsn}大 商场\end{CJK} (big market) : 3 &
          \begin{CJK}{UTF8}{gbsn}干净 总体\end{CJK} (clean overall) : 4 &
           &
           &
           \\ \hline
        \multirow{5}{*}{\textbf{\begin{tabular}[c]{@{}c@{}}4: 10,000 to\\  15,000 yen\end{tabular}}} &
          \begin{CJK}{UTF8}{gbsn}不错\end{CJK} (not bad) : 601 &
          \begin{CJK}{UTF8}{gbsn}大\end{CJK} (big) : 348 &
          \begin{CJK}{UTF8}{gbsn}干净\end{CJK} (clean) : 455 &
          \begin{CJK}{UTF8}{gbsn}近\end{CJK} (near) : 323 &
          \begin{CJK}{UTF8}{gbsn}新\end{CJK} (new) : 37 &
          \begin{CJK}{UTF8}{gbsn}棒\end{CJK} (great) : 73 \\
         &
          \begin{CJK}{UTF8}{gbsn}不错 位置\end{CJK} (nice location) : 72 &
          \begin{CJK}{UTF8}{gbsn}大 房间\end{CJK} (big room) : 76 &
          \begin{CJK}{UTF8}{gbsn}干净 房间\end{CJK} (clean room) : 66 &
          \begin{CJK}{UTF8}{gbsn}近 酒店\end{CJK} (near hotel) : 27 &
          \begin{CJK}{UTF8}{gbsn}新 设施\end{CJK} (new facility) : 9 &
          \begin{CJK}{UTF8}{gbsn}棒 位置\end{CJK} (great position) : 6 \\
         &
          \begin{CJK}{UTF8}{gbsn}不错 酒店\end{CJK} (nice hotel) : 37 &
          \begin{CJK}{UTF8}{gbsn}大 床\end{CJK} (big bed) : 30 &
          \begin{CJK}{UTF8}{gbsn}干净 卫生\end{CJK} (clean and hygienic) : 52 &
          \begin{CJK}{UTF8}{gbsn}近 站\end{CJK} (near station) : 14 &
          \begin{CJK}{UTF8}{gbsn}新 装修\end{CJK} (new decoration) : 2 &
          \begin{CJK}{UTF8}{gbsn}棒 房间\end{CJK} (great room) : 3 \\
         &
          \begin{CJK}{UTF8}{gbsn}不错 服务\end{CJK} (nice service) : 34 &
          \begin{CJK}{UTF8}{gbsn}大 社\end{CJK} (big club) : 26 &
          \begin{CJK}{UTF8}{gbsn}干净 酒店\end{CJK} (clean hotel) : 48 &
          \begin{CJK}{UTF8}{gbsn}近 地铁\end{CJK} (near subway) : 12 &
          \begin{CJK}{UTF8}{gbsn}新 酒店\end{CJK} (new hotel) : 2 &
          \begin{CJK}{UTF8}{gbsn}棒 水平\end{CJK} (great level) : 3 \\
         &
          \begin{CJK}{UTF8}{gbsn}不错 早餐\end{CJK} (nice breakfast) : 26 &
          \begin{CJK}{UTF8}{gbsn}大 空间\end{CJK} (big space) : 16 &
          \begin{CJK}{UTF8}{gbsn}干净 打扫\end{CJK} (clean up) : 9 &
          \begin{CJK}{UTF8}{gbsn}近 车站\end{CJK} (near the station) : 10 &
           &
          \begin{CJK}{UTF8}{gbsn}棒 温泉\end{CJK} (great hot spring) : 3 \\ \hline
        \multirow{5}{*}{\textbf{\begin{tabular}[c]{@{}c@{}}5: 15,000 to\\  20,000 yen\end{tabular}}} &
          \begin{CJK}{UTF8}{gbsn}不错\end{CJK} (not bad) : 1925 &
          \begin{CJK}{UTF8}{gbsn}大\end{CJK} (big) : 1277 &
          \begin{CJK}{UTF8}{gbsn}干净\end{CJK} (clean) : 1348 &
          \begin{CJK}{UTF8}{gbsn}近\end{CJK} (near) : 1016 &
          \begin{CJK}{UTF8}{gbsn}新\end{CJK} (new) : 234 &
          \begin{CJK}{UTF8}{gbsn}棒\end{CJK} (great) : 241 \\
         &
          \begin{CJK}{UTF8}{gbsn}不错 位置\end{CJK} (nice location) : 207 &
          \begin{CJK}{UTF8}{gbsn}大 房间\end{CJK} (big room) : 316 &
          \begin{CJK}{UTF8}{gbsn}干净 房间\end{CJK} (clean room) : 234 &
          \begin{CJK}{UTF8}{gbsn}近 酒店\end{CJK} (near hotel) : 82 &
          \begin{CJK}{UTF8}{gbsn}新 设施\end{CJK} (new facility) : 47 &
          \begin{CJK}{UTF8}{gbsn}棒 位置\end{CJK} (great position) : 33 \\
         &
          \begin{CJK}{UTF8}{gbsn}不错 酒店\end{CJK} (nice hotel) : 168 &
          \begin{CJK}{UTF8}{gbsn}大 床\end{CJK} (big bed) : 140 &
          \begin{CJK}{UTF8}{gbsn}干净 酒店\end{CJK} (clean hotel) : 161 &
          \begin{CJK}{UTF8}{gbsn}近 站\end{CJK} (near station) : 35 &
          \begin{CJK}{UTF8}{gbsn}新 酒店\end{CJK} (new hotel) : 25 &
          \begin{CJK}{UTF8}{gbsn}棒 酒店\end{CJK} (great hotel) : 25 \\
         &
          \begin{CJK}{UTF8}{gbsn}不错 服务\end{CJK} (nice service) : 131 &
          \begin{CJK}{UTF8}{gbsn}大 超市\end{CJK} (big supermarket) : 73 &
          \begin{CJK}{UTF8}{gbsn}干净 卫生\end{CJK} (clean and hygienic) : 92 &
          \begin{CJK}{UTF8}{gbsn}近 地铁站\end{CJK} (near subway station) : 34 &
          \begin{CJK}{UTF8}{gbsn}新 装修\end{CJK} (new decoration) : 15 &
          \begin{CJK}{UTF8}{gbsn}棒 服务\end{CJK} (great service) : 22 \\
         &
          \begin{CJK}{UTF8}{gbsn}不错 早餐\end{CJK} (nice breakfast) : 109 &
          \begin{CJK}{UTF8}{gbsn}大 酒店\end{CJK} (big hotel) : 49 &
          \begin{CJK}{UTF8}{gbsn}干净 设施\end{CJK} (clean facilities) : 19 &
          \begin{CJK}{UTF8}{gbsn}近 桥\end{CJK} (near bridge) : 29 &
          \begin{CJK}{UTF8}{gbsn}新 房间\end{CJK} (new room) : 10 &
          \begin{CJK}{UTF8}{gbsn}棒 早餐\end{CJK} (great breakfast) : 8 \\ \hline
        \multirow{5}{*}{\textbf{\begin{tabular}[c]{@{}c@{}}6: 20,000 to\\  30,000 yen\end{tabular}}} &
          \begin{CJK}{UTF8}{gbsn}不错\end{CJK} (not bad) : 3110 &
          \begin{CJK}{UTF8}{gbsn}大\end{CJK} (big) : 2245 &
          \begin{CJK}{UTF8}{gbsn}干净\end{CJK} (clean) : 1940 &
          \begin{CJK}{UTF8}{gbsn}近\end{CJK} (near) : 1433 &
          \begin{CJK}{UTF8}{gbsn}新\end{CJK} (new) : 517 &
          \begin{CJK}{UTF8}{gbsn}棒\end{CJK} (great) : 440 \\
         &
          \begin{CJK}{UTF8}{gbsn}不错 位置\end{CJK} (nice location) : 409 &
          \begin{CJK}{UTF8}{gbsn}大 房间\end{CJK} (big room) : 680 &
          \begin{CJK}{UTF8}{gbsn}干净 房间\end{CJK} (clean room) : 360 &
          \begin{CJK}{UTF8}{gbsn}近 酒店\end{CJK} (near hotel) : 164 &
          \begin{CJK}{UTF8}{gbsn}新 设施\end{CJK} (new facility) : 89 &
          \begin{CJK}{UTF8}{gbsn}棒 酒店\end{CJK} (great hotel) : 51 \\
         &
          \begin{CJK}{UTF8}{gbsn}不错 酒店\end{CJK} (nice hotel) : 326 &
          \begin{CJK}{UTF8}{gbsn}大 床\end{CJK} (big bed) : 198 &
          \begin{CJK}{UTF8}{gbsn}干净 酒店\end{CJK} (clean hotel) : 203 &
          \begin{CJK}{UTF8}{gbsn}近 地铁\end{CJK} (near subway) : 34 &
          \begin{CJK}{UTF8}{gbsn}新 酒店\end{CJK} (new hotel) : 51 &
          \begin{CJK}{UTF8}{gbsn}棒 位置\end{CJK} (great position) : 45 \\
         &
          \begin{CJK}{UTF8}{gbsn}不错 服务\end{CJK} (nice service) : 206 &
          \begin{CJK}{UTF8}{gbsn}大 酒店\end{CJK} (big hotel) : 102 &
          \begin{CJK}{UTF8}{gbsn}干净 卫生\end{CJK} (clean and hygienic) : 137 &
          \begin{CJK}{UTF8}{gbsn}近 地铁站\end{CJK} (near subway station) : 31 &
          \begin{CJK}{UTF8}{gbsn}新 装修\end{CJK} (new decoration) : 24 &
          \begin{CJK}{UTF8}{gbsn}棒 服务\end{CJK} (great service) : 23 \\
         &
          \begin{CJK}{UTF8}{gbsn}不错 环境\end{CJK} (nice environment) : 183 &
          \begin{CJK}{UTF8}{gbsn}大 空间\end{CJK} (big space) : 64 &
          \begin{CJK}{UTF8}{gbsn}干净 环境\end{CJK} (clean environment) : 21 &
          \begin{CJK}{UTF8}{gbsn}近 车站\end{CJK} (near the station) : 27 &
          \begin{CJK}{UTF8}{gbsn}新 房间\end{CJK} (new room) : 10 &
          \begin{CJK}{UTF8}{gbsn}棒 早餐\end{CJK} (great breakfast) : 20 \\ \hline
        \multirow{5}{*}{\textbf{\begin{tabular}[c]{@{}c@{}}7: 30,000 to\\  50,000 yen\end{tabular}}} &
          \begin{CJK}{UTF8}{gbsn}不错\end{CJK} (not bad) : 2291 &
          \begin{CJK}{UTF8}{gbsn}大\end{CJK} (big) : 1913 &
          \begin{CJK}{UTF8}{gbsn}干净\end{CJK} (clean) : 1159 &
          \begin{CJK}{UTF8}{gbsn}近\end{CJK} (near) : 935 &
          \begin{CJK}{UTF8}{gbsn}新\end{CJK} (new) : 260 &
          \begin{CJK}{UTF8}{gbsn}棒\end{CJK} (great) : 448 \\
         &
          \begin{CJK}{UTF8}{gbsn}不错 位置\end{CJK} (nice location) : 277 &
          \begin{CJK}{UTF8}{gbsn}大 房间\end{CJK} (big room) : 643 &
          \begin{CJK}{UTF8}{gbsn}干净 房间\end{CJK} (clean room) : 224 &
          \begin{CJK}{UTF8}{gbsn}近 酒店\end{CJK} (near hotel) : 80 &
          \begin{CJK}{UTF8}{gbsn}新 设施\end{CJK} (new facility) : 63 &
          \begin{CJK}{UTF8}{gbsn}棒 酒店\end{CJK} (great hotel) : 68 \\
         &
          \begin{CJK}{UTF8}{gbsn}不错 酒店\end{CJK} (nice hotel) : 274 &
          \begin{CJK}{UTF8}{gbsn}大 床\end{CJK} (big bed) : 141 &
          \begin{CJK}{UTF8}{gbsn}干净 酒店\end{CJK} (clean hotel) : 146 &
          \begin{CJK}{UTF8}{gbsn}近 站\end{CJK} (near station) : 24 &
          \begin{CJK}{UTF8}{gbsn}新 酒店\end{CJK} (new hotel) : 25 &
          \begin{CJK}{UTF8}{gbsn}棒 位置\end{CJK} (great position) : 34 \\
         &
          \begin{CJK}{UTF8}{gbsn}不错 服务\end{CJK} (nice service) : 140 &
          \begin{CJK}{UTF8}{gbsn}大 超市\end{CJK} (big supermarket) : 74 &
          \begin{CJK}{UTF8}{gbsn}干净 卫生\end{CJK} (clean and hygienic) : 71 &
          \begin{CJK}{UTF8}{gbsn}近 桥\end{CJK} (near bridge) : 20 &
          \begin{CJK}{UTF8}{gbsn}新 装修\end{CJK} (new decoration) : 15 &
          \begin{CJK}{UTF8}{gbsn}棒 服务\end{CJK} (great service) : 24 \\
         &
          \begin{CJK}{UTF8}{gbsn}不错 环境\end{CJK} (nice environment) : 140 &
          \begin{CJK}{UTF8}{gbsn}大 酒店\end{CJK} (big hotel) : 66 &
          \begin{CJK}{UTF8}{gbsn}干净 环境\end{CJK} (clean environment) : 16 &
          \begin{CJK}{UTF8}{gbsn}近 山\end{CJK} (near mountain) : 12 &
          \begin{CJK}{UTF8}{gbsn}新 房间\end{CJK} (new room) : 11 &
          \begin{CJK}{UTF8}{gbsn}棒 早餐\end{CJK} (great breakfast) : 14 \\ \hline
        \multirow{5}{*}{\textbf{\begin{tabular}[c]{@{}c@{}}8: 50,000 to\\  100,000 yen\end{tabular}}} &
          \begin{CJK}{UTF8}{gbsn}不错\end{CJK} (not bad) : 4451 &
          \begin{CJK}{UTF8}{gbsn}大\end{CJK} (big) : 3670 &
          \begin{CJK}{UTF8}{gbsn}干净\end{CJK} (clean) : 1577 &
          \begin{CJK}{UTF8}{gbsn}近\end{CJK} (near) : 1354 &
          \begin{CJK}{UTF8}{gbsn}新\end{CJK} (new) : 1634 &
          \begin{CJK}{UTF8}{gbsn}棒\end{CJK} (great) : 1626 \\
         &
          \begin{CJK}{UTF8}{gbsn}不错 酒店\end{CJK} (nice hotel) : 587 &
          \begin{CJK}{UTF8}{gbsn}大 房间\end{CJK} (big room) : 1340 &
          \begin{CJK}{UTF8}{gbsn}干净 房间\end{CJK} (clean room) : 310 &
          \begin{CJK}{UTF8}{gbsn}近 酒店\end{CJK} (near hotel) : 88 &
          \begin{CJK}{UTF8}{gbsn}新 设施\end{CJK} (new facility) : 141 &
          \begin{CJK}{UTF8}{gbsn}棒 酒店\end{CJK} (great hotel) : 281 \\
         &
          \begin{CJK}{UTF8}{gbsn}不错 位置\end{CJK} (nice location) : 415 &
          \begin{CJK}{UTF8}{gbsn}大 床\end{CJK} (big bed) : 238 &
          \begin{CJK}{UTF8}{gbsn}干净 酒店\end{CJK} (clean hotel) : 161 &
          \begin{CJK}{UTF8}{gbsn}近 桥\end{CJK} (near bridge) : 76 &
          \begin{CJK}{UTF8}{gbsn}新 酒店\end{CJK} (new hotel) : 123 &
          \begin{CJK}{UTF8}{gbsn}棒 早餐\end{CJK} (great breakfast) : 112 \\
         &
          \begin{CJK}{UTF8}{gbsn}不错 服务\end{CJK} (nice service) : 328 &
          \begin{CJK}{UTF8}{gbsn}大 酒店\end{CJK} (big hotel) : 144 &
          \begin{CJK}{UTF8}{gbsn}干净 卫生\end{CJK} (clean and hygienic) : 101 &
          \begin{CJK}{UTF8}{gbsn}近 地铁站\end{CJK} (near subway station) : 35 &
          \begin{CJK}{UTF8}{gbsn}新 装修\end{CJK} (new decoration) : 57 &
          \begin{CJK}{UTF8}{gbsn}棒 位置\end{CJK} (great position) : 96 \\
         &
          \begin{CJK}{UTF8}{gbsn}不错 早餐\end{CJK} (nice breakfast) : 251 &
          \begin{CJK}{UTF8}{gbsn}大 商场\end{CJK} (big market) : 88 &
          \begin{CJK}{UTF8}{gbsn}干净 服务\end{CJK} (clean service) : 13 &
          \begin{CJK}{UTF8}{gbsn}近 铁\end{CJK} (Kintetsu) : 24 &
          \begin{CJK}{UTF8}{gbsn}新 斋\end{CJK} (new) : 22 &
          \begin{CJK}{UTF8}{gbsn}棒 服务\end{CJK} (great service) : 86 \\ \hline
        \multirow{5}{*}{\textbf{\begin{tabular}[c]{@{}c@{}}9: 100,000 to\\  200,000 yen\end{tabular}}} &
          \begin{CJK}{UTF8}{gbsn}不错\end{CJK} (not bad) : 375 &
          \begin{CJK}{UTF8}{gbsn}大\end{CJK} (big) : 315 &
          \begin{CJK}{UTF8}{gbsn}干净\end{CJK} (clean) : 72 &
          \begin{CJK}{UTF8}{gbsn}近\end{CJK} (near) : 65 &
          \begin{CJK}{UTF8}{gbsn}新\end{CJK} (new) : 77 &
          \begin{CJK}{UTF8}{gbsn}棒\end{CJK} (great) : 189 \\
         &
          \begin{CJK}{UTF8}{gbsn}不错 酒店\end{CJK} (nice hotel) : 53 &
          \begin{CJK}{UTF8}{gbsn}大 房间\end{CJK} (big room) : 131 &
          \begin{CJK}{UTF8}{gbsn}干净 房间\end{CJK} (clean room) : 9 &
          \begin{CJK}{UTF8}{gbsn}近 酒店\end{CJK} (near hotel) : 8 &
          \begin{CJK}{UTF8}{gbsn}新 酒店\end{CJK} (new hotel) : 19 &
          \begin{CJK}{UTF8}{gbsn}棒 酒店\end{CJK} (great hotel) : 36 \\
         &
          \begin{CJK}{UTF8}{gbsn}不错 位置\end{CJK} (nice location) : 30 &
          \begin{CJK}{UTF8}{gbsn}大 面积\end{CJK} (large area) : 19 &
          \begin{CJK}{UTF8}{gbsn}干净 酒店\end{CJK} (clean hotel) : 8 &
          \begin{CJK}{UTF8}{gbsn}近 地铁站\end{CJK} (near subway station) : 3 &
          \begin{CJK}{UTF8}{gbsn}新 设施\end{CJK} (new facility) : 13 &
          \begin{CJK}{UTF8}{gbsn}棒 体验\end{CJK} (great experience) : 10 \\
         &
          \begin{CJK}{UTF8}{gbsn}不错 环境\end{CJK} (nice environment) : 27 &
          \begin{CJK}{UTF8}{gbsn}大 床\end{CJK} (big bed) : 15 &
          \begin{CJK}{UTF8}{gbsn}干净 卫生\end{CJK} (clean and hygienic) : 5 &
          \begin{CJK}{UTF8}{gbsn}近 市场\end{CJK} (near market) : 3 &
          \begin{CJK}{UTF8}{gbsn}新 装修\end{CJK} (new decoration) : 3 &
          \begin{CJK}{UTF8}{gbsn}棒 服务\end{CJK} (great service) : 10 \\
         &
          \begin{CJK}{UTF8}{gbsn}不错 服务\end{CJK} (nice service) : 22 &
          \begin{CJK}{UTF8}{gbsn}大 卫生间\end{CJK} (big toilet) : 13 &
           &
           &
          \begin{CJK}{UTF8}{gbsn}新 位置\end{CJK} (new location) : 2 &
          \begin{CJK}{UTF8}{gbsn}棒 早餐\end{CJK} (great breakfast) : 8 \\ \hline
        \end{tabular}%
        }
    \end{table}
    \end{landscape}

    %%%%%%%%%%%%% NEGATIVE CHINESE ADJECTIVE PAIRS
    % \usepackage{multirow}
    % \usepackage{graphicx}
    % \usepackage{lscape}
    \begin{landscape}
    \begin{table}[p]
      \centering
      \caption{Top 4 words related to the mainly used adjectives in negative texts.}
      \label{tab:adj_zh_neg}
      \resizebox{0.7\linewidth}{!}{%
      \begin{tabular}{|c|l|l|l|}
        \hline
        \multicolumn{1}{|l|}{\textbf{Price range}} &
          \multicolumn{1}{c|}{\textbf{\begin{CJK}{UTF8}{gbsn}一般\end{CJK} (general)}} &
          \multicolumn{1}{c|}{\textbf{\begin{CJK}{UTF8}{gbsn}陈旧\end{CJK} (obsolete)}} &
          \multicolumn{1}{c|}{\textbf{\begin{CJK}{UTF8}{gbsn}老\end{CJK} (old)}} \\ \hline
        \multirow{5}{*}{\textbf{0: All Prices}} &
          \begin{CJK}{UTF8}{gbsn}一般\end{CJK} (general) : 1713 &
          \begin{CJK}{UTF8}{gbsn}陈旧\end{CJK} (obsolete) : 319 &
          \begin{CJK}{UTF8}{gbsn}老\end{CJK} (old) : 297 \\
         &
          \begin{CJK}{UTF8}{gbsn}一般 设施\end{CJK} (general facilities) : 137 &
          \begin{CJK}{UTF8}{gbsn}陈旧 设施\end{CJK} (obsolete facilities) : 184 &
          \begin{CJK}{UTF8}{gbsn}老 酒店\end{CJK} (old hotel) : 74 \\
         &
          \begin{CJK}{UTF8}{gbsn}一般 服务\end{CJK} (general service) : 115 &
          \begin{CJK}{UTF8}{gbsn}陈旧 设备\end{CJK} (obsolete equipment) : 18 &
          \begin{CJK}{UTF8}{gbsn}老 设施\end{CJK} (old facility) : 58 \\
         &
          \begin{CJK}{UTF8}{gbsn}一般 酒店\end{CJK} (average hotel) : 106 &
          \begin{CJK}{UTF8}{gbsn}陈旧 房间\end{CJK} (outdated room) : 10 &
          \begin{CJK}{UTF8}{gbsn}老 店\end{CJK} (old shop) : 15 \\
         &
          \begin{CJK}{UTF8}{gbsn}一般 早餐\end{CJK} (average breakfast) : 97 &
          \begin{CJK}{UTF8}{gbsn}陈旧 酒店\end{CJK} (outdated hotel) : 10 &
          \begin{CJK}{UTF8}{gbsn}老 装修\end{CJK} (old decoration) : 11 \\ \hline
        \multirow{5}{*}{\textbf{3: 5000 to 10,000 yen}} &
          \begin{CJK}{UTF8}{gbsn}一般\end{CJK} (general) : 28 &
           &
          \begin{CJK}{UTF8}{gbsn}老\end{CJK} (old) : 2 \\
         &
          \begin{CJK}{UTF8}{gbsn}一般 设施\end{CJK} (general facilities) : 5 &
           &
           \\
         &
          \begin{CJK}{UTF8}{gbsn}一般 早餐\end{CJK} (average breakfast) : 3 &
           &
           \\
         &
          \begin{CJK}{UTF8}{gbsn}一般 味道\end{CJK} (general taste) : 2 &
           &
           \\
         &
          \begin{CJK}{UTF8}{gbsn}一般 效果\end{CJK} (general effect) : 2 &
           &
           \\ \hline
        \multirow{5}{*}{\textbf{4: 10,000 to 15,000 yen}} &
          \begin{CJK}{UTF8}{gbsn}一般\end{CJK} (general) : 91 &
          \begin{CJK}{UTF8}{gbsn}陈旧\end{CJK} (obsolete) : 34 &
          \begin{CJK}{UTF8}{gbsn}老\end{CJK} (old) : 30 \\
         &
          \begin{CJK}{UTF8}{gbsn}一般 设施\end{CJK} (general facilities) : 10 &
          \begin{CJK}{UTF8}{gbsn}陈旧 设施\end{CJK} (obsolete facilities) : 17 &
          \begin{CJK}{UTF8}{gbsn}老 酒店\end{CJK} (old hotel) : 8 \\
         &
          \begin{CJK}{UTF8}{gbsn}一般 位置\end{CJK} (general location) : 8 &
          \begin{CJK}{UTF8}{gbsn}陈旧 家具\end{CJK} (obsolete furniture) : 2 &
          \begin{CJK}{UTF8}{gbsn}老 设施\end{CJK} (old facility) : 7 \\
         &
          \begin{CJK}{UTF8}{gbsn}一般 酒店\end{CJK} (average hotel) : 6 &
          \begin{CJK}{UTF8}{gbsn}陈旧 设备\end{CJK} (obsolete equipment) : 2 &
          \begin{CJK}{UTF8}{gbsn}老 建筑\end{CJK} (old building) : 3 \\
         &
          \begin{CJK}{UTF8}{gbsn}一般 早餐\end{CJK} (average breakfast) : 5 &
           &
           \\ \hline
        \multirow{5}{*}{\textbf{5: 15,000 to 20,000 yen}} &
          \begin{CJK}{UTF8}{gbsn}一般\end{CJK} (general) : 218 &
          \begin{CJK}{UTF8}{gbsn}陈旧\end{CJK} (obsolete) : 43 &
          \begin{CJK}{UTF8}{gbsn}老\end{CJK} (old) : 26 \\
         &
          \begin{CJK}{UTF8}{gbsn}一般 设施\end{CJK} (general facilities) : 23 &
          \begin{CJK}{UTF8}{gbsn}陈旧 设施\end{CJK} (obsolete facilities) : 25 &
          \begin{CJK}{UTF8}{gbsn}老 酒店\end{CJK} (old hotel) : 11 \\
         &
          \begin{CJK}{UTF8}{gbsn}一般 酒店\end{CJK} (average hotel) : 21 &
          \begin{CJK}{UTF8}{gbsn}陈旧 设备\end{CJK} (obsolete equipment) : 3 &
          \begin{CJK}{UTF8}{gbsn}老 设施\end{CJK} (old facility) : 7 \\
         &
          \begin{CJK}{UTF8}{gbsn}一般 早餐\end{CJK} (average breakfast) : 14 &
          \begin{CJK}{UTF8}{gbsn}陈旧 酒店\end{CJK} (outdated hotel) : 2 &
          \begin{CJK}{UTF8}{gbsn}老 外观\end{CJK} (old appearance) : 2 \\
         &
          \begin{CJK}{UTF8}{gbsn}一般 卫生\end{CJK} (general hygiene) : 8 &
           &
           \\ \hline
        \multirow{5}{*}{\textbf{6: 20,000 to 30,000 yen}} &
          \begin{CJK}{UTF8}{gbsn}一般\end{CJK} (general) : 504 &
          \begin{CJK}{UTF8}{gbsn}陈旧\end{CJK} (obsolete) : 75 &
          \begin{CJK}{UTF8}{gbsn}老\end{CJK} (old) : 55 \\
         &
          \begin{CJK}{UTF8}{gbsn}一般 设施\end{CJK} (general facilities) : 42 &
          \begin{CJK}{UTF8}{gbsn}陈旧 设施\end{CJK} (obsolete facilities) : 42 &
          \begin{CJK}{UTF8}{gbsn}老 酒店\end{CJK} (old hotel) : 9 \\
         &
          \begin{CJK}{UTF8}{gbsn}一般 酒店\end{CJK} (average hotel) : 37 &
          \begin{CJK}{UTF8}{gbsn}陈旧 设备\end{CJK} (obsolete equipment) : 7 &
          \begin{CJK}{UTF8}{gbsn}老 设施\end{CJK} (old facility) : 8 \\
         &
          \begin{CJK}{UTF8}{gbsn}一般 服务\end{CJK} (general service) : 34 &
          \begin{CJK}{UTF8}{gbsn}陈旧 装修\end{CJK} (old decoration) : 3 &
          \begin{CJK}{UTF8}{gbsn}老 店\end{CJK} (old shop) : 3 \\
         &
          \begin{CJK}{UTF8}{gbsn}一般 早餐\end{CJK} (average breakfast) : 21 &
          \begin{CJK}{UTF8}{gbsn}陈旧 酒店\end{CJK} (outdated hotel) : 2 &
          \begin{CJK}{UTF8}{gbsn}老 房间\end{CJK} (old room) : 3 \\ \hline
        \multirow{5}{*}{\textbf{7: 30,000 to 50,000 yen}} &
          \begin{CJK}{UTF8}{gbsn}一般\end{CJK} (general) : 311 &
          \begin{CJK}{UTF8}{gbsn}陈旧\end{CJK} (obsolete) : 71 &
          \begin{CJK}{UTF8}{gbsn}老\end{CJK} (old) : 45 \\
         &
          \begin{CJK}{UTF8}{gbsn}一般 设施\end{CJK} (general facilities) : 23 &
          \begin{CJK}{UTF8}{gbsn}陈旧 设施\end{CJK} (obsolete facilities) : 43 &
          \begin{CJK}{UTF8}{gbsn}老 酒店\end{CJK} (old hotel) : 11 \\
         &
          \begin{CJK}{UTF8}{gbsn}一般 服务\end{CJK} (general service) : 22 &
          \begin{CJK}{UTF8}{gbsn}陈旧 设备\end{CJK} (obsolete equipment) : 5 &
          \begin{CJK}{UTF8}{gbsn}老 设施\end{CJK} (old facility) : 7 \\
         &
          \begin{CJK}{UTF8}{gbsn}一般 早餐\end{CJK} (average breakfast) : 19 &
          \begin{CJK}{UTF8}{gbsn}陈旧 房间\end{CJK} (outdated room) : 3 &
          \begin{CJK}{UTF8}{gbsn}老 店\end{CJK} (old shop) : 3 \\
         &
          \begin{CJK}{UTF8}{gbsn}一般 酒店\end{CJK} (average hotel) : 15 &
           &
          \begin{CJK}{UTF8}{gbsn}老 房间\end{CJK} (old room) : 2 \\ \hline
        \multirow{5}{*}{\textbf{8: 50,000 to 100,000}} &
          \begin{CJK}{UTF8}{gbsn}一般\end{CJK} (general) : 510 &
          \begin{CJK}{UTF8}{gbsn}陈旧\end{CJK} (obsolete) : 90 &
          \begin{CJK}{UTF8}{gbsn}老\end{CJK} (old) : 134 \\
         &
          \begin{CJK}{UTF8}{gbsn}一般 服务\end{CJK} (general service) : 39 &
          \begin{CJK}{UTF8}{gbsn}陈旧 设施\end{CJK} (obsolete facilities) : 53 &
          \begin{CJK}{UTF8}{gbsn}老 酒店\end{CJK} (old hotel) : 34 \\
         &
          \begin{CJK}{UTF8}{gbsn}一般 设施\end{CJK} (general facilities) : 32 &
          \begin{CJK}{UTF8}{gbsn}陈旧 房间\end{CJK} (outdated room) : 5 &
          \begin{CJK}{UTF8}{gbsn}老 设施\end{CJK} (old facility) : 26 \\
         &
          \begin{CJK}{UTF8}{gbsn}一般 早餐\end{CJK} (average breakfast) : 30 &
          \begin{CJK}{UTF8}{gbsn}陈旧 感觉\end{CJK} (Stale feeling) : 2 &
          \begin{CJK}{UTF8}{gbsn}老 装修\end{CJK} (old decoration) : 9 \\
         &
          \begin{CJK}{UTF8}{gbsn}一般 酒店\end{CJK} (average hotel) : 25 &
           &
          \begin{CJK}{UTF8}{gbsn}老 店\end{CJK} (old shop) : 7 \\ \hline
        \multirow{5}{*}{\textbf{9: 100,000 to 200,000}} &
          \begin{CJK}{UTF8}{gbsn}一般\end{CJK} (general) : 51 &
          \begin{CJK}{UTF8}{gbsn}陈旧\end{CJK} (obsolete) : 6 &
          \begin{CJK}{UTF8}{gbsn}老\end{CJK} (old) : 5 \\
         &
          \begin{CJK}{UTF8}{gbsn}一般 服务\end{CJK} (general service) : 7 &
          \begin{CJK}{UTF8}{gbsn}陈旧 设施\end{CJK} (obsolete facilities) : 4 &
          \begin{CJK}{UTF8}{gbsn}老 设施\end{CJK} (old facility) : 2 \\
         &
          \begin{CJK}{UTF8}{gbsn}一般 早餐\end{CJK} (average breakfast) : 5 &
           &
           \\
         &
          \begin{CJK}{UTF8}{gbsn}一般 位置\end{CJK} (general location) : 2 &
           &
           \\
         &
          \begin{CJK}{UTF8}{gbsn}一般 房间\end{CJK} (average room) : 2 &
           &
           \\ \hline
        \end{tabular}%
        }
    \end{table}
    \end{landscape}

    %%%%%%%%%%%%% POSITIVE ENGLISH ADJECTIVE PAIRS
    % \usepackage{multirow}
    % \usepackage{graphicx}
    % \usepackage[normalem]{ulem}
    % \usepackage{lscape}
    \begin{landscape}
    \begin{table}[p]
      \centering
      \caption{Top 4 words related to the mainly used adjectives in positive English texts.}
      \label{tab:adj_en_pos}
      \resizebox{\linewidth}{!}{%
      \begin{tabular}{|c|l|l|l|l|l|l|l|l|}
        \hline
        \textbf{Price range} &
          \multicolumn{1}{c|}{\textbf{good}} &
          \multicolumn{1}{c|}{\textbf{clean}} &
          \multicolumn{1}{c|}{\textbf{comfortable}} &
          \multicolumn{1}{c|}{\textbf{helpful}} &
          \multicolumn{1}{c|}{\textbf{free}} &
          \multicolumn{1}{c|}{\textbf{large}} &
          \multicolumn{1}{c|}{\textbf{firendly}} &
          \multicolumn{1}{c|}{\textbf{great}} \\ \hline
        \multirow{5}{*}{\textbf{0: All Prices}} &
          good : 19148 &
          clean : 9064 &
          comfortable : 5625 &
          helpful : 5846 &
          free : 4318 &
          large : 4104 &
          friendly : 5606 &
          great : 16127 \\
         &
          good location : 1985 &
          clean room : 3596 &
          comfortable bed : 1919 &
          helpful staff : 2927 &
          free wifi : 773 &
          large room : 1256 &
          friendly staff : 3819 &
          great location : 2313 \\
         &
          good service : 1042 &
          clean hotel : 969 &
          comfortable room : 1098 &
          helpful concierge : 304 &
          free shuttle : 286 &
          large hotel : 268 &
          friendly service : 169 &
          great view : 1099 \\
         &
          good breakfast : 942 &
          clean bathroom : 282 &
          comfortable stay : 272 &
          helpful desk : 110 &
          free drink : 234 &
          large bathroom : 202 &
          friendly hotel : 73 &
          great service : 841 \\
         &
          good hotel : 874 &
          clean everything : 200 &
          comfortable hotel : 238 &
          helpful service : 74 &
          free bus : 225 &
          larger room : 192 &
          friendly person : 63 &
          great hotel : 802 \\ \hline
        \multirow{5}{*}{\textbf{\begin{tabular}[c]{@{}c@{}}3: 5000 to\\ 10,000 yen\end{tabular}}} &
          good : 206 &
          clean : 174 &
          comfortable : 79 &
          helpful : 70 &
          free : 35 &
          large : 31 &
          friendly : 64 &
          great : 143 \\
         &
          good location : 30 &
          clean room : 55 &
          comfortable bed : 21 &
          helpful staff : 36 &
          free wifi : 10 &
          large room : 7 &
          friendly staff : 53 &
          great location : 21 \\
         &
          good value : 19 &
          clean bathroom : 14 &
          comfortable room : 9 &
           &
          free tea : 4 &
          large area : 2 &
          friendly everyone : 2 &
          great view : 14 \\
         &
          good english : 10 &
          clean place : 12 &
          comfortable futon : 8 &
           &
          free raman : 2 &
          large size : 2 &
          friendly service : 2 &
          great place : 13 \\
         &
          good place : 7 &
          clean hotel : 6 &
          comfortable stay : 3 &
           &
          free toothbrush : 2 &
           &
           &
          great experience : 5 \\ \hline
        \multirow{5}{*}{\textbf{\begin{tabular}[c]{@{}c@{}}4: 10,000 to\\  15,000 yen\end{tabular}}} &
          good : 1399 &
          clean : 656 &
          comfortable : 391 &
          helpful : 393 &
          free : 271 &
          large : 250 &
          friendly : 400 &
          great : 961 \\
         &
          good location : 159 &
          clean room : 247 &
          comfortable bed : 123 &
          helpful staff : 206 &
          free wifi : 53 &
          large room : 84 &
          friendly staff : 292 &
          great location : 158 \\
         &
          good breakfast : 87 &
          clean hotel : 74 &
          comfortable room : 90 &
          helpful concierge : 20 &
          free breakfast : 15 &
          large bathroom : 20 &
          friendly service : 15 &
          great service : 51 \\
         &
          good hotel : 71 &
          clean bathroom : 20 &
          comfortable hotel : 26 &
          helpful desk : 10 &
          free service : 12 &
          larger room : 12 &
          friendly hotel : 7 &
          great hotel : 43 \\
         &
          good service : 67 &
          clean everything : 14 &
          comfortable stay : 20 &
          helpful service : 4 &
          free drink : 11 &
          large hotel : 10 &
          friendly person : 6 &
          great place : 35 \\ \hline
        \multirow{5}{*}{\textbf{\begin{tabular}[c]{@{}c@{}}5: 15,000 to\\  20,000 yen\end{tabular}}} &
          good : 2242 &
          clean : 1204 &
          comfortable : 621 &
          helpful : 552 &
          free : 581 &
          large : 349 &
          friendly : 615 &
          great : 1414 \\
         &
          good location : 242 &
          clean room : 440 &
          comfortable bed : 219 &
          helpful staff : 301 &
          free wifi : 109 &
          large room : 85 &
          friendly staff : 444 &
          great location : 199 \\
         &
          good hotel : 116 &
          clean hotel : 133 &
          comfortable room : 99 &
          helpful desk : 11 &
          free shuttle : 35 &
          large suitcase : 18 &
          friendly hotel : 12 &
          great view : 81 \\
         &
          good breakfast : 113 &
          clean bathroom : 38 &
          comfortable stay : 30 &
          helpful concierge : 9 &
          free bus : 30 &
          larger room : 18 &
          friendly service : 8 &
          great hotel : 68 \\
         &
          good service : 108 &
          clean everything : 26 &
          comfortable hotel : 20 &
          helpful reception : 5 &
          free breakfast : 27 &
          large hotel : 17 &
          friendly most : 7 &
          great place : 61 \\ \hline
        \multirow{5}{*}{\textbf{\begin{tabular}[c]{@{}c@{}}6: 20,000 to\\  30,000 yen\end{tabular}}} &
          good : 6550 &
          clean : 3364 &
          comfortable : 1941 &
          helpful : 1970 &
          free : 1186 &
          large : 1257 &
          friendly : 1915 &
          great : 5074 \\
         &
          good location : 703 &
          clean room : 1379 &
          comfortable bed : 658 &
          helpful staff : 1019 &
          free wifi : 269 &
          large room : 329 &
          friendly staff : 1311 &
          great location : 881 \\
         &
          good service : 331 &
          clean hotel : 379 &
          comfortable room : 359 &
          helpful concierge : 79 &
          free breakfast : 68 &
          large hotel : 87 &
          friendly service : 51 &
          great service : 249 \\
         &
          good english : 304 &
          clean bathroom : 95 &
          comfortable stay : 100 &
          helpful desk : 42 &
          free coffee : 57 &
          larger room : 81 &
          friendly person : 21 &
          great hotel : 232 \\
         &
          good breakfast : 303 &
          clean everything : 77 &
          comfortable hotel : 82 &
          helpful receptionist : 17 &
          free drink : 38 &
          large bed : 43 &
          friendly hotel : 19 &
          great view : 220 \\ \hline
        \multirow{5}{*}{\textbf{\begin{tabular}[c]{@{}c@{}}7: 30,000 to\\  50,000 yen\end{tabular}}} &
          good : 3407 &
          clean : 1750 &
          comfortable : 1000 &
          helpful : 1147 &
          free : 933 &
          large : 580 &
          friendly : 1001 &
          great : 2620 \\
         &
          good location : 380 &
          clean room : 725 &
          comfortable bed : 345 &
          helpful staff : 607 &
          free drink : 145 &
          large room : 174 &
          friendly staff : 715 &
          great location : 393 \\
         &
          good breakfast : 191 &
          clean hotel : 197 &
          comfortable room : 193 &
          helpful concierge : 53 &
          free wifi : 129 &
          larger room : 32 &
          friendly service : 24 &
          great view : 162 \\
         &
          good service : 182 &
          clean bathroom : 61 &
          comfortable hotel : 49 &
          helpful service : 20 &
          free coffee : 45 &
          large hotel : 30 &
          friendly hotel : 13 &
          great hotel : 134 \\
         &
          good english : 155 &
          clean everything : 36 &
          comfortable stay : 47 &
          helpful desk : 17 &
          free bus : 38 &
          large bed : 28 &
          friendly person : 13 &
          great service : 114 \\ \hline
        \multirow{5}{*}{\textbf{\begin{tabular}[c]{@{}c@{}}8: 50,000 to\\  100,000 yen\end{tabular}}} &
          good : 4350 &
          clean : 1655 &
          comfortable : 1246 &
          helpful : 1313 &
          free : 1072 &
          large : 1233 &
          friendly : 1238 &
          great : 4425 \\
         &
          good location : 406 &
          clean room : 648 &
          comfortable bed : 425 &
          helpful staff : 589 &
          free shuttle : 181 &
          large room : 442 &
          friendly staff : 810 &
          great location : 506 \\
         &
          good service : 296 &
          clean hotel : 156 &
          comfortable room : 266 &
          helpful concierge : 108 &
          free wifi : 172 &
          large hotel : 109 &
          friendly service : 51 &
          great view : 436 \\
         &
          good hotel : 196 &
          clean bathroom : 48 &
          comfortable stay : 56 &
          helpful service : 28 &
          free bus : 127 &
          large bathroom : 58 &
          friendly hotel : 20 &
          great service : 267 \\
         &
          good breakfast : 191 &
          cleanliness : 40 &
          comfortable hotel : 51 &
          helpful desk : 26 &
          free service : 65 &
          larger room : 38 &
          friendly person : 12 &
          great hotel : 241 \\ \hline
        \multirow{5}{*}{\textbf{\begin{tabular}[c]{@{}c@{}}9: 100,000 to\\  200,000 yen\end{tabular}}} &
          good : 994 &
          clean : 261 &
          comfortable : 347 &
          helpful : 401 &
          free : 240 &
          large : 404 &
          friendly : 370 &
          great : 1488 \\
         &
          good location : 65 &
          clean room : 102 &
          comfortable bed : 128 &
          helpful staff : 169 &
          free wifi : 31 &
          large room : 135 &
          friendly staff : 194 &
          great location : 155 \\
         &
          good service : 56 &
          clean hotel : 24 &
          comfortable room : 82 &
          helpful concierge : 35 &
          free breakfast : 19 &
          large bathroom : 38 &
          friendly service : 18 &
          great view : 155 \\
         &
          good breakfast : 53 &
          cleanliness : 8 &
          comfortable stay : 16 &
          helpful everyone : 7 &
          free drink : 16 &
          large hotel : 15 &
          friendly everyone : 7 &
          great service : 101 \\
         &
          good hotel : 40 &
          clean place : 7 &
          comfortable hotel : 10 &
          helpful team : 5 &
          free bus : 14 &
          large bed : 12 &
          friendly person : 4 &
          great hotel : 80 \\ \hline
        \end{tabular}%
        }
    \end{table}
    \end{landscape}

    %%%%%%%%%%%%% NEGATIVE ENGLISH ADJECTIVE PAIRS
    % \usepackage{multirow}
    % \usepackage{graphicx}
    % \usepackage{lscape}
    \begin{landscape}
    \begin{table}[p]
      \centering
      \caption{Top 4 words related to the mainly used adjectives in negative English texts.}
      \label{tab:adj_en_neg}
      \resizebox{0.8\linewidth}{!}{%
      \begin{tabular}{|c|l|l|l|l|l|}
        \hline
        \textbf{Price range} &
          \multicolumn{1}{c|}{\textbf{poor}} &
          \multicolumn{1}{c|}{\textbf{dated}} &
          \multicolumn{1}{c|}{\textbf{worst}} &
          \multicolumn{1}{c|}{\textbf{dirty}} &
          \multicolumn{1}{c|}{\textbf{uncomfortable}} \\ \hline
        \multirow{5}{*}{\textbf{0: All Prices}} &
          poor : 460 &
          dated : 431 &
          worst : 327 &
          dirty : 188 &
          uncomfortable : 253 \\
         &
          poor service : 55 &
          outdated : 128 &
          worst hotel : 43 &
          dirty carpet : 34 &
          uncomfortable bed : 63 \\
         &
          poor breakfast : 41 &
          outdated room : 20 &
          worst experience : 18 &
          dirty room : 23 &
          uncomfortable pillow : 20 \\
         &
          poor quality : 27 &
          outdated hotel : 10 &
          worst part : 15 &
          not dirty : 7 &
          uncomfortable mattress : 8 \\
         &
          poor english : 24 &
          outdated bathroom : 7 &
          worst service : 10 &
          dirty bathroom : 6 &
          uncomfortable night : 8 \\ \hline
        \multirow{5}{*}{\textbf{3: 5000 to 10,000 yen}} &
          poor : 3 &
           &
          worst : 6 &
          dirty : 3 &
          uncomfortable : 2 \\
         &
           &
           &
          worst room : 2 &
           &
           \\
         &
           &
           &
           &
           &
           \\
         &
           &
           &
           &
           &
           \\
         &
           &
           &
           &
           &
           \\ \hline
        \multirow{5}{*}{\textbf{4: 10,000 to 15,000 yen}} &
          poor : 29 &
          dated : 40 &
          worst : 24 &
          dirty : 11 &
          uncomfortable : 23 \\
         &
          poor breakfast : 3 &
          outdated : 11 &
          worst hotel : 4 &
          dirty floor : 2 &
          uncomfortable bed : 4 \\
         &
          poor service : 3 &
          outdated decor : 2 &
          worst experience : 2 &
           &
          not uncomfortable : 2 \\
         &
          poor conditioning : 2 &
          outdated room : 2 &
           &
           &
          uncomfortable night : 2 \\
         &
          poor view : 2 &
           &
           &
           &
          uncomfortable pillow : 2 \\ \hline
        \multirow{5}{*}{\textbf{5: 15,000 to 20,000 yen}} &
          poor : 57 &
          dated : 41 &
          worst : 36 &
          dirty : 14 &
          uncomfortable : 26 \\
         &
          poor service : 10 &
          outdated : 8 &
          worst hotel : 8 &
          dirty room : 2 &
          uncomfortable bed : 7 \\
         &
          poor breakfast : 6 &
           &
          worst experience : 3 &
           &
          uncomfortable pillow : 2 \\
         &
          poor hotel : 5 &
           &
          worst part : 2 &
           &
           \\
         &
          poor experience : 3 &
           &
          worst service : 2 &
           &
           \\ \hline
        \multirow{5}{*}{\textbf{6: 20,000 to 30,000 yen}} &
          poor : 136 &
          dated : 131 &
          worst : 86 &
          dirty : 67 &
          uncomfortable : 103 \\
         &
          poor breakfast : 15 &
          outdated : 31 &
          worst hotel : 11 &
          dirty room : 10 &
          uncomfortable bed : 24 \\
         &
          poor service : 14 &
          outdated room : 6 &
          worst part : 7 &
          dirty carpet : 8 &
          uncomfortable pillow : 11 \\
         &
          poor english : 9 &
          outdated hotel : 2 &
          worst breakfast : 5 &
          dirty bathroom : 3 &
          uncomfortable night : 4 \\
         &
          poor quality : 9 &
           &
          worst experience : 5 &
          dirty chair : 2 &
          uncomfortable experience : 3 \\ \hline
        \multirow{5}{*}{\textbf{7: 30,000 to 50,000 yen}} &
          poor : 92 &
          dated : 65 &
          worst : 64 &
          dirty : 51 &
          uncomfortable : 55 \\
         &
          poor service : 8 &
          outdated : 17 &
          worst hotel : 10 &
          dirty carpet : 11 &
          uncomfortable bed : 20 \\
         &
          poor breakfast : 7 &
          outdated hotel : 4 &
          worst room : 3 &
          dirty room : 7 &
          uncomfortable mattress : 6 \\
         &
          poor english : 7 &
          outdated bathroom : 2 &
          worst service : 3 &
          dirty clothe : 2 &
          uncomfortable pillow : 5 \\
         &
          poor connection : 5 &
          outdated decor : 2 &
          worst part : 2 &
          dirty luggage : 2 &
          uncomfortable room : 5 \\ \hline
        \multirow{5}{*}{\textbf{8: 50,000 to 100,000 yen}} &
          poor : 124 &
          dated : 150 &
          worst : 98 &
          dirty : 36 &
          uncomfortable : 33 \\
         &
          poor service : 16 &
          outdated : 58 &
          worst hotel : 9 &
          dirty carpet : 12 &
          uncomfortable bed : 7 \\
         &
          poor breakfast : 9 &
          outdated room : 9 &
          worst experience : 5 &
          dirty room : 3 &
           \\
         &
          poor quality : 9 &
          outdated furniture : 6 &
          worst part : 3 &
          dirty cup : 2 &
           \\
         &
          poor english : 6 &
          outdated hotel : 4 &
           &
          dirty rug : 2 &
           \\ \hline
        \multirow{5}{*}{\textbf{9: 100,000 to 200,000 yen}} &
          poor : 19 &
          dated : 3 &
          worst : 12 &
          dirty : 6 &
          uncomfortable : 8 \\
         &
          poor service : 4 &
          outdated : 2 &
          worst experience : 2 &
           &
          little uncomfortable : 2 \\
         &
          poor choice : 2 &
           &
           &
           &
           \\
         &
          poor experience : 2 &
           &
           &
           &
           \\
         &
           &
           &
           &
           &
           \\ \hline
        \end{tabular}%
        }
    \end{table}
    \end{landscape}

  \subsection{Determining hard and soft attribute usage}\label{det_hard_soft}

    To further understand the differences in satisfaction and dissatisfaction in Chinese and Western customers of Japanese hotels, we classified these factors as either hard or soft attributes of a hotel. We define hard attributes as matters regarding the hotel's physical or environmental aspects, such as facilities, location, or infrastructure. Some of these aspects would be impractical for the hotel to change, such as its surroundings and location. Others can be expensive to change, such as matters requiring construction costs, which are possible but would require significant infrastructure investment. On the other hand, soft attributes are the non-physical attributes of the hotel service and staff behavior that are practical to change through management. For example, the hotel's services or the cleanliness of the rooms are soft attributes. For our purposes, amenities, clean or good quality bed sheets or curtains, and other physical attributes that are part of the service and not the hotel's physical structure are considered soft attributes. Thus, we can observe the top 10 satisfaction and dissatisfaction keywords and determine whether they are soft or hard attributes.

\begin{figure}[ht]
      \centering
      \begin{subfigure}[b]{0.45\textwidth}
          \includegraphics[width=\textwidth]{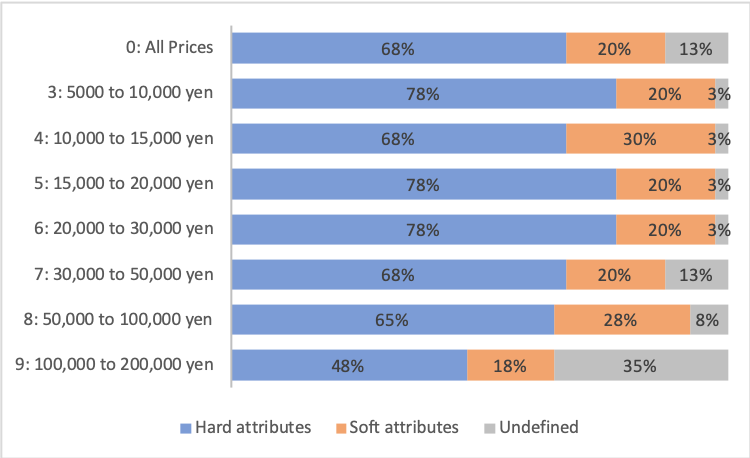}
          \caption{Positive keywords}
      \end{subfigure}
      \begin{subfigure}[b]{0.45\textwidth}
          \includegraphics[width=\textwidth]{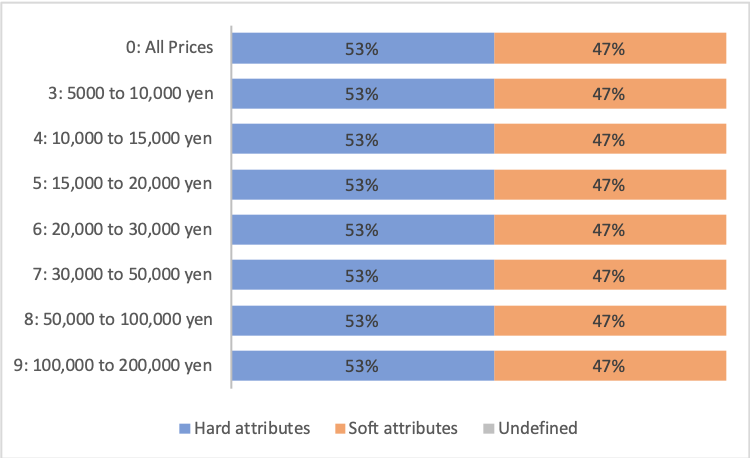}
          \caption{Negative keywords}
      \end{subfigure}
  \caption{Hard and soft attributes from the top Chinese keywords for all price ranges}
  \label{fig:hard_soft_zh}
  \end{figure}

\begin{figure}[ht]
      \centering
      \begin{subfigure}[b]{0.45\textwidth}
          \includegraphics[width=\textwidth]{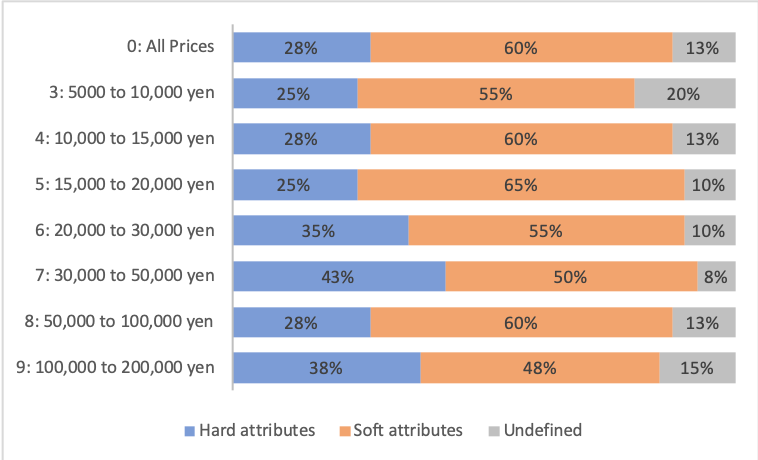}
          \caption{Positive keywords}
      \end{subfigure}
      \begin{subfigure}[b]{0.45\textwidth}
          \includegraphics[width=\textwidth]{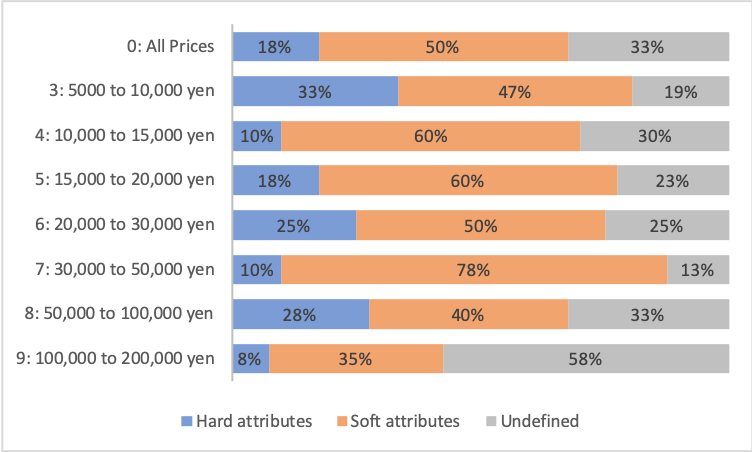}
          \caption{Negative keywords}
      \end{subfigure}
  \caption{Hard and soft attributes from the top English keywords for all price ranges}
  \label{fig:hard_soft_en}
  \end{figure}

    We manually labeled each language's top keywords into either hard or soft by considering how the word would be used when writing a review. If the word described unchangeable physical factors by the staff or management, we consider them hard. If the word implied an issue that could be solved or managed by the hotel staff or management, we consider it soft. For adjectives, we looked at the top four adjective and noun pairings used in the entire dataset and counted the usage percentage in each context. If it was not clear from the word or the pairing alone, we declared it undefined. Then, we added the counts of these words in each category. A single word with no pairing is always deemed 100\% in the category it corresponds to. We add the partial percentages for each category when an adjective includes various contexts. The interpretation of these keywords is shown in the Tables \ref{tab:zh_hard_soft_keywords} and \ref{tab:en_hard_soft_keywords}. We can see the summarized results for the hard and soft percentages of positive and negative Chinese keywords in Figure \ref{fig:hard_soft_zh}. For the English keywords, see Figure \ref{fig:hard_soft_en}.

  % \usepackage{multirow}
  % \usepackage{graphicx}
  \begin{table}[ht]
    \centering
    \caption{Determination of hard and soft attributes for Chinese keywords. }
    \label{tab:zh_hard_soft_keywords}
    \resizebox{0.7\textwidth}{!}{%
    \begin{tabular}{|c|l|l|}
      \hline
      \textbf{Keyword Emotion}                     & \multicolumn{1}{c|}{\textbf{Keyword}} & \multicolumn{1}{c|}{\textbf{Attribute Category}} \\ \hline
      \multirow{19}{*}{\textbf{Positive Keywords}} & \begin{CJK}{UTF8}{gbsn}不错\end{CJK}    & 50\% hard, 25\% soft, 25\% undefined             \\ \cline{2-3} 
       & \begin{CJK}{UTF8}{gbsn}大\end{CJK}   & 100\% hard                           \\ \cline{2-3} 
       & \begin{CJK}{UTF8}{gbsn}干净\end{CJK}  & 25\% hard, 75\% soft                 \\ \cline{2-3} 
       & \begin{CJK}{UTF8}{gbsn}早餐\end{CJK}  & 100\% soft                           \\ \cline{2-3} 
       & \begin{CJK}{UTF8}{gbsn}交通\end{CJK}  & 100\% hard                           \\ \cline{2-3} 
       & \begin{CJK}{UTF8}{gbsn}棒\end{CJK}   & 25\% hard, 50\% soft, 25\% undefined \\ \cline{2-3} 
       & \begin{CJK}{UTF8}{gbsn}近\end{CJK}   & 100\% hard                           \\ \cline{2-3} 
       & \begin{CJK}{UTF8}{gbsn}购物\end{CJK}  & 100\% hard                           \\ \cline{2-3} 
       & \begin{CJK}{UTF8}{gbsn}环境\end{CJK}  & 100\% hard                           \\ \cline{2-3} 
       & \begin{CJK}{UTF8}{gbsn}地铁\end{CJK}  & 100\% hard                           \\ \cline{2-3} 
       & \begin{CJK}{UTF8}{gbsn}卫生\end{CJK}  & 100\% soft                           \\ \cline{2-3} 
       & \begin{CJK}{UTF8}{gbsn}新\end{CJK}   & 50\% hard, 25\% soft, 25\% undefined \\ \cline{2-3} 
       & \begin{CJK}{UTF8}{gbsn}推荐\end{CJK}  & 100\% undefined                      \\ \cline{2-3} 
       & \begin{CJK}{UTF8}{gbsn}选择\end{CJK}  & 100\% undefined                      \\ \cline{2-3} 
       & \begin{CJK}{UTF8}{gbsn}地铁站\end{CJK} & 100\% hard                           \\ \cline{2-3} 
       & \begin{CJK}{UTF8}{gbsn}远\end{CJK}   & 100\% hard                           \\ \cline{2-3} 
       & \begin{CJK}{UTF8}{gbsn}附近\end{CJK}  & 100\% hard                           \\ \cline{2-3} 
       & \begin{CJK}{UTF8}{gbsn}周边\end{CJK}  & 100\% hard                           \\ \cline{2-3} 
       & \begin{CJK}{UTF8}{gbsn}赞\end{CJK}   & 100\% undefined                      \\ \hline
      \multirow{8}{*}{\textbf{Negative Keywords}}  & \begin{CJK}{UTF8}{gbsn}价格\end{CJK}    & 100\% soft                                       \\ \cline{2-3} 
       & \begin{CJK}{UTF8}{gbsn}一般\end{CJK}  & 50\% hard, 50\% soft                 \\ \cline{2-3} 
       & \begin{CJK}{UTF8}{gbsn}中文\end{CJK}  & 100\% soft                           \\ \cline{2-3} 
       & \begin{CJK}{UTF8}{gbsn}距离\end{CJK}  & 100\% hard                           \\ \cline{2-3} 
       & \begin{CJK}{UTF8}{gbsn}地理\end{CJK}  & 100\% hard                           \\ \cline{2-3} 
       & \begin{CJK}{UTF8}{gbsn}陈旧\end{CJK}  & 100\% hard                           \\ \cline{2-3} 
       & \begin{CJK}{UTF8}{gbsn}老\end{CJK}   & 75\% hard, 25\% soft                 \\ \cline{2-3} 
       & \begin{CJK}{UTF8}{gbsn}华人\end{CJK}  & 100\% soft                           \\ \hline
      \end{tabular}%
      }
  \end{table}

  % \usepackage{multirow}
  % \usepackage{graphicx}
  % \usepackage[normalem]{ulem}
  \begin{table}[ht]
    \centering
    \caption{Determination of hard and soft attributes for English keywords. }
    \label{tab:en_hard_soft_keywords}
    \resizebox{0.7\textwidth}{!}{%
    \begin{tabular}{|c|l|l|}
      \hline
      \textbf{Keyword Emotion}                     & \multicolumn{1}{c|}{\textbf{Keyword}} & \multicolumn{1}{c|}{\textbf{Attribute Category}} \\ \hline
      \multirow{18}{*}{\textbf{Positive Keywords}} & good                                  & 25\% hard, 50\% soft, 25\% undefined             \\ \cline{2-3} 
       & great          & 50\% hard, 25\% soft, 25\% undefined \\ \cline{2-3} 
       & staff          & 100\% soft                           \\ \cline{2-3} 
       & clean          & 100\% soft                           \\ \cline{2-3} 
       & location       & 100\% hard                           \\ \cline{2-3} 
       & nice           & 50\% hard, 25\% soft, 25\% undefined \\ \cline{2-3} 
       & excellent      & 25\% hard, 50\% soft, 25\% undefined \\ \cline{2-3} 
       & helpful        & 100\% soft                           \\ \cline{2-3} 
       & comfortable    & 25\% hard, 50\% soft, 25\% undefined \\ \cline{2-3} 
       & shopping       & 100\% hard                           \\ \cline{2-3} 
       & beautiful      & 25\% hard, 75\% soft                 \\ \cline{2-3} 
       & friendly       & 100\% soft                           \\ \cline{2-3} 
       & train          & 100\% hard                           \\ \cline{2-3} 
       & large          & 100\% hard                           \\ \cline{2-3} 
       & free           & 100\% soft                           \\ \cline{2-3} 
       & subway         & 100\% hard                           \\ \cline{2-3} 
       & recommend      & 100\% undefined                      \\ \cline{2-3} 
       & wonderful      & 50\% soft, 50\% undefined            \\ \hline
      \multirow{24}{*}{\textbf{Negative Keywords}} & pricey                                & 100\% soft                                       \\ \cline{2-3} 
       & worst          & 25\% hard, 50\% soft, 25\% undefined \\ \cline{2-3} 
       & dated          & 75\% hard, 25\% undefined            \\ \cline{2-3} 
       & poor           & 100\% soft                           \\ \cline{2-3} 
       & walkway        & 100\% hard                           \\ \cline{2-3} 
       & sense          & 100\% undefined                      \\ \cline{2-3} 
       & unable         & 100\% soft                           \\ \cline{2-3} 
       & disappointing  & 50\% soft, 50\% undefined            \\ \cline{2-3} 
       & minor          & 100\% undefined                      \\ \cline{2-3} 
       & worse          & 100\% undefined                      \\ \cline{2-3} 
       & annoying       & 75\% hard, 25\% undefined            \\ \cline{2-3} 
       & lighting       & 100\% soft                           \\ \cline{2-3} 
       & uncomfortable  & 100\% soft                           \\ \cline{2-3} 
       & carpet         & 100\% soft                           \\ \cline{2-3} 
       & dirty          & 75\% soft, 25\% undefined            \\ \cline{2-3} 
       & cigarette      & 100\% soft                           \\ \cline{2-3} 
       & funny smell    & 100\% soft                           \\ \cline{2-3} 
       & rude           & 100\% soft                           \\ \cline{2-3} 
       & smallest       & 75\% hard, 25\% undefined            \\ \cline{2-3} 
       & mixed          & 100\% undefined                      \\ \cline{2-3} 
       & renovation     & 100\% hard                           \\ \cline{2-3} 
       & paper          & 100\% undefined                      \\ \cline{2-3} 
       & disappointment & 100\% undefined                      \\ \cline{2-3} 
       & outdated       & 75\% hard, 25\% undefined            \\ \hline
      \end{tabular}%
      }
  \end{table}

\section{Results}\label{results}

  \subsection{Experimental results and answers to research questions}

    Our research questions were related to two issues. Based on research questions \ref{rsq:hospitality} and \ref{rsq:hospitality_both}, the objective of this study was to determine the differences in how Chinese and Western tourists perceive Japanese hotels, whose hospitality and service are influenced by the \textit{omotenashi} culture. 

    Observing the top-ranking positive keywords in Chinese reviews, as shown in Tables \ref{tab:freq_res_pos} and Table \ref{tab:adj_zh_pos}, it was revealed that, while service, cleanliness, and breakfast were praised in most hotels, the location was more important when observing the pairings. Hard attributes were abundant lower on the lists. The negative keywords in Table \ref{tab:freq_res_neg} indicate that a lack of a Chinese-friendly environment was perceived, although there were more complaints about hard attributes such as the building's age and the distance from other convenient spots. However, most complaints were about the hotel's price, which included all of the price ranges; therefore, the price was the primary concern for Chinese customers with different travel purposes.

    On the other hand, the word ``staff'' is the second or third in the lists of satisfaction factors in English-written reviews in all the price ranges. This word is followed by a few other keywords lower in the top 10 list, such as ``helpful'' or ``friendly''. When we look at the pairings of the top-ranked keyword ``good'' in Table \ref{tab:adj_en_pos}, we find that customers mostly praise the location, service, breakfast, or English availability. When we look at the negative keyword ``poor'' and its pairings in Table \ref{tab:adj_en_neg}, we see that it is also service-related concepts that the Western tourists are disappointed with. 

    We can also observe some keywords that are not considered by their counterparts. For example, English-speaking customers mentioned tobacco smell in many reviews. However, it was not statistically identified as a problem for their Chinese counterparts. On the other hand, although they appear in both English and Chinese lists, references to ``\begin{CJK}{UTF8}{gbsn}购物\end{CJK} (shopping)'' are more common in the Chinese lists across hotels of 15,000 yen to 200,000 yen per night. Meanwhile, the term ``shopping'' appeared solely in the top 10 positive keywords list for English speakers who stayed in rooms priced 20,000–30,000 yen per night.

    With these results, we can observe that both Chinese and English-speaking tourists in Japan have different priorities. However, both populations consider the hotel's location and transport availability (subways and trains) nearby as secondary but still essential points in their satisfaction with a hotel. The Chinese customers are primarily satisfied with the room quality in spaciousness and cleanliness and the service of breakfast.

    For research questions \ref{rsq:hard_soft} and \ref{rsq:hard_soft_diff}, we considered how customers of both cultural backgrounds evaluated the hard and soft attributes of hotels. Our study discovered that Chinese tourists mostly positively react to the hotel's hard attributes, albeit the negative evaluations are more uniform than the positive evaluations, with a tendency of 53 \% towards hard attributes. On the other hand, English-speaking tourists were more responsive to soft attributes, either positively or negatively. In the case of negative keywords, they were more concerned about the hotel's soft attributes.

    One factor that both populations had in common is that, when perceiving the hotel negatively, the ``\begin{CJK}{UTF8}{gbsn}老\end{CJK} (old),'' ``dated,'' ``outdated,'' or ``\begin{CJK}{UTF8}{gbsn}陈旧\end{CJK} (obsolete)'' aspects of the room or the hotel were surprisingly criticized across most price ranges. However, this is a hard attribute and is unlikely to change for most hotels.
    
  \subsection{Chinese tourists: A big and clean space}\label{disc:zh}

    We found that mainland Chinese tourists were mainly satisfied by big and clean spaces in Japanese hotels. The adjectival pairings extracted with dependency parsing and POS tagging (Table \ref{tab:adj_zh_pos})  imply big and clean rooms. Other mentions included big markets nearby or a big bed. Across different price ranges, the usage of the word ``\begin{CJK}{UTF8}{gbsn}大\end{CJK} (big)'' increased with the increasing price of the hotel. When inspecting closer by taking random samples of the pairs of ``\begin{CJK}{UTF8}{gbsn}大 空间\end{CJK} (big space)'' or ``\begin{CJK}{UTF8}{gbsn}大 面积\end{CJK} (large area),'' we notice that there were also many references to the public bathing facilities in the hotel. Such references were also implied by a word pairing ``\begin{CJK}{UTF8}{gbsn}棒 温泉\end{CJK} (great hot spring).'' 

    In Japan, there are the so-called ``\begin{CJK}{UTF8}{min}銭湯\end{CJK} (\textit{sent\=o}),'' which are artificially constructed public bathing facilities, including saunas and baths with unique qualities. On the other hand, there are natural hot springs, called ``\begin{CJK}{UTF8}{gbsn}温泉\end{CJK} (\textit{onsen}).'' However, they are interchangeable if natural hot spring water is used in artificially made tiled bath facilities. It is a Japanese custom that all customers first clean themselves in a shower and afterward use the baths nude. It could be a cultural shock for many tourists but a fundamental attraction for many others. 

    Chinese customers are satisfied with the size of the room or bed; however, it is not trivial to change this. In contrast, cleanliness is mostly related to soft attributes when we observe its adjectival pairings. We can observe pairs such as ``\begin{CJK}{UTF8}{gbsn}干净 房间\end{CJK} (clean room)'' at the top rank of all price ranges and thereupon ``\begin{CJK}{UTF8}{gbsn}干净 酒店\end{CJK} (clean hotel),'' ``\begin{CJK}{UTF8}{gbsn}干净 总体\end{CJK} (clean overall),'' ``\begin{CJK}{UTF8}{gbsn}干净 环境\end{CJK} (clean environment),'' and ``\begin{CJK}{UTF8}{gbsn}干净 设施\end{CJK} (clean facilities),'' among other examples. In negative reviews, there was a mention of criticizing the ``\begin{CJK}{UTF8}{gbsn}一般 卫生\end{CJK} (general hygiene)'' of the hotel, although it was an uncommon pair. Therefore, we can assert that cleanliness was an important soft attribute for Chinese customers, and they were mostly pleased when their expectations were fulfilled. 

    A key soft satisfaction factor was the inclusion of breakfast within the hotel. While other food-related words were extracted, most of them were general, such as ``food'' or ``eating,'' and were lower-ranking. In contrast, the word ``\begin{CJK}{UTF8}{gbsn}早餐\end{CJK} (breakfast),'' referring to the hotel commodities, was frequently used in positive texts compared to other food-related words across all price ranges, albeit at different priorities in each of them. For this reason, we regard it as an important factor. From the word pairs of the positive Chinese keywords in Table \ref{tab:adj_zh_pos}, we can also note that ``\begin{CJK}{UTF8}{gbsn}不错\end{CJK} (not bad)'' is paired with ``\begin{CJK}{UTF8}{gbsn}不错 早餐\end{CJK} (nice breakfast)'' in four of the seven price ranges with reviews available as part of the top four pairings. It is only slightly lower in other categories, although it is not depicted in the table. Thus, we consider that a recommended strategy for hotel management is to invest in the inclusion or improvement of hotel breakfast to increase good reviews.

  \subsection{Western tourists: A friendly face and absolutely clean}\label{disc:en}

    From the satisfaction factors of English-speaking tourists, we observed at least three words were related to staff friendliness and services in the general database: ``staff,'' ``helpful,'' and ``friendliness.'' The word ``staff'' is the highest-ranked of these three, ranking second for satisfied customers across most price ranges and only third in one of them. The word ``good'' mainly refers to the location, service, breakfast, or English availability in Table \ref{tab:adj_en_pos}. Similar to Chinese customers, Western customers also seemed to enjoy the included breakfasts regarding their satisfaction keyword pairings. However, the relevant word does not appear in the top 10 list directly, in contrast to their Chinese counterparts. The words ``helpful'' and ``friendly'' are mostly paired with ``staff,'' ``concierge,'' ``desk,'' and ``service.'' By considering the negative keyword ``poor'' and its pairings in Table \ref{tab:adj_en_neg}, we realized once again that Western tourists were disappointed with service-related concepts and reacted negatively.

    Another soft attribute that is high on the list for most of the price ranges is the word ``clean'', so we examined its word pairings. Customers largely praised ``clean rooms'' and ``clean bathrooms'' and also referred to the hotel in general. When observing the negative keyword frequencies for English speakers, we can find words such as ``dirty'' and ``carpet'' as well as word pairings such as ``dirty carpet,'' ``dirty room,'' and ``dirty bathroom.'' Along with complaints about off-putting smells, we could conclude that Western tourists had high expectations about cleanliness when traveling in Japan.

    An interesting detail of the keyword ranking is that the word ``comfortable'' was high on the satisfaction factors, and ``uncomfortable'' was high on the dissatisfaction factors. The words were paired with nouns such as ``bed,'' ``room,'' ``pillow,'' and ``mattress,'' when they generally referred to their sleep conditions in the hotel.
    It seems that Western tourists were particularly sensitive about the hotels’ comfort levels and whether they reached their expectations. The ranking for the negative keyword ``uncomfortable'' is similar across most price ranges except the two most expensive ones, where this keyword disappears from the top 10 list.

    Albeit lower in priority, the price range of 15,000 to 20,000 yen hotels also includes ``free'' as one of the top 10 positive keywords, mainly paired with ``Wi-Fi.'' This price range corresponds to business hotels, where users would expect this feature the most. 

  \subsection{Tobacco, an unpleasant smell in the room}\label{disc:tobacco}

    A concern for Western tourists was uncleanliness and the smell of cigarettes in their room, which can be regarded as soft attributes. Cigarette smell was an issue even in the middle- and high-class hotels, of which the rooms were priced at more than 30,000 yen per night. For hotels with rooms priced above 50,000 yen per night, however, this problem seemed to disappear from the list of top 10 concerns. Tobacco was referenced singularly as ``cigarette'', but also in word pairs in Table \ref{tab:adj_en_neg} as ``funny smell.'' By manually inspecting a sample of reviews with this keyword, we noticed that the room was often advertised as non-smoking; however, the smell permeated the room and curtains. Another common complaint was that there were no nonsmoking facilities available. The smell of smoke can completely ruin some customers’ stay, leading to bad reviews, thereby lowering the number of future customers.

    In contrast, Chinese customers seemed not to be bothered by this. Previous research has stated that 49–60 \% of Chinese men (and 2.0–2.8 \% of women) currently smoke or smoked in the past. This was derived from a sample of 170,000 Chinese adults in 2013–2014, which is high compared to many English-speaking countries \cite[][]{zhang2019tobacco,who2015tobacco}.

    Japan has a polarized view on the topic of smoking. Although it has one of the world’s largest tobacco markets, tobacco use has decreased in recent years. Smoking in public spaces is prohibited in some wards of Tokyo (namely Chiyoda, Shinjuku, and Shibuya). However, it is generally only suggested and not mandatory to lift smoking restrictions in restaurants, bars, hotels, and public areas. Many places have designated smoking rooms to keep the smoke in an enclosed area and avoid bothering others.

    Nevertheless, businesses, especially those who cater to certain customers, are generally discouraged by smoking restrictions if they want to maintain their clientele. To cater to all kinds of customers, including Western and Asian, Japanese hotels must provide spaces without tobacco smell. Even if the smoke does not bother a few customers, the lack of such a smell will make it an appropriate space for all customers.

  \subsection{Location, location, location}\label{disc:location}

    The hotel's location, closeness to the subway and public transportation, and availability of nearby shops proved to be of importance to both Chinese and English-speaking tourists. In positive word pairings in Tables \ref{tab:adj_zh_pos} and \ref{tab:adj_en_pos}, we can find pairs such as ``\begin{CJK}{UTF8}{gbsn}不错 位置\end{CJK} (nice location),'' ``\begin{CJK}{UTF8}{gbsn}近 地铁站\end{CJK} (near subway station),'' ``\begin{CJK}{UTF8}{gbsn}近 地铁\end{CJK} (near subway)'' in Chinese texts and ``good location,'' ``great location,'' and ``great view'' as well as single keywords ``location'' and ``shopping'' for English speakers, and ``\begin{CJK}{UTF8}{gbsn}交通\end{CJK} (traffic),'' ``\begin{CJK}{UTF8}{gbsn}购物\end{CJK} (shopping),'' ``\begin{CJK}{UTF8}{gbsn}地铁\end{CJK} (subway),'' and ``\begin{CJK}{UTF8}{gbsn}环境\end{CJK} (environment or surroundings)'' for Chinese speakers. All of these keywords and their location in each population's priorities across the price ranges signify that the hotel's location was a secondary but still important point for their satisfaction. However, since this is a hard attribute, it is not often considered in the literature. By examining examples from the data, we recognized that most customers were satisfied if the hotel was near at least two of the following facilities: subway, train, and convenience stores. 

    Japan is a country with a peculiar public transportation system. During rush hour, the subway is crowded with commuters, and trains and subway stations create a confusing public transportation map for a visitor in Tokyo. Buses are also available, albeit less used than rail systems in metropolitan cities. These three means of transportation are usually affordable in price. There are more expensive means, such as the bullet train \textit{shinkansen} for traveling across the country and taxis. The latter is a luxury in Japan compared to other countries. In Japan, taxis provide a high-quality experience with a matching price. Therefore, for people under a budget, subway availability and maps or GPS applications, as well as a plan to travel the city, are of utmost necessity for tourists, using taxis only as a last resort. 

    Japanese convenience stores are also famous worldwide because they offer a wide range of services and products, from drinks and snacks to full meals, copy and scanning machines, alcohol, cleaning supplies, personal hygiene items, underwear, towels, and international ATMs. If some trouble occurs, or a traveler forgot to pack a particular item, it is most certain that they can find it. 

    Therefore, considering that both transportation systems and nearby shops are points of interest for Chinese and Western tourists, and perhaps offering guide maps and information about these as an appeal point could result in greater satisfaction.

\section{Discussion}\label{discussion}

  \subsection{Western and Chinese tourists in the Japanese hospitality environment}\label{disc:omotenashi}

    To date, scholars have been correcting our historical bias towards the West. Studies have determined that different cultural backgrounds lead to different expectations, which influence tourists' satisfaction. In other words, tourists of a particular culture have different leading satisfaction factors across different destinations. However, Japan presents a particular environment; the spirit of hospitality and service, \textit{omotenashi}, which is considered to be of the highest standard across the world. Our study explores whether such an environment can affect different cultures equally or whether it is attractive only to certain cultures.

    Our results indicate that Western tourists are more satisfied with soft attributes than Chinese tourists. As explained earlier in this paper, Japan is well known for its customer service. Respectful language and bowing are not exclusive to high-priced hotels or businesses; these are met in convenience stores as well. Even in the cheapest convenience store, the level of hospitality is starkly different from Western culture and perhaps unexpected. In higher-priced hotels, the adjectives used to praise the service ranged from normal descriptors like ``good'' to higher levels of praise like ``wonderful staff,'' ``wonderful experience,'' ``excellent service,'' and ``excellent staff.'' Furthermore, \cite{kozak2002} and \cite{shanka2004} have also proven that hospitality and staff friendliness are two determinants of Western tourists' satisfaction.

    However, the negative English keywords indicate that a large part of the dissatisfaction with Japanese hotels stemmed from a lack of hygiene and room cleanliness. Although Chinese customers had solely positive keywords about cleanliness, English-speaking customers deemed many places unacceptable to their standards, particularly hotels with rooms priced below 50,000 yen per night. The most common complaint regarding cleanliness was about the carpet, followed by complaints about cigarette smell and lack of general hygiene. \cite{kozak2002} also proved that hygiene and cleanliness were essential satisfaction determinants for Western tourists. However, in the previous literature, this was linked merely to satisfaction. In contrast, our research revealed that words related to cleanliness were mostly linked to dissatisfaction. We could assert that Westerners had a high standard of room cleanliness compared to their Chinese counterparts.

    According to previous research, Western tourists are already inclined to appreciate hospitality for their satisfaction. When presented with Japanese hospitality, this expectation is met and overcome. In contrast, according to our results, Chinese tourists were more concerned about room quality rather than hospitality, staff, or service. However, when analyzing the word pairs for ``\begin{CJK}{UTF8}{gbsn}不错\end{CJK} (not bad)'' and ``\begin{CJK}{UTF8}{gbsn}棒\end{CJK} (great),'' we can see that they praise staff, service, and breakfast. By observing the percentage of hard to soft attributes in Figure \ref{fig:hard_soft_zh}, however, we discover that Chinese customers were more satisfied with hard attributes compared to Western tourists, who seemed to be meeting more than their expectations.

    It could be considered that Chinese culture does not expect high-level service initially. When an expectation that is not held is met, the satisfaction derived is less than that if it was expected. In contrast, some tourists report a ``nice surprise'': when an unknown need is unexpectedly met, there is more satisfaction. It is necessary to note the difference between these two reactions. The ``nice surprise'' reaction fulfills a need unexpectedly. Perhaps the hospitality grade in Japan does not fulfill a need high enough for the Chinese population, thereby resulting in less satisfaction. For greater satisfaction, a need must be met. However, the word ``not bad'' is at the top of the list in most price ranges, and one of the uses is related to service. Thus, we cannot conclude that they were not satisfied with the service. Instead, they held other factors at a higher priority; thus, the keyword frequency was higher for other pairings.

    Another possibility occurs when we observe the Chinese tourists’ dissatisfaction factors. Chinese tourists may have expectations about their treatment that are not being met, even in this high-standard hospitality environment. This could be because Japan is monolingual and has a relatively large language barrier to tourists \cite[][]{heinrich2012making,coulmas2002japan}. While the Japanese effort to accommodate English speakers is slowly developing, efforts for Chinese accommodations can be lagging. Chinese language pamphlets and Chinese texts on instructions for the hotel room and its appliances and features (e.g., T.V. channels, Wi-Fi setup, etc.), or the treatment towards Chinese people, could be examples of these accommodations. \cite{ryan2001} also found that communication difficulty was one of the main reasons Chinese customers would state for not visiting again. However, this issue is not exclusive to Japan.

    Our initial question was whether the environment of high-grade hospitality would affect both cultures equally. This study attempted to determine the answer. It is possible that Chinese customers had high-grade hospitality and were equally satisfied with Westerners. In that case, it appears that the difference in perception stems from a psychological source; expectation leads to satisfaction and a lack of expectation results in lesser satisfaction. There is also a possibility that Chinese customers are not receiving the highest grade of hospitality because of cultural friction between Japan and China.

    It is unclear which of these two is most likely from our results. However, competing in hospitality and service includes language services, especially in the international tourism industry. Better multilingual support can only improve the hospitality standard in Japan. Considering that most of the tourists in Japan come from other countries in Asia, multilingual support is beneficial. Proposals for this endeavor include hiring Chinese-speaking staff, preparing pamphlets in Chinese, or having a translator application readily available with staff trained in interacting through an electronic translator.

  \subsection{Hard vs. soft satisfaction factors}\label{disc:hard_soft}

    As stated in section \ref{theory_satisfaction}, previous research has mostly focused on the hotel's soft attributes and their influence on customer satisfaction \cite[e.g.,][]{shanka2004,choi2001}. Examples of soft attributes include staff behavior, commodities, amenities, and appliances that can be improved within the hotel. However, hard attributes are not usually analyzed in satisfaction studies. It is important to consider both kinds of attributes. If the satisfaction was based on soft attributes, a hotel can improve its services to attract more customers in the future. Otherwise, if the satisfaction was related more to hard attributes overall, hotels should be built considering the location while minimizing other costs. Because the satisfaction factors were decided statistically in our study via customers’ online reviews, we can see the importance of the hard or soft attributes in their priorities.

    Figure \ref{fig:hard_soft_zh} shows that, in regards to Chinese customer satisfaction, in general, 68 \% of the top 10 keywords are hard factors; in contrast, only 20 \% are soft factors. The rates are similar for most price ranges except the highest-priced hotels. However, two of these soft attributes are all concentrated at the top of the list (``\begin{CJK}{UTF8}{gbsn}不错\end{CJK} (not bad),'' ``\begin{CJK}{UTF8}{gbsn}干净\end{CJK} (clean)''), and the adjective pairs related to soft attributes of ``\begin{CJK}{UTF8}{gbsn}不错\end{CJK} (not bad)'' are also at the top in most price ranges. Chinese tourists may expect spaciousness and cleanliness when coming to Japan. The expectation may be due to reputation, previous experiences, or cultural backgrounds. We can compare these results with previous literature, where traveling Chinese tourists choose their destination based on several factors, including cleanliness, nature, architecture, and scenery \cite[][]{ryan2001}. These factors found in previous literature could be linked to the keyword ``\begin{CJK}{UTF8}{gbsn}环境\end{CJK} (environment or surroundings)'' as well. This keyword was found for hotels priced at more than 20,000 yen per night. 

    In contrast, English speakers are mostly satisfied with the hotels' soft attributes. Figure \ref{fig:hard_soft_en} shows that soft attributes are above 48 \% in all price ranges, the highest being 65 \% in the price range of 15,000 to 20,000 yen per night, which corresponds to, for example, affordable business hotels. The exception to this is the hard attribute that is the hotels' location, which is consistently around the middle of the top 10 lists for all price ranges. 

    For both customer groups, the main reason for dissatisfaction was pricing, which can be interpreted as a concern about value for money. However, English-speaking customers complained less about the price in lower-priced hotels. In contrast, Chinese customers consistently had ``\begin{CJK}{UTF8}{gbsn}价格\end{CJK} (price)'' as the first or second-most concern across all price ranges. A study on Chinese tourists found that they had this concern \cite[][]{truong2009}. However, our results indicate that this has more to do with the pricing of hotels in Japan than with Chinese culture. In general, Japan is an expensive place to visit, thereby impacting this placement in the ranking. Space is scarce in Japan, and capsule hotels with cramped spaces of 2 x 1 meters cost around 3000 to 6000 yen per night. Bigger business hotel rooms are relatively expensive, ranging from 5000 to 12,000 yen per night. For comparison, hotels in the USA with a similar quality can charge half the price.

    Around half of the dissatisfaction factors for both Chinese and Western customers are caused by issues that could be improved; this is true for all price ranges. The improvements could be staff training (perhaps in language), hiring professional cleaning services for rooms with cigarette smoke smells, or improving the bedding; however, these considerations can be costly. However, once the hotel's location and construction are set, only a few changes can be made to satisfy Chinese customers further. As mentioned previously, Chinese language availability is a soft attribute that can be improved with staff and training investment.

    Western tourists are mainly dissatisfied with soft attributes. This is revealed by a low satisfaction level of 35 \% in the highest price range where undefined factors are the majority and a maximum of 78 \% in the price range of 30,000 to 50,000 yen per night in a hotel. Improvement scope for Western tourists is more extensive than that for their Chinese counterparts. As such, it presents a larger investment opportunity. 

  \subsection{Satisfaction across different price ranges}\label{disc:price}

    In previous sections of this paper, we mentioned the differences reflected in hotel price ranges. The most visible change across differently priced hotels is the change in voice when describing satisfaction. We noticed this by observing the adjective-noun pairs and finding pairs with different adjectives for the same nouns. For example, in English, words describing nouns such as ``location'' or ``hotel'' are ``good'' or ``nice'' in lower-priced hotels. In contrast, the adjectives that pair with the same nouns for higher-priced hotels are ``wonderful'' and ``excellent.'' In Chinese, the change ranges from ``\begin{CJK}{UTF8}{gbsn}不错\end{CJK} (not bad)'' to ``\begin{CJK}{UTF8}{gbsn}棒\end{CJK} (great)'' or ``\begin{CJK}{UTF8}{gbsn}赞\end{CJK} (awesome).'' We can infer that the level of satisfaction is high and influences how customers write their reviews. Regarding the negative keywords, however, the change ranges from ``annoying'' or ``disappointing'' to ``worst.''

    In this paper, we follow the definition of satisfaction by \cite{hunt1975}, where meeting or exceeding expectations produces satisfaction. Conversely, the failure of meeting expectations causes dissatisfaction. We can assume that a customer that pays more for a higher-class experience has higher expectations. For example, in a highly-priced hotel, any lack of cleanliness can lead to disappointment. In the case of English-speaking customers in the 30,000-–50,000 yen per night price range, cigarette smell is particularly disappointing. However, we consistently see customers with high expectations for high-class hotels reacting even more positively when satisfied. In the positive case, expectations appear to be exceeded in most cases, judging from their reactions. 

    We argue that these are two different kinds of expectations: logical and emotional. In the first case, customers are determined that the service must not fall below a specific standard; for example, they can be disappointed with unhygienic rooms or cigarette smell. In contrast, in the second case, customers have a vague idea of having a positive experience but do not measure it against any standard. For example, they expect a pleasant customer service experience or a hospitable treatment by the staff at a high-class hotel. Regardless of their knowledge in advance, positive emotions offer them a perception of exceeded expectations and high satisfaction. Thus, hospitality and service enhance the experience of the customers. 

    There are interesting differences between Chinese and English-speaking tourists in their satisfaction to differently priced hotels. For example, Chinese tourists have ``\begin{CJK}{UTF8}{gbsn}购物\end{CJK} (shopping)'' as a top keyword in all the price ranges. In contrast, English-speaking tourists mention it only as a top keyword in the 20,000–-30,000 yen price range. It is widely known in Japan that many Chinese tourists visit Japan for shopping. \cite{tsujimoto2017purchasing} analyzed the souvenir purchasing behavior of Chinese tourists in Japan and showed that common products besides food and drink are: electronics, cameras, cosmetics, and medicine, among \textit{souvenir} items representative of the culture or places that they visit \cite{japan2014consumption}. Furthermore, Chinese tourists’ choice to shop in Japan is more related to the quality of the items rather than their relation to the tourist attractions. Our results suggested that Western tourists were engaging more in tourist attractions rather than shopping activities compared to Chinese tourists. 

    Another interesting difference is that English-speaking tourists start using negative keywords about the hotel's price only if it concerns hotels of 15,000 yen or more; thereafter, the more expensive the hotel, the higher the ranking. In contrast, for Chinese customers, this keyword is the top keyword across all price ranges. Previous research suggests that value for money is a key concern for Chinese and Asian tourists \cite[][]{choi2000,choi2001,truong2009}, whereas Western customers are more concerned about hospitality \cite[][]{kozak2002}.

    While some attributes' value changes depending on the hotel's price range, some other attributes remain constant for each culture's customers. For example, appreciation for staff from English-speaking tourists is ranked close to the top satisfaction factor in all the price ranges. Satisfaction for cleanliness by both cultures constantly remains part of the top 10 keywords, except for the most expensive one, where other keywords replace keywords related to satisfaction or cleanliness in the ranking; however, they remain still high on the list. Chinese tourists have a high ranking for the word ``\begin{CJK}{UTF8}{gbsn}早餐\end{CJK} (breakfast)'' across all price ranges as well. As discussed in section \ref{disc:location}, transportation and location are also important for hotels of all classes and prices. While the ranking of attributes might differ between price ranges, hard and soft attribute proportions also appear to be constant within a 13 \% margin of error per attribute. This suggests that, from a cultural aspect, the customers have a particular bias to consider some attributes more than others.

  \subsection{Cross-culture analysis of expectations and satisfaction}\label{disc:culture}

    The basic premise of this study is that different cultures lead to different expectations and satisfaction factors. This premise also plays a role in the differentiation between the preferences of hard or soft attributes.

    In \cite{donthu1998cultural}, subjects from 10 different countries were compared with respect to their expectations of service quality and analyzed based on Hofstede's typology of culture \cite[][]{hofstede1984culture}. The previous study states that, although culture has no specific index, five dimensions of culture can be used to analyze or categorize a country in comparison to others. These are \textit{power distance}, \textit{uncertainty avoidance}, \textit{individualism–-collectivism}, \textit{masculinity–femininity}, and \textit{long-term-–short-term orientation}. In each of these dimensions, at least one element of service expectations was found to be significantly different for countries grouped under contrasting attributes (e.g., individualistic countries vs. collectivist countries, high uncertainty avoidance countries vs. low uncertainty avoidance countries). 

    However, Hofstede's typology has received criticism from academics, particularly for the fifth dimension that Hofstede proposed, which was later added with the alternative name \textit{Confucian dynamic}. Academics with a Chinese background criticized Hofstede for being misinformed on the philosophical aspects of Confucianism as well as considering a difficult dimension to measure \cite[][]{fang2003critique}. Other models, such as the GLOBE model, also consider some of Hofstede's dimensions and replace them with others, making a total of nine dimensions \cite[][]{house1999cultural}. The \textit{masculinity–-femininity} dimension, for example, is proposed to be instead of two dimensions: \textit{gender egalitarianism} and \textit{assertiveness}. This addition of dimensions avoids assuming that assertiveness is either masculine or feminine, which stems from outdated gender stereotypes. Such gender stereotypes have also been the subject of critique on Hofstede's model\cite[][]{jeknic2014gender}. We agree with these critiques and thus avoid considering such stereotypes in our discussion.

    For our purposes of contrasting Western vs. Chinese satisfaction stemming from expectations, these dimensions could explain why Chinese customers are generally satisfied more often with hard factors while Westerners are satisfied or dissatisfied with soft factors. 

    The backgrounds of collectivism in China and individualism in Western countries have been studied previously \cite[][]{gao2017chinese, kim2000}. These backgrounds as well as the differences in these cultural dimensions could be the underlying cause for differences in expectations. Regardless of the cause, however, measures in the past have proven that such differences exist \cite[][]{armstrong1997importance}. 

    The cultural background of Chinese tourists emphasizes their surroundings and their place in nature and the environment. Chinese historical backgrounds of Confucianism, Taoism, and Buddhism permeate the thought processes of Chinese populations. However, scholars argue that the changes in generations and their economic and recent history attaches less importance to these concepts in their lives \cite[][]{gao2017chinese}. Nevertheless, one could argue a Chinese cultural attribute emphasizes that the environment and the location affect satisfaction rather than the treatment they receive. 

    A more anthropocentric and individualistic Western culture could correlate more of their expectations and priorities to the treatment in social circumstances rather than the environment. According to \cite{donthu1998cultural}, highly individualistic customers, in contrast to collectivist customers, have a higher expectation of empathy and assurance from the provider, which are aspects of service, a soft attribute of a hotel.

    Among other dimensions in both models, we can consider uncertainty avoidance. Customers of high uncertainty avoidance carefully plan their travel and thus have higher expectations towards service. In contrast, customers of lower uncertainty avoidance do not take risks in their decisions and thus face less disappointment with different expectations. However, according to \cite{xiumei2011cultural}, the difference between China and the USA in uncertainty avoidance is not clear when measuring with the Hofstede typology and the GLOBE typology. While the USA is not representative of Western society, uncertainty avoidance may not cause the difference in hard-soft attribute satisfaction between Chinese and Western cultures. Differences in another factor, power distance, were also noted when using  Hofstede's method compared to the GLOBAL method; therefore, power distance was not considered for comparison.

  \subsection{Implications for hotel managers}\label{disc:implications}

    Our study reached two important conclusions: one about hospitality and cultural differences and another about managerial decisions towards two different populations. Overall, Chinese tourists did not attach much importance to hospitality and service factors. Instead, they focused on the hard attributes of a hotel. In particular, they were not satisfied with hospitality as much as Western tourists were; otherwise, they felt that basic language and communication needs were not met; thereby, they were not much satisfied. Western tourists were highly satisfied with Japanese hospitality and preferred soft attributes to hard ones. 

    The other conclusion is that managerial decisions could mostly benefit Western tourists, except for language improvements and breakfast inclusion could satisfy both groups. As mentioned earlier in this paper, Westerners are ``long-haul'' customers, spending more of their budget on lodging than Asian tourists \cite[][]{choi2000}. With bigger returns on managerial improvements, we recommend investing in improving attributes that dissatisfy Western customers, such as cleanliness and removing tobacco smell. In addition, breaking the language barrier is one of the few strategies to satisfy both groups. Recently, Japan has been facing an increase in Chinese students as well as students of Western universities. Hiring students as part-time workers could increase the language services of a hotel.

    To satisfy both customer types, hotel managers need to invest in cleanliness, deodorizing, and making hotel rooms tobacco-free. It could also be recommended to invest in breakfast inclusion and multilingual services and staff preparedness to deal with Chinese and English speakers. Western tourists were also observed to have high comfort standards, which could be managerially improved for better reviews. Perhaps it could be suggested to perform surveys of the bedding that is most comfortable for Western tourists. However, not all hotels can invest in all of these factors simultaneously. Our results suggest that satisfying cleanliness needs could satisfy both customer types. We suggest investing in making the facilities tobacco-free. Our results are also divided by price ranges; thereby, a hotel manager could consider which analysis suits their hotel the most. Hard attributes are difficult to change; however, improvements in service can be made to accompany these attributes. For example, transportation guides for foreigners that might not know the area could increase satisfaction.

    The managers must consider their business model for implementing the next strategy. One option could be attracting more Chinese customers with their observed low budgeting. Another could be attracting more big-budget Western customers. For example, investing more in cleanliness could improve Western customers looking for high-quality lodging satisfaction, even for an increased price per night. On the other hand, hotels might be deemed costly by Chinese customers wherever such an investment is made.

\section{Limitations and Future Work}\label{limitations}

  In this study, we analyzed keywords based on whether they appeared on satisfied reviews or dissatisfied ones. Following that, we attempted to understand these words' context by using a dependency parser and observing the related nouns. However, a limitation is that it analyzed solely the words directly related to each keyword and did not search for further connections. This means that if the words were used in combination with other keywords, we did not trace the effects of multiple contradicting statements. For example, in the sentence ``The room is good, but the food is lacking,'' we extracted ``good room'' and ``lacking food'' but did not consider the fact that both occurred in the same sentence.

  This study analyzed the differences in customers' expectations at different levels of hospitality and service factors by dividing our data into price ranges. However, in the same price range, for example, the highest one, we can find both a Western-style five-star resort and a high-end Japanese style \textit{ryokan}. Services offered in these hotels are of high quality, albeit very different. Nevertheless, most of our database was focused on the middle range priced hotels, the services of which are comparably less varied. 

  An essential aspect of this study is that we focused on the satisfaction and dissatisfaction towards the expectations of individual aspects of the hotels. This gave us insight into the factors that hotel managers can consider. However, each customer's overall satisfaction was not measured since it would require methods that are out of the scope of this paper. Another limitation is that further typology analysis could not be made because of the nature of the data collected (for example, Chinese men and women of different ages or their Westerner counterparts).

  In future work, we plan to investigate these topics further. We plan to extend our data to research different trends and regions of Japan, different kinds of hotels, and customers traveling alone or in groups, whether for fun or for work. Another point of interest in this study's future work is to use word clusters with similar meanings instead of single words. 

\section{Conclusion}\label{conclusion}

  In this study, we analyzed the differences in satisfaction and dissatisfaction between Chinese and English-speaking customers of Japanese hotels, particularly in the context of Japanese hospitality, \textit{omotenashi}. We extracted keywords from their online reviews on \textit{Ctrip} and \textit{TripAdvisor} using Shannon's entropy calculations. We used these keywords for sentiment classification via an SVC. We then used dependency parsing and part of speech tagging to extract common pairs of adjectives and nouns as well as single words. We divided these data by sentiment and hotel price range (most expensive room/night). 

  We found that Western tourists were most satisfied with staff behavior, cleanliness, and other soft attributes. However, Chinese customers had other concerns for their satisfaction; they were more inclined to praise the room, location, and hotel's convenience. We found that the two cultures had different reactions to the hospitality environment and the prices. Thus, we discussed two possible theories on why Chinese tourists responded differently from Westerners in the environment of \textit{omotenashi}. One theory is that, although they were treated well, their experience was deteriorated by language or culture barriers. The second possible theory is that they reacted to hospitality differently since they did not have the same expectations. We theorized that a lack of expectations could result in lessened satisfaction than that to the same service if expected. On the other hand, even when they held high expectations in a high-priced hotel, Japanese hospitality exceeded Western tourists’ expectations, judging by their vocabulary for expressing their satisfaction. We considered that Western tourists were more reactive to hospitality and service factors Chinese tourists.

  Lastly, we measured the satisfaction and dissatisfaction factors, that is, a hotel's hard and soft attributes. Hard attributes are physical and environmental elements, and as such, are impractical elements to change. In contrast, soft attributes can be changed via management and staff by an improvement in services or amenities. We found that, for satisfaction, Western tourists favored soft attributes in contrast to Chinese tourists, who were more interested in the hard attributes of hotels across all the price ranges consistently. For dissatisfaction, Western tourists were also highly inclined to criticize soft attributes, such as cleanliness or cigarette smell in rooms. In contrast, Chinese tourists' dissatisfaction derived from both hard and soft attributes evenly.

  One approach for hotel managers is to work to satisfy Chinese tourists more, who dedicate a lower percentage of their budget to hotels but are more numerous. They are less satisfied with soft attributes but have an identifiable method for improving satisfaction by lessening language barriers and providing a satisfactory breakfast. Another approach was focused on the cleanliness, comfort, and tobacco-free space expected by Western tourists. ``Long-haul'' Western tourists, who spend almost half of their budget on hotels with this strategy, were favored. Although Westerners are less in number than Chinese tourists, it could be proven that they have more substantial returns. This is because Chinese customers also favor cleanliness as a satisfaction factor, and both populations could be pleased.

\begin{acknowledgements}

  During our research, we received the commentary and discussion by our dear colleagues necessary to understand particular cultural aspects that could influence the data's interpretation. We would like to show gratitude to Mr. Liangyuan Zhou, Ms. Min Fan, and Ms. Eerdengqiqige for this. 

  We would also like to show gratitude to Ms. Aleksandra Jajus, from whom we also received notes on the editing and commentary on the content of our manuscript.

  Funding: This work was supported by the Japan Construction Information Center Foundation (JACIC).

  Conflict of interest: none

\end{acknowledgements}

% \section*{References}

\bibliography{bibfile-emotion}

\end{document}